\documentclass[a4paper,UKenglish,cleveref, autoref, thm-restate]{tgdk-v2021}
\usepackage[utf8]{inputenc}

\PassOptionsToPackage{dvipsnames}{xcolor}

\usepackage{amsmath}  % Required for matrix environments
\usepackage{etoolbox} % Optional but sometimes needed by tcolorbox
\usepackage{xparse}   % For advanced macros
\usepackage[most]{tcolorbox} % Activates all tcolorbox features

\usepackage{algorithm}
\usepackage{algpseudocode}
\usepackage{graphicx}

\usepackage{textcomp}
\usepackage{xcolor}
\usepackage{listings}

\usepackage{scalerel}

\usepackage{pdflscape}

\usepackage{caption}
\usepackage{subcaption}

\usepackage{amsmath}

\usepackage{tikz}
\usetikzlibrary{arrows}
\usetikzlibrary{backgrounds}
\usetikzlibrary{matrix}
\usetikzlibrary{positioning}
\usetikzlibrary{svg.path}
\usetikzlibrary{shapes}
\usetikzlibrary{plotmarks}
\usetikzlibrary{shapes.geometric}
\usetikzlibrary{calc}
\usetikzlibrary{spy}

\usepackage{adjustbox}
\usepackage{bbding}

\usepackage[T1]{fontenc}

\usepackage{upquote}
\usepackage{amssymb}
\usepackage{booktabs}
\usepackage{makecell}
\usepackage[normalem]{ulem}

\usepackage{collcell}

\usepackage{xurl}

\usepackage{siunitx}

\makeatletter
\@ifundefined{citep}{%
  \DeclareRobustCommand{\citep}{\cite}%
}{}
\makeatother
\graphicspath{{./figures/}}%helpful if your graphic files are in another directory

\newcommand{\FeaturesReferenceDate}{2026-01-01}

\newcommand{\rot}[1]{\rotatebox[origin=c]{90}{\scriptsize\strut #1}}

\providecommand{\rowcat}[1]{%
  \rotatebox[origin=c]{90}{\scriptsize\bfseries\shortstack[c]{#1}}%
}
\providecommand{\rowsubcat}[1]{%
  \rotatebox[origin=c]{90}{\scriptsize\shortstack[c]{#1}}%
}
\newcommand{\pie}[1]{%
\begin{tikzpicture}
  \draw (0,0) circle (1ex);
  \ifnum#1=90
    \fill (0,0) -- (1ex,0) arc (0:90:1ex) -- cycle;
  \else\ifnum#1=180
    \fill (0,0) -- (1ex,0) arc (0:180:1ex) -- cycle;
  \else\ifnum#1=270
    \fill (0,0) -- (1ex,0) arc (0:270:1ex) -- cycle;
  \else\ifnum#1=360
    \fill (0,0) -- (1ex,0) arc (0:360:1ex) -- cycle;
  \fi\fi\fi\fi
\end{tikzpicture}%
}

\newlength\MAX  
\newcommand*\Chart[2]{#2 #1~\rlap{\textcolor{black!20}{\rule{\MAX}{2ex}}}\rule{#1\MAX}{2ex}}

\newcommand{\dbAllegroGraph}{\texttt{Al\-le\-gro\-Gra\-ph}}
\newcommand{\dbAlibabaGraphDatabase}{\texttt{Ali\-ba\-ba\- Gra\-ph\- Da\-ta\-ba\-se\- (GD)}} 
\newcommand{\dbAmazonNeptune}{\texttt{Ama\-zon\- Nep\-tu\-ne}}
\newcommand{\dbAnzoGraphDB}{\texttt{An\-zo\-Gra\-ph\- DB}}
\newcommand{\dbAltairGraphLakehouse}{\texttt{Al\-ta\-ir\- Gra\-ph\- La\-ke\-hou\-se}}
\newcommand{\dbArangoDB}{\texttt{A\-ran\-go\-DB}}
\newcommand{\dbArcadeDB}{\texttt{Ar\-ca\-de\-DB}}
\newcommand{\dbAzureCosmosDB}{\texttt{Azu\-re\- Cos\-mos\- DB}}

\newcommand{\dbBangDB}{\texttt{Bang\-DB}}
\newcommand{\dbMillenniumDB}{\texttt{Millen\-ni\-um\-DB}}

\newcommand{\dbBlazegraph}{\texttt{Bla\-ze\-gra\-ph}}
\newcommand{\dbBrightstarDB}{\texttt{Bri\-ght\-star\-DB}}
\newcommand{\dbByteGraph}{\texttt{By\-te\-Gra\-ph}}
\newcommand{\dbCassandra}{\texttt{Cas\-san\-dra}}
\newcommand{\dbChronoGraph}{\texttt{Chro\-no\-Gra\-ph}}
\newcommand{\dbDataStaxEnterprise}{\texttt{Da\-ta\-Stax\- En\-ter\-pri\-se\- Gra\-ph\- (DSE)}}

\newcommand{\dbCrayGraphEngine}{\texttt{Cray\- Gra\-ph\- En\-gi\-ne\- (CGE)}}

\newcommand{\dbDgraph}{\texttt{D\-gra\-ph}}
\newcommand{\dbFaunaDB}{\texttt{Fau\-na\-DB}}
\newcommand{\dbGaffer}{\texttt{Gaf\-fer}}
\newcommand{\dbGoogleCayley}{\texttt{Goo\-gle\- Cay\-ley}}
\newcommand{\dbGraphflow}{\texttt{Gra\-ph\-flow}}
\newcommand{\dbGTran}{\texttt{G-Tran}}
\newcommand{\dbHBase}{\texttt{H\-Ba\-se}}

\newcommand{\dbHugeGraph}{\texttt{Hu\-ge\-Gra\-ph}}
\newcommand{\dbStellarDB}{\texttt{Stel\-larDB}}

\newcommand{\dbKatanaGraph}{\texttt{Ka\-ta\-na\-Gra\-ph}}

\newcommand{\dbFluree}{\texttt{Flu\-ree}}

\newcommand{\dbOntotextGraphDB}{\texttt{Onto\-text\- Gra\-ph\-DB}}

\newcommand{\dbIBMSystemG}{\texttt{IBM\- Sys\-tem\- G}}

\newcommand{\dbHyperGraphDB}{\texttt{Hy\-per\-Gra\-ph\-DB}}
\newcommand{\dbJanusGraph}{\texttt{Ja\-nus\-Gra\-ph}}
\newcommand{\dbLiveGraph}{\texttt{Li\-ve\-Gra\-ph}}
\newcommand{\dbMemgraph}{\texttt{Mem\-gra\-ph}}
\newcommand{\dbNebulaGraph}{\texttt{Ne\-bu\-la\- Gra\-ph}}
\newcommand{\dbNeoFRj}{\texttt{Neo\-4j}}

\newcommand{\dbGalaxybase}{\texttt{Ga\-la\-xy\-ba\-se}}

\newcommand{\dbTAO}{\texttt{TAO}}

\newcommand{\dbObjectivityDB}{\texttt{Ob\-jec\-ti\-vi\-ty\-/DB}}

\newcommand{\dbOrientDB}{\texttt{Orient\-DB}}
\newcommand{\dbPandaDB}{\texttt{Pan\-da\-DB}}
\newcommand{\dbRedisGraph}{\texttt{Re\-dis\-Gra\-ph}}
\newcommand{\dbRocksDB}{\texttt{Ro\-cks\-DB}}
\newcommand{\dbSAPHanaGraph}{\texttt{SAP\- Ha\-na\- Gra\-ph}}
\newcommand{\dbSparksee}{\texttt{Spar\-ksee}}
\newcommand{\dbStardog}{\texttt{Star\-dog}}

\newcommand{\dbSurrealDB}{\texttt{Sur\-re\-al\-DB}}

\newcommand{\dbWeaver}{\texttt{Wea\-ver}}

\newcommand{\dbOracleSpatialAndGraph}{\texttt{Oracle\- Spa\-tial\- An\-d\- Gra\-ph\- (OSG)}}

\newcommand{\dbTigerGraph}{\texttt{Ti\-ger\-Gra\-ph}}
\newcommand{\dbTerminusDB}{\texttt{Ter\-mi\-nus\-DB}}

\newcommand{\dbTypeDB}{\texttt{Ty\-pe\-DB}}

\newcommand{\dbKuzu}{\texttt{Ku\-zu}}

\newcommand{\dbUltipa}{\texttt{Ul\-ti\-pa}}

\newcommand{\dbHGraphDB}{\texttt{H\-Gra\-phDB}}

\newcommand{\dbVirtuoso}{\texttt{Vir\-tu\-oso}}
\newcommand{\dbZipG}{\texttt{Zip\-G}}

\newcommand{\dbAgensGraph}{\texttt{Ag\-ens\-Gra\-ph}}

\newcommand{\gqlCypher}{\texttt{Cy\-pher}}
\newcommand{\gqlGremlin}{\texttt{Grem\-lin}}
\newcommand{\gqlSPARQL}{\texttt{SPAR\-QL}}

\newcommand{\gpeGraphFrames}{\texttt{Gra\-ph\-Fra\-mes}}
\newcommand{\gpeGRADOOP}{\texttt{GRA\-DOOP}}
\newcommand{\gpeApacheGiraph}{\texttt{Apa\-che\- Gi\-ra\-ph}}
\newcommand{\gpeApacheSpark}{\texttt{A\-pa\-che\- Spa\-rk}}

\newcommand{\frameworkTinkerPop}{\texttt{Tin\-ker\-Pop} }

\newcommand{\langClojure}{\texttt{Clo\-ju\-re}}
\newcommand{\langHaskell}{\texttt{Has\-kell}}
\newcommand{\langJava}{\texttt{Ja\-va}}
\newcommand{\langJavaScript}{\texttt{Ja\-va\-Scri\-pt}}
\newcommand{\langPython}{\texttt{Py\-thon}}
\newcommand{\langRuby}{\texttt{Ru\-by}}
\newcommand{\langScala}{\texttt{Sca\-la}}

\newcommand{\licenseAffero}{\texttt{GNU\- Af\-fe\-ro\- GE\-NE\-RAL\- PU\-BLIC\- LI\-CEN\-SE\- ver\-si\-on\- 3.0}}
\newcommand{\licenseApacheTwo}{\texttt{A\-pa\-che\- Li\-cen\-se\- 2.0}}

\newcommand{\licenseEclipseTwo}{\texttt{E\-cli\-pse\- Pu\-blic\- Li\-cen\-se\- \- 2.0}}
\newcommand{\licenseGPLTwo}{\texttt{GNU\- Ge\-ne\-ral\- Pu\-blic\- Li\-cen\-se\- \- 2.0}}
\newcommand{\licenseMIT}{\texttt{MIT\- Li\-cen\-se}}

\makeatletter
\tcbset{commonstyle/.style={boxrule=0pt,sharp corners,enhanced jigsaw,nobeforeafter,boxsep=0pt,left=\fboxsep,right=\fboxsep}}
\makeatother

\newtcolorbox{mycolorbox}[1][]{commonstyle,#1}

\newlength\myboxwidth

\definecolor{GreenReview}{RGB}{102,204,0}
\definecolor{YellowReview}{RGB}{255, 255, 51}
\definecolor{OrangeReview}{RGB}{255, 128, 0}

\makeatletter

\def\@highlightttpeeknext{\futurelet\@nexttoken\@highlightttaux}
\def\@highlighttt #1.{%
    \def\@highlightttaux{\ifx\@nexttoken\egroup
       \myhighlightmethod {#1}\else
       \myhighlightmethod {#1.}\linebreak[2]%
       \expandafter\@highlighttt\fi}%
    \@highlightttpeeknext}

\def\@plaintt {\futurelet\@nexttoken\@plainttaux}
\def\@plainttaux {\ifx\@nexttoken\egroup\else
                  \ifx\@nexttoken\bgroup
                  \expandafter\expandafter\expandafter\@plaintta\else
                  \expandafter\expandafter\expandafter\@plainttb\fi\fi}
\def\@plaintta #1{{#1}\@plaintt}
\def\@plainttb #1{\ifcat\@nexttoken a\penalty\hyphenpenalty \plaintthook
  #1\else \plaintthook{#1}\linebreak[2]\fi\@plaintt}

\newcommand{\highlighttt}[1]{{\fontfamily{txtt}\selectfont
   \@highlighttt #1.}}

\newcommand\plaintt{\bgroup\fontfamily{txtt}\selectfont
   \afterassignment\@plaintt\let\next= }

\makeatother

\newcommand{\myhighlightmethod}[1]{\fboxsep0pt\colorbox{yellow}{\strut#1}}
\newcommand{\plaintthook}{}

\definecolor{alizarin}{rgb}{0.82, 0.1, 0.26}
\definecolor{amethyst}{rgb}{0.6, 0.4, 0.8}
\definecolor{arylideyellow}{rgb}{0.91, 0.84, 0.42}
\definecolor{azure(colorwheel)}{rgb}{0.0, 0.5, 1.0}
\definecolor{ballblue}{rgb}{0.13, 0.67, 0.8}
\definecolor{bananayellow}{rgb}{1.0, 0.88, 0.21}

\definecolor{OIblack}{HTML}{000000}
\definecolor{OIorange}{HTML}{E69F00}
\definecolor{OIskyblue}{HTML}{56B4E9}
\definecolor{OIgreen}{HTML}{009E73}
\definecolor{OIyellow}{HTML}{F0E442}
\definecolor{OIblue}{HTML}{0072B2}
\definecolor{OIvermillion}{HTML}{D55E00}
\definecolor{OIpurple}{HTML}{CC79A7}
\definecolor{keyword}{HTML}{2771a3}
\definecolor{pattern}{HTML}{b53c2f}
\definecolor{string}{HTML}{be681c}
\definecolor{relation}{HTML}{7e4894}
\definecolor{variable}{HTML}{107762}
\definecolor{comment}{HTML}{8d9094}

\lstdefinestyle{CypherStyle}{
	numbers=none,
	stepnumber=1,
	numbersep=5pt,
	basicstyle=\small\ttfamily,
	keywordstyle=\color{keyword}\bfseries\ttfamily,
	commentstyle=\color{comment}\ttfamily,
	stringstyle=\color{string}\ttfamily,
	identifierstyle=,
	showstringspaces=false,
	aboveskip=3pt,
	belowskip=3pt,
	columns=flexible,
	keepspaces=true,
	breaklines=true,	
	captionpos=b,
	tabsize=2,
	frame=none,
}

\lstdefinelanguage{cypher}
{
	morekeywords={
		MATCH, OPTIONAL, WHERE, NOT, AND, OR, XOR, RETURN, DISTINCT, ORDER, BY, ASC, ASCENDING, DESC, DESCENDING, UNWIND, AS, UNION, WITH, ALL, CREATE, DELETE, DETACH, REMOVE, SET, MERGE, SET, SKIP, LIMIT, IN, CASE, WHEN, THEN, ELSE, END,
		INDEX, DROP, UNIQUE, CONSTRAINT, EXPLAIN, PROFILE, START,
	}
}

\newcommand{\mycdots}{\cdot\!\cdot\!\cdot}
\definecolor{dkgreen}{rgb}{0,0.6,0}
\definecolor{gray}{rgb}{0.5,0.5,0.5}
\definecolor{mauve}{rgb}{0.58,0,0.82}

\makeatletter
\renewcommand*\env@matrix[1][*\c@MaxMatrixCols c]{%
  \hskip -\arraycolsep
  \let\@ifnextchar\new@ifnextchar
  \array{#1}}
\makeatother

\definecolor{LicOpen}{HTML}{009E73}   % green
\definecolor{LicProp}{HTML}{E69F00}   % orange
\definecolor{LicMixed}{HTML}{CC79A7}  % purple
\definecolor{LicOther}{HTML}{7A7A7A}  % fallback

\definecolor{TernPG}{HTML}{0072B2}     % Property Graph Model
\definecolor{TernRDF}{HTML}{E69F00}    % RDF Data Model
\definecolor{TernMulti}{HTML}{009E73}  % Multiple Data Models
\definecolor{TernAlt}{HTML}{CC79A7}    % Alternative Data Models
\definecolor{TernOther}{HTML}{7A7A7A}  % fallback

\tikzset{
  modelshape/PG/.style   ={regular polygon, regular polygon sides=3}, % up triangle
  modelshape/RDF/.style  ={regular polygon, regular polygon sides=3, shape border rotate=180}, % down triangle
  modelshape/Multi/.style={circle}, % circle
  modelshape/Alt/.style  ={rectangle, rounded corners=0pt}, % sharp square
  modelshape/Other/.style={circle},
}

\def\ternaryColor{LicOther}
\def\ternaryShapeStyle{modelshape/Other}

\newcommand{\ternarySetColorFromLicense}[1]{%
  \def\ternaryColor{LicOther}%
  \ifnum\pdfstrcmp{#1}{Open source}=0  \def\ternaryColor{LicOpen}\fi
  \ifnum\pdfstrcmp{#1}{Proprietary}=0  \def\ternaryColor{LicProp}\fi
  \ifnum\pdfstrcmp{#1}{Mixed}=0        \def\ternaryColor{LicMixed}\fi
}

\newcommand{\ternarySetShapeFromModel}[1]{%
  \def\ternaryShapeStyle{modelshape/Other}%
  \ifnum\pdfstrcmp{#1}{Property Graph Model}=0   \def\ternaryShapeStyle{modelshape/PG}\fi
  \ifnum\pdfstrcmp{#1}{RDF Data Model}=0         \def\ternaryShapeStyle{modelshape/RDF}\fi
  \ifnum\pdfstrcmp{#1}{Multiple Data Models}=0   \def\ternaryShapeStyle{modelshape/Multi}\fi
  \ifnum\pdfstrcmp{#1}{Alternative Data Models}=0\def\ternaryShapeStyle{modelshape/Alt}\fi
}

\newcommand{\ternaryplotpoint}[6]{%
  \ternarySetColorFromLicense{#6}%
  \ternarySetShapeFromModel{#5}%
  \pgfmathsetlengthmacro{\ms}{2*#4pt}%
  \node[
    \ternaryShapeStyle,
    minimum size=\ms,
    inner sep=0pt,
    draw=\ternaryColor!85!black,
    fill=\ternaryColor!18,
    line width=0.7pt,
    fill opacity=0.55,
    draw opacity=0.90
  ] at (#2,#3) {};
}

\newcommand{\triLegScale}{0.125} % <-- change this (e.g., 1.2) to scale triangle icons
\newcommand{\adjustedLegScale}{0.125} % <-- change this (e.g., 1.2) to scale square/circle icons

\newcommand{\legIconUpTri}{%
  \tikz[baseline=-0.6ex, scale=\triLegScale]{
    \path[use as bounding box] (-1,-1) rectangle (1,1);
    \draw[draw=black!65, fill=black!10, line width=0.4pt]
      (0, 0.82) -- (-0.78, -0.62) -- (0.78, -0.62) -- cycle;
  }%
}
\newcommand{\legIconDownTri}{%
  \tikz[baseline=-0.6ex, scale=\triLegScale]{
    \path[use as bounding box] (-1,-1) rectangle (1,1);
    \draw[draw=black!65, fill=black!10, line width=0.4pt]
      (0, -0.82) -- (-0.78, 0.62) -- (0.78, 0.62) -- cycle;
  }%
}
\newcommand{\legIconCircle}{%
  \tikz[baseline=-0.6ex, scale=\adjustedLegScale]{
    \path[use as bounding box] (-1,-1) rectangle (1,1);
    \draw[draw=black!65, fill=black!10, line width=0.4pt]
      (0,0) circle (0.78);
  }%
}
\newcommand{\legIconSquare}{%
  \tikz[baseline=-0.6ex, scale=\adjustedLegScale]{
    \path[use as bounding box] (-1,-1) rectangle (1,1);
    \draw[rounded corners=0pt, draw=black!65, fill=black!10, line width=0.4pt]
      (-0.78,-0.78) rectangle (0.78,0.78);
  }%
}

\newcommand{\ternaryplotlabel}[4]{%
  \node[
    font=\scriptsize,
    text=black!85,
    fill=white,
    fill opacity=0.85,
    text opacity=1,
    inner sep=1.2pt,
    rounded corners=1pt,
    #4
  ] at (#2,#3) {#1};
}

\newcommand{\ternaryplotcentroidcalloutpos}[7]{%
  \node[modelshape/#4, minimum size=6.0pt, inner sep=0pt, draw=#5!85!black, fill=white, line width=1.0pt] (centroid) at (#2,#3) {};%
  \node[overlay, font=\scriptsize\bfseries, text=#5!85!black, fill=white, fill opacity=0.85, text opacity=1, inner sep=1.2pt, rounded corners=1pt] (centroidlbl) at (#6,#7) {#1};%
  \draw[overlay, draw=#5!70!black, line width=0.5pt] (centroidlbl) -- (centroid);%
}

\newcommand{\ternaryplotcentroidcallout}[7]{%
  \ternaryplotcentroidcalloutpos{#1}{#2}{#3}{#4}{#5}{#6}{#7}%
}

\newcommand{\stripplotpoint}[5]{%
  \ternarySetColorFromLicense{#4}%
  \ternarySetShapeFromModel{#3}%
  \pgfmathsetlengthmacro{\ms}{2*#5pt}%
  \node[
    \ternaryShapeStyle,
    minimum size=\ms,
    inner sep=0pt,
    draw=\ternaryColor!85!black,
    fill=\ternaryColor!18,
    line width=0.6pt,
    fill opacity=0.55,
    draw opacity=0.90
  ] at (#1,#2) {};
}

\newcommand{\stripaxis}[3]{%
  \draw[black!60, line width=0.5pt] (0,#1) -- (#3,#1);
  \foreach \t/\lab in {0/0,0.5/0.5,1/1} {
    \draw[black!60, line width=0.5pt] (\t*#3,#1-0.01) -- (\t*#3,#1+0.01);
    \node[font=\scriptsize, text=black!70, below] at (\t*#3,#1-0.01) {\lab};
  }
  \node[font=\scriptsize\bfseries, text=black!80, anchor=east] at (-0.03,#1) {#2};
}

\newcommand{\xyzplotpoint}[7]{}
\hypersetup{%colorlinks,
   debug=false,
   linkcolor=blue,  %%% cor do tableofcontents, \ref, \footnote, etc
   citecolor=blue,  %%% cor do \citep
   urlcolor=blue,   %%% cor do \url e \href
   bookmarksopen=true,
   breaklinks=true,
   pdfauthor={Miguel E. Coimbra},
   pdfsubject={Survey: Graph Databases},
}

\title{Survey: On the Landscape of Graph Databases} %TODO Please add

\author{Miguel E. Coimbra\footnote{Corresponding author}}{INESC-ID, [R. Alves Redol 9, 1000-029 Lisboa], Portugal \and IST, [Av. Rovisco Pais 1, 1049-001 Lisboa], Portugal}{miguel.e.coimbra@tecnico.ulisboa.pt}{https://orcid.org/0000-0002-7191-5895}{}

\author{Lucie Svitáková}{Faculty of Information Technology, CTU in Prague, [Thákurova 9, 160 00, Prague], Czech Republic}{svitaluc@fit.cvut.cz}{https://orcid.org/0000-0002-6961-5041}{}

\author{Domagoj Vrgo\v{c}}{Department of Computer Science (DVRG), Pontificia Universidad Católica de Chile, [Av. Vicuña Mackenna 4860, Edificio Hernán Briones, 2do piso, 7820436, Macul, Santiago], Chile \and \url{http://dvrgoc.ing.puc.cl/}}{vrdomagoj@uc.cl}{https://orcid.org/0000-0001-5854-2652}{}

\author{Alexandre P. Francisco}{INESC-ID, [R. Alves Redol 9, 1000-029 Lisboa], Portugal \and IST, [Av. Rovisco Pais 1, 1049-001 Lisboa], Portugal}{aplf@inesc-id.pt}{https://orcid.org/0000-0003-4852-1641}{}

\author{Luís Veiga}{INESC-ID, [R. Alves Redol 9, 1000-029 Lisboa], Portugal \and IST, [Av. Rovisco Pais 1, 1049-001 Lisboa], Portugal \and \url{https://www.dpss.inesc-id.pt/~lveiga/}}{luis.veiga@inesc-id.pt}{https://orcid.org/0000-0002-9285-0736}{}

\authorrunning{M.E. Coimbra et al.} %TODO mandatory. First: Use abbreviated first/middle names. Second (only in severe cases): Use first author plus 'et al.'

\Copyright{Miguel E. Coimbra and Lucie Svitáková and Domagoj Vrgo\v{c} and Alexandre P. Francisco and Luís Veiga} %TODO mandatory, please use full first names. TGDK license is "CC-BY";  http://creativecommons.org/licenses/by/4.0/


 %TODO mandatory; please add comma-separated list of keywords

\category{} %optional, e.g. invited paper

\relatedversion{This is the publication version of:} %optional, e.g. full version hosted on arXiv, HAL, or other respository/website
\relatedversiondetails[linktext={Survey: Graph Databases, arXiv}, cite=coimbra2025survey]{arXiv Version}{https://arxiv.org/abs/2505.24758} %linktext and cite are optional

\funding{This work was supported by national funds through Fundação para a Ciência e a Tecnologia, I.P. (FCT) under projects UID/500\-21/20\-25 (DOI: https://doi.org/10.54499/UID/50021/2025) and UID/PRR/500\-21/2025 (DOI: https://doi.org/10.54499/UID/PRR/50021/2025), and 2023.169\-86.ICDT. This work was supported by the CloudStars project, funded by the European Union’s Horizon research and innovation program under grant agreement number 101086248.}%optional, to capture a funding statement, which applies to all authors. Please enter author specific funding statements as fifth argument of the \author macro.

\begin{document}

\maketitle

\begin{abstract}
Graph databases have become essential tools for managing complex and interconnected data, which is common in areas like social networks, bioinformatics, and recommendation systems. 
Unlike traditional relational databases, graph databases offer a more natural way to model and query intricate relationships, making them particularly effective for applications that demand flexibility and efficiency in handling interconnected data.

Despite their increasing use, graph databases face notable challenges. 
One significant issue is the irregular nature of graph data, often marked by structural sparsity, such as in its adjacency matrix representation, which can lead to inefficiencies in data read and write operations. 
Other obstacles include the high computational demands of traversal-based queries, especially within large-scale networks, and complexities in managing transactions in distributed graph environments. 
Additionally, the reliance on traditional centralized architectures limits the scalability of Online Transaction Processing (OLTP), creating bottlenecks due to contention, CPU overhead, and network bandwidth constraints.

The growing relevance of graph databases in handling large-scale datasets across diverse fields—such as healthcare, industry, artificial intelligence (AI), life sciences, social networks, and software engineering—drives the motivation for this study. 
Addressing the challenges they face requires efficient query languages and optimized storage solutions to improve their overall performance.

This paper presents a thorough survey of graph databases. 
It begins by examining property models, query languages, and storage architectures, outlining the foundational aspects that users and developers typically engage with. 
Following this, it provides a detailed analysis of recent advancements in graph database technologies, evaluating these in the context of key aspects such as architecture, deployment, usage, and development, which collectively define the capabilities of graph database solutions.

We conducted a comprehensive analysis of numerous graph database systems and surveys to shed light on the most critical intricacies within this ecosystem. 
Alongside this, we outlined a range of features to assist developers and researchers in identifying the specific requirements of their use cases and the technologies suited to their needs. 

The next decade will likely reveal fascinating trends among commercial players and prioritized features. 
We anticipate that graph databases will continue to advance as the volume of data grows, particularly within the realm of AI.
\end{abstract}

%%%%%%%%%%%%%%%%%%%%%%%%%%%%%%%%%%%%%%%%%%%%%%%%
%%%%%%%%%%%%%%%%%%%%%%%%%%%%%%%%%%%%%%%%%%%%%%%% INTRODUCTION
%%%%%%%%%%%%%%%%%%%%%%%%%%%%%%%%%%%%%%%%%%%%%%%%

% BEGIN inlined from sections/introduction.tex
\section{Introduction}~\label{chapter:publications:survey:sec:introduction}

Graph databases have become increasingly vital in handling complex, interconnected data, prevalent in domains like social networks, bioinformatics, or recommendation systems~\citep{knowledge-graphs,10.1145/3434642}. 
Unlike traditional relational databases, graph databases offer a more intuitive model for complex and interrelated data, making them particularly suited for applications that require efficient and flexible modeling and querying of complex relationships. 

%UNCOMMENT
%MOTIV
The motivation for this study lies in the growing relevance of graph databases in handling large-scale data with diverse use-cases in multiple domains (e.g. health~\citep{graph-db-clinical}, industry~\citep{graph-db-industry}, artificial intelligence (AI)~\citep{graph-db-isec}, life sciences~\citep{graph-db-life}, social networks~\citep{graph-db-social}, network analysis~\citep{survey-subgraphs-networking-21} and software engineering~\citep{graph-db-se}), and the need for efficient query languages and storage solutions in this context.

The complexity of the graph database ecosystem has increased in the last decade, with the ecosystem flourishing with new commercial offerings often benefiting from community engagement. 
Choosing the appropriate solution requires an understanding of the knowledge on graph storage, graph representations, query languages and system architectures. 
Such knowledge is essential to properly guide the questions which will help the technological decision-makers of both industry and academia in identifying their use-cases. 

Performance and lower processing costs are relevant, but so is the ability to control data on-premises, or the offering of quality documentation. 
An open-source community, with whose members it is possible to quickly iterate ideas and troubleshoot problems, can reduce technological friction and significantly speed up development tasks. 
We set out to clarify these and other aspects in this paper, and observe if and how they are considered by different graph database systems which are part of an exhaustive list we have identified from the ecosystem.

%\textcolor{Cerulean}{DONE? - extend the motivation to specifically fit the content of the paper. }\textcolor{purple}{The motivation is now awesome! It fits the document content perfectly.}

%SHORTCOMINGS
Despite the growing popularity and application of graph databases, several challenges persist. 
For example, the irregularity in graph data such as the sparseness of its structure (e.g., manifested in the concept of its adjacency matrix) may lead to inefficiencies in data reads and writes. 
%\textcolor{Cerulean}{DONE? - describe better (what is irregularity in graph data? }\textcolor{purple}{Great!} 
Other challenges include the high cost of traversal-based queries, especially in large-scale networks, and issues related to transaction processing in distributed graph environments. Moreover, the traditional centralized system architectures often limit the scalability of Online Transaction Processing (OLTP) in graph databases, presenting bottlenecks in terms of contention likelihood, CPU overheads, and network bandwidth.

%CONTRIBS
This paper presents a comprehensive survey of graph databases, focusing initially on property models, query languages and storage architectures, to frame the main aspects that users and developers typically have to deal with. Following, we make an in-depth analysis of recent and relevant graph database works, according to the previous aspects, and also across a number of broader dimensions (or areas), each embodying classes of key features (or aspects) of graph database solutions' architecture, deployment, usage, and development. 
These include support, licensing, programming model, operating system, deployment, data representation, concurrency, performance, security, scalability, fault tolerance, or integration, to name a few.

The rest of the paper is organized as follows. In Section~\ref{chapter:publications:survey:sec:property_model_gql}, we present relevant graph models and query languages that a potential user of a graph database might encounter because the models and languages play a decisive role in the selection of a graph database to match the consumer's use case and capabilities properly. 
Section~\ref{chapter:publications:survey:sec:data_rep_and_indexing} describes and shows the advantages of various storage architectures, from unstructured to advanced ones, applied in actual graph database solutions. 
We follow with introductions of each surveyed graph database in Section~\ref{chapter:publications:survey:sec:graph-databases}, opening with features that were evaluated for each one and classified into their respective categories. 
We also analyze and discuss the previously listed graph databases and their capabilities. 
Since the field of graph database research has already become rich, we recommend related work in the following Section~\ref{chapter:publications:survey:sec:related-work}, concluding our survey in the last Section~\ref{chapter:publications:survey:sec:conclusions}.
% END inlined from sections/introduction.tex

%%%%%%%%%%%%%%%%%%%%%%%%%%%%%%%%%%%%%%%%%%%%%%%%
%%%%%%%%%%%%%%%%%%%%%%%%%%%%%%%%%%%%%%%%%%%%%%%% PROPERTY MODEL AND QUERY LANGUAGES
%%%%%%%%%%%%%%%%%%%%%%%%%%%%%%%%%%%%%%%%%%%%%%%%

% Tiny pie-circle glyphs (outline always visible)
\newcommand{\scorecircle}[1]{%
  \tikz[baseline=-0.6ex]{
    \def\r{0.85ex}
    \draw (0,0) circle (\r);
    \ifdim #1pt>0pt
      \fill (0,0) -- (\r,0) arc (0:{360*#1}:\r) -- cycle;
    \fi
  }%
}
\newcommand{\scoreNone}{\scorecircle{0}}
\newcommand{\scoreQuarter}{\scorecircle{0.25}}
\newcommand{\scoreHalf}{\scorecircle{0.50}}
\newcommand{\scoreThreeQuarter}{\scorecircle{0.75}}
\newcommand{\scoreFull}{\scorecircle{1.00}}
\newcommand{\scorehdr}[2]{%
  \multicolumn{2}{c}{\shortstack[c]{#1\\[-2pt]\scriptsize #2}}%
}

% BEGIN inlined from sections/model_and_query_languages/model_and_query_languages.tex

%\section{\textcolor{purple}{\sout{Property} Graph} Models and Query Languages}~\label{chapter:publications:survey:sec:property_model_gql}

\section{Graph Models and Query Languages}~\label{chapter:publications:survey:sec:property_model_gql}

% \textcolor{purple}{
% \sout{In this section we detail models, languages, and subsequent standards for querying graph-based data.}
% This section introduces the most common graph models used in graph database systems. For these models, several graph query languages have been developed, which are presented afterward. These languages often serve not only for basic create-read-update-delete operations but also for sophisticated querying of the graphs.}

This section introduces the most common graph models used in graph database systems. 
For these models, several graph query languages have been developed, which are presented afterward. 
These languages often serve not only for basic create-read-update-delete (CRUD) operations but also for sophisticated querying of the graphs.

% \textcolor{purple}{
% \subsection{Graph Data Models}\label{chapter:publications:survey:sec:property_model_gql:sec:gql}
% }

\subsection{Graph Data Models}\label{chapter:publications:survey:sec:property_model_gql:sec:gdm}

Two most common graph data models in use today are \emph{property graphs} and \emph{edge-labeled graphs}. 
The former are widely adopted in commercial systems, while the latter form the backbone of many open-source solutions in the Semantic Web ecosystem. 
Several generalizations extending or mixing the two models have also been developed, but are not as popular. 
Next we further elaborate on each data model.

%\textcolor{purple}{Nowadays, a widely accepted general model for graph data is the property graph (PG) model.} 
\subsubsection{Property graphs} Nowadays, a widely accepted general model for graph data is the property graph (PG) model. 
The property graph model defines the organization of interrelated data as nodes (vertices), relationships (edges), and the properties of both types 
of elements~\citep{neo4j_graph_db}. 
Properties themselves are usually defined as key-value pairs and by default do not need to conform to a specific schema, allowing for fully flexible data representation. 
The model is also often associated with the adjectives "labeled" or "directed" since the nodes or edges often have assigned labels, and the relationships can be directed from a source node to a target node. 
More recent incarnations also support undirected edges~\citep{deutsch2022graph}.  
Whether just one label per node/edge is required (so-called "type"), the assignment of any number or at least one of the labels is optional, or the labeling ability is even provided, varies from one particular graph database management system to another. 
%\textcolor{purple}{It is also often associated with the adjectives "labeled" or "directed" since the nodes or edges often have assigned labels, and the relationships can be directed from a source node to a target node. 
% Whether just one label per node/edge is required (so-called "type"), the assignment of any number or at least one of the labels is optional, or the labeling ability is even provided varies from one particular graph database management system to another.} Instances of this model are usually found coupled to the design of graph databases and at the level of graph query languages~\citep{rodriguez2015gremlin,van2016pgql,angles2018property,green2018opencypher}.
Instances of this model are usually found coupled to the design of graph query languages~\citep{rodriguez2015gremlin,van2016pgql,angles2018property,green2018opencypher}.
Figure~\ref{fig:property-graph-model} depicts the properties that may be associated with vertices and edges in the context of the property graph model.

\begin{figure}[!t]
\centering
\begin{tikzpicture}[
  font=\small,
  >=latex, % legacy arrow tip (works widely)
  node distance=3.6cm,
  every node/.style={align=center},
  gnode/.style={draw, very thick, circle, minimum size=18mm, inner sep=0pt},
  prof/.style={gnode, draw=OIblue, fill=OIblue!10},
  camp/.style={gnode, draw=OIblue, fill=OIblue!10},
  uni/.style={gnode, draw=OIorange, fill=OIorange!10},
  edge/.style={very thick, ->},
  etype/.style={font=\ttfamily\scriptsize, inner sep=1pt, fill=white},
  eprops/.style={font=\ttfamily\scriptsize, inner sep=2pt, rounded corners=1.5pt, draw=black!35, fill=black!3},
  nprops/.style={font=\ttfamily\scriptsize, inner sep=3pt, rounded corners=2pt, draw=black!35, fill=black!3},
  callout/.style={font=\scriptsize, text=black!70, align=left},
  callline/.style={->, thin, draw=black!55}
]

% Nodes
\node[prof] (p) {Professor};
\node[uni, right=of p] (u) {University};
\node[camp, right=of u] (c) {Campus};

% Edges
\draw[edge] (u) -- (p);
\draw[edge] (u) -- (c);

% Edge type labels
\node[etype] at ($(u)!0.5!(p) + (0,6mm)$) {:HAS\_PROFESSOR};
\node[etype] (lblHQ) at ($(u)!0.5!(c) + (0,6mm)$) {:HEADQUARTERED\_IN};

% Edge properties
\node[eprops] (epprops) at ($(u)!0.5!(p) + (0,-8mm)$) {start\_date: 1938};

% Node properties
\node[nprops, below=7mm of p] (pprops) {name: Erd\H{o}s\\date\_of\_birth: 1913-03-26};

% Callouts
\node[callout, above=12mm of p, xshift=-10mm] (cl1) {Node label};
\draw[callline] (cl1) -- (p.north);

\node[callout, below=11mm of pprops, xshift=-6mm] (cl2) {Node properties\\(key/value pairs)};
\draw[callline] (cl2) -- (pprops.south);

\node[callout, above=15mm of u] (cl3) {Directed edge\\with a type label};
\draw[callline] (cl3) -- ($(lblHQ.north) + (0,1.2mm)$);

\node[callout, below=14mm of u] (cl4) {Edge properties\\(key/value pairs)};
\draw[callline] (cl4) -- (epprops.south);

\end{tikzpicture}
\caption{Illustration of the property graph model, inspired by the memory of Professor Paul Erd\H{o}s.}
\label{fig:property-graph-model}
\end{figure}

%\textcolor{purple}{An even more general graph model is an \texttt{RDF} - Resource Description Framework. Its central concept is a triple of a subject-predicate-object that can be naturally mapped to a directed labeled graph. As the W3C standard of the \texttt{RDF} dates back twenty years, the \texttt{RDF} is supported by various graph database management systems as well.~\citep{w3cRDF}}

\subsubsection{Directed edge-labeled graphs (\texttt{RDF})} The second widely used graph model is \texttt{RDF} - Resource Description Framework~\citep{w3cRDF}. 
Its central concept is a triple of a subject-predicate-object statements that can be naturally mapped to a directed edge-labeled graph. 
While conceptually simple, the \texttt{RDF} data model supports a rich hierarchy of data types,  identifiers and inference rules, and can in fact be shown to be more expressive than the property graph data model~\citep{AnglesHLRSSV22}. 
As the W3C standard of the \texttt{RDF} dates back twenty years, the \texttt{RDF} is supported by various graph database management systems (see~\citep{AliSYHN22} for an overview)  and was thoroughly explored in the academic literature~\citep{sagi2022design}.

\subsubsection{Beyond property graphs and \texttt{RDF}} While graph data models are suitable for expressing binary relations, they tend to struggle with $n$-ary relationships. 
\texttt{RDF} can handle these through complex reification schemes~\citep{HernandezHK15}, but the solution is rather cumbersome to use due to its verbose nature. 
Similarly, models such as hypergraphs (extending the notion of an edge to a hyperedge) or hypernodes (nesting graphs by allowing nodes to be graphs) are considered too complex to enable intuitive manipulation, so their usage as a core graph representation in graph database systems is rare~\citep{fletcher2018graphDataManagement}. 
That being said, a simplified version of hypergraphs has recently been developed~\citep{AnglesHLRSSV22} and is being deployed as an intermediary model unifying both \texttt{RDF} and property graphs in several solutions~\citep{VrgocRAAABHNRR23,BroekemaELLNS0S24,IlievskiGCDYRLL20,LassilaSHBBBKLL23}.

%\textcolor{purple}{More complex models such as hypergraphs (extending the notion of an edge to a hyperedge) or hypernodes (nesting graphs by allowing nodes to be graphs) are considered too complex to enable intuitive manipulation, so their usage as a core graph representation in graph database systems is rare~\citep{fletcher2018graphDataManagement}.}

\subsubsection{Terminology} We note that across the bodies of literature touching upon the use of graphs, the terms vertex/node are used interchangeably to a certain degree. 
In this document we adhere to this traditional practice, though referring preferably to a "vertex" when mentioning a graph element (but also "node" may be used when more adequate). 
When used to refer to a member of a group of machines (e.g., a cluster) used for processing, we will make that explicit (e.g. server node).

%%%%%%%%%%%%%%%%%%%%%%%%%%%%%%%%%%%%%%
%%%%%%%%%%%%%%%%%%%%%%%%%%%%%%%%%%%%%% GRAPH QUERY LANGUAGES
%%%%%%%%%%%%%%%%%%%%%%%%%%%%%%%%%%%%%%

\subsection{Graph Query Languages}\label{chapter:publications:survey:sec:property_model_gql:sec:gql}

The access to elements offered by graph databases is generally performed using specific query languages which vary depending on the underlying data model and system architecture. 
When it comes to \texttt{RDF}, the situation is rather clear, with all existing engines implementing \gqlSPARQL~\citep{sparql11}, which is a W3C standard, or minor variations thereof. 
On the other hand, the lack of a common standard in the property graph world meant that every engine implemented its own variant of a graph query language. 
Historically, these came in the form of an access and traversal API, Gremlin~\citep{rodriguez2015gremlin} being the most prominent example, or since the early 2010s as fully fledged  declarative languages with \dbNeoFRj's \gqlCypher~\citep{francis2018cypher} being the most representative example.

While the openCypher project~\citep{opencypher_site}, established in 2015, attempted to define a common language for graph querying, it did not succeed in establishing a  standard. %, leading to different vendors defining custom languages, with Oracle's \texttt{PGQL}~\citep{PGQL} or \dbTigerGraph's \texttt{GQSL}~\citep{gsql} being some of the more commonly used ones.
With the expansion of the graph database market the lack of a standard query language became apparent, and in 2019 the Joint Technical Committee 1 of ISO/IEC started work on the \texttt{GQL} (Graph Query Language) standard, which was released in 2024~\citep{GQL2024}. 
In parallel, the same ISO/IEC committee developed the \texttt{SQL/\-PGQ} standard (PGQ standing for \emph{property graph queries}), which extends \texttt{SQL} with the ability to define graph views over tables and run read-only queries over such views. 
The two standards provide the same graph pattern matching capabilities, but serve different purposes, \texttt{SQL/\-PGQ} being an \texttt{SQL} extension for graph querying, while \texttt{GQL} is a stand alone graph language. 
\texttt{SQL/\-PGQ} was released in 2023 as part of the new version of the \texttt{SQL} standard~\citep{SQLPGQ2023}.

Fundamental query features common to most graph query languages~\citep{angles2017foundations,gcore} can be roughly classified as follows:
\begin{itemize}
    \item \emph{Basic graph patterns}, which amount to matching a smaller graph-shaped pattern to the database;\item \emph{Complex graph patterns}, which extend basic graph pattern with filters and data tests;
    \item \emph{Path queries}, which search for paths of potentially unbounded length;
    \item \emph{Navigational graph patterns}, which allow replacing a simple edge in a basic graph pattern with a path specification; and
    \item \emph{Aggregate graph queries}, which allow for aggregation on top of the previous pattern classes.
\end{itemize}
Some other atomic tests~\citep{van2016pgql,gubichev2014graph} include matches on the vertices that are adjacent and edges that are incident to a given vertex, or triangles, which look for three vertices adjacent to each other. 
Generally, when defining constructs of a graph query language, it is important to consider if it shall be declarative or not, and how easy it will be to translate the use of the language into actual computational execution and performance.

Here, we address some of the most relevant graph query languages (and proposals in the scope of graph query languages), projects that structure 
the access to graph-based data in graph databases as well as open-source constructs that form their basis.

\begin{itemize}

\item \texttt{Cy\-pher}.
An evolving graph query language~\citep{francis2018cypher} which debuted with \dbNeoFRj's entrance in the field of graph databases.
There have been efforts to adopt and develop the language in an open-source approach~\citep{green2018opencypher}. 
\gqlCypher\ has been an open and evolving language as part of the \texttt{Open\-Cy\-pher} project~\citep{opencypher_repo}. 
The project has members involved in aspects such as synergy of engineering efforts with \gpeApacheSpark~\citep{Zaharia:2010:SCC:1863103.1863113}, 
a language group, and even interoperability features for systems that use \gqlGremlin.
This graph query language heavily influenced the ISO project for creating a standard graph query language~\citep{opencypher_site} and has a syntax 
familiar to developers with knowledge of \texttt{SQL}.
Examples of databases where \gqlCypher\ queries may be run include \dbNeoFRj, \dbGraphflow~\citep{kankanamge2017graphflow}, \dbRedisGraph~\citep{cailliau2019redisgraph}, \dbSAPHanaGraph~\citep{rudolf2013graph}, and all databases where \gqlGremlin\ is supported~\citep{opencypher_4_gremlin}.
This language is also used to express computation in \gpeGRADOOP~\citep{DBLP:conf/grades/JunghannsKAPR17} and the \texttt{Py\-thon\- Ru\-ru\-ki}~\citep{python_ruruki} lightweight in-memory graph database.

\item \texttt{GQL} and \texttt{SQL/PGQ}. 
The two recent standards draw heavily on the \gqlCypher\ syntax and share common pattern matching  capabilities~\citep{deutsch2022graph}. 
Given their recent publication, there are not many systems that are fully compliant, but multiple vendors started offering support for \texttt{GQL} and \texttt{SQL/PGQ}. 
For example, \texttt{Oracle Database 23ai} is the first commercially available system with \texttt{SQL/PGQ}  support~\citep{oracle} and several engines such as \texttt{PostgreSQL}~\citep{postgesPGQ}, \texttt{DuckDB}~\citep{duckPGQ} or \texttt{Goo\-gle\- Span\-ner\- Gra\-ph}~\citep{spanner} offer partial coverage of the standard. 
In the case of \texttt{GQL}, given the closeness of the language with \gqlCypher\, the coverage is more immediate, with multiple engines claiming full or partial support for the standard. 
These include \dbNeoFRj~\citep{gqlNeo}, \dbNebulaGraph~\citep{gqlNebula}, \dbUltipa~\citep{ultipa_graph}, \dbMillenniumDB~\citep{VrgocRAAABHNRR23}, \texttt{Goo\-gle\- Span\-ner\- Gra\-ph}, and \texttt{Mi\-cro\-soft\- Fa\-bric}~\citep{fabric} among others.

\item \texttt{Grem\-lin}. 
% https://en.wikipedia.org/wiki/Gremlin_(programming_language)
A graph traversal machine and language developed in the scope of the \texttt{Apa\-che\- Tin\-ker\-Pop} project~\citep{tinkerpop_site}.
This project's development and growth were promoted by the now-defunct \texttt{Ti\-tan} graph database~\citep{titan}, which was forked into the 
open-source \dbJanusGraph\ database~\citep{janus} and the commercial \dbDataStaxEnterprise\ solution~\citep{datastax}.
The traversal machine of \gqlGremlin\ is defined as a set of three components~\citep{rodriguez2015gremlin}: the data represented by a graph $G$, 
a traversal $\Psi$ (instructions) which consists of a tree of functions called \textit{steps}; a set of traversers $T$ (read/write heads).
With this composition, the \gqlGremlin\ and its traversal machine enable the exploration of multi-dimensional structures that \textit{model a heterogeneous 
set of ``things'' related to each other in a heterogeneous set of ways} as detailed in~\citep{rodriguez2015gremlin}.
A traversal evaluated against a graph may generate billions of traversers, even on small graphs, due to the exponential growth of the number of paths 
that exist with each step the traversers take.
Examples of databases supporting \gqlGremlin\ include: \dbOrientDB~\citep{DBLP:conf/bncod/0001DLT21}, \dbNeoFRj~\citep{Webber:2012:PIN:2384716.2384777}, \texttt{Da\-ta\-Stax\- En\-ter\-pri\-se}~\citep{datastax}, \texttt{In\-fi\-ni\-te\-Gra\-ph}~\citep{objectivitydb}, \dbJanusGraph~\citep{janus}, \dbAzureCosmosDB~\citep{paz2018introduction} or \dbAmazonNeptune~\citep{bebee2018amazon}, while the graph processing systems that allow its use are \gpeApacheGiraph~\citep{ching2013scaling} or \gpeApacheSpark~\citep{Armbrust:2015:SSR:2723372.2742797}.

\item \gqlSPARQL. 
% https://en.wikipedia.org/wiki/List_of_SPARQL_implementations
The standard query language for \texttt{RDF} data (triples), also known as the query language for the semantic web~\citep{sparql11}. 
We mention it due to its graph pattern matching capability~\citep{perez2009semantics} and scalability potential of querying large \texttt{RDF} graphs~\citep{huang2011scalable}.
As \texttt{RDF} is a directed labeled graph data format, \gqlSPARQL\ becomes a language for graph-matching.
Its queries have three components: a \textit{pattern matching part} allowing for pattern unions, nesting, filtering values of matchings, and choosing the data 
source to match by a pattern; a \textit{solution modifier} to allow modifications to the computed output of the pattern, such as applying operators as projections, 
orderings, limits and distinct; the \textit{output}, which can be binary answers, selections of values for variables that matched the patterns, construction of new 
\texttt{RDF} data from the values or descriptions of resources.
While graph databases may not necessarily be triple stores, the graph query languages they support may allow, for example, the \texttt{RDF}-specific 
semantics of a \gqlSPARQL\ query to be translated to \gqlCypher, \gqlGremlin\, or another language.
For example, \gqlSPARQL\ is also supported (analytics) over \texttt{GraphX}~\citep{schatzle2015s2x} and the higher-level graph analytics tool \gpeGraphFrames~\citep{bahrami2017efficient}. 
Most prominent systems supporting \gqlSPARQL\ are \texttt{Virtuoso}~\citep{erling2012virtuoso}, \texttt{Jena TDB}~\citep{JenaTDB}, \texttt{BlazeGraph}~\citep{blazegraph}, \dbAmazonNeptune~\citep{bebee2018amazon} and  \dbAllegroGraph~\citep{allegrograph}. 
Some new engines like \texttt{Stardog}~\citep{stardog}, \texttt{MillenniumDB}~\citep{VrgocRAAABHNRR23} or \texttt{QLever}~\citep{BastB17} are also gaining traction, most notably as contenders for large public service endpoints such as Wikidata~\citep{peter}. 
For a detailed overview of over a hundred \gqlSPARQL\ engines, please consult~\citep{AliSYHN22}.

\item \texttt{Gra\-ph\-QL}.
A framework developed and internally used at Facebook for years before its reference implementation was released as open-source~\citep{fb_graphql}.
It introduced a new type of web-based interface for data access.
As a framework, one of its core components is a query language for expressing data retrieval requests sent to web servers that are \texttt{Gra\-ph\-QL}-aware.
The queries are syntactically similar to \texttt{Ja\-va\-Scri\-pt\- Ob\-ject\- No\-ta\-tion (JSON)}.
The \texttt{Gra\-ph\-QL} specification implicitly assumes a logical data model implemented as a virtual, graph-based view over an underlying database 
management system~\citep{hartig2017initial}.
It has been studied with the semantics of its queries formalized as a labeled-graph data model, and the total size of a \texttt{Gra\-ph\-QL} response has been shown 
to be computable in polynomial time~\citep{hartig2018semantics}.
\texttt{Gra\-ph\-QL} is more than a query language - it defines a contract between the back-end and front-end over an agreed-upon type system, forming an 
application data model as a graph.
It is helpful as it proposes a decoupling between the back-end and front-end, allowing each component to be changed independently of the other.
For example, to serve queries in the graph, the data in the back-end could come from databases (e.g., \dbNeoFRj\ as the back-end serving \texttt{Gra\-ph\-QL} 
queries received at a web endpoint~\citep{neo4j_graphql}), in-memory representations, or other APIs.

%\item \textcolor{purple}{\sout{\texttt{PGQL}.
%The \texttt{Pro\-per\-ty\- Gra\-ph\- Que\-ry\- Lan\-gua\-ge}, based on the paradigm of graph pattern matching~\citep{van2016pgql}.
%It closely follows the syntactic structures of \texttt{SQL}, providing regular path queries with conditions on labels and properties to enable reachability 
%and path-finding queries. 
%The data types it defines are the intrinsic \textit{vertex}, \textit{edge}, \textit{path}, and intrinsic \textit{graph} type, allowing for graph construction 
%and query composition.
%It was motivated by the fact that \gqlSPARQL\ is the \texttt{RDF} standard query language, thus imposing that graphs be represented as a set of triples 
%(or edges), and by \gqlCypher's lack of support for regular path queries and graph construction as fundamental graph querying functionalities.
%\texttt{PGQL} also provides tabular output, allowing its queries to be naturally nested inside \texttt{SQL} queries, allowing for easy integration into existing database technology.
%It was developed by Oracle and used in Oracle's products, but its specification is open source.}} %\footnote{\url{https://analyticsanddatasummit2019.sched.com/event/Kyxe/pgql-a-query-language-for-property-graphs}}

\item \texttt{Academic languages}. 
There is a rich history of graph query languages coming from the academic community, some of these with a direct impact on query standards. 
Early works focused on extracting paths~\citep{CruzMW87} and introduced regular path queries (RPQs), which form the core of virtually all standards when it comes to path extraction. 
More complex patterns allowing to combine paths, relational joins and aggregation were soon introduced~\citep{ConsensM90} and adopted in the literature. 
Another example is \texttt{G-Log},  
a declarative query language on graphs, which was designed to combine the expressiveness of logic, the modeling of complex objects with identity, and the 
representations enabled by graphs~\citep{paredaens1995g}. 
In recent years many academic proposals directly influenced features of concrete query languages~\citep{GQLinfluences}. 
For example RPQs~\citep{CruzMW87} and their extensions with data tests~\citep{LibkinMV16} influenced the design of Oracle's \texttt{PGQL} and together with \texttt{STRU\-QL}~\citep{FernandezFLS97} and regular queries~\citep{ReutterRV17}, inspired the \texttt{G-CORE} proposal~\citep{gcore} from the Graph Data Council (formerly known as the Linked Data Benchmark Council).\footnote{Website:~\url{https://ldbcouncil.org/}}
\texttt{G-CORE} incorporates multiple query facets from academic proposals and unifies them in a single setting, including complex path patterns, relational operations and the ability to manipulate path as atomic objects. 
\texttt{G-CORE} is considered a fundamental influence on the design of \texttt{GQL} and \texttt{SQL/\-PGQ} patterns.

%\item \texttt{G-Log}. 
% A declarative query language on graphs, which was designed to combine the expressiveness of logic, the modeling of complex objects with identity, and the representations enabled by graphs~\citep{paredaens1995g}. 
% The authors describe it as a \textit{deductive language for complex objects with identity}, with a data model that captures the modeling capabilities of object-oriented languages, although lacking their typical data abstraction features related to system dynamism. 
% They claim that \texttt{G-Log} may be seen as a graph-based symbolism for first-order logic, and they prove that all sentences of first-order logic may be written in \texttt{G-Log}. 
% Secondly, they define the semantics of the language for database query evaluation. 
% Lastly, the authors state that its computation could be unnecessarily inefficient in the most general case because it is \textit{a very powerful language}. 
% We mention \texttt{G-Log} as it is historically relevant due to highlighting the importance and expressiveness of graph-based data models in  manipulating the relationships between data.\\

\end{itemize}

These different query languages were proposed across decades, some inspired by the lessons of their elders, others branching off from the same idea. 
They were influenced by the graph data models that were created to express the representation of graphs.
The standardization efforts for a graph query language are helpful to integrate the lessons from across the literature.

Together with the mentioned main languages, a non-negligible amount of proprietary languages has arisen. 
Often, a particular graph database product develops a custom flavor of an existing language to properly fit its specific cases. 
An example of a language based on \texttt{Cy\-pher} can be \texttt{Cy\-pher\-Plus} - a proprietary language of \texttt{Pan\-da\-DB}, or \texttt{Trans\-warp\- Ex\-ten\-ded} - a \texttt{Cy\-pher} extension of Stellar DB. 
Gizmo, on the other hand, is a \texttt{Grem\-lin}-like language of \texttt{Goo\-gle\- Cay\-ley} database, and \texttt{UQL} (\texttt{Ul\-ti\-pa\- Que\-ry\- Lan\-gua\-ge}) a \texttt{GQL}'s extension of Ultipa database.

Multi-model databases also like to extend \texttt{SQL} that is still the most familiar language for their users (\texttt{AQL} of \texttt{A\-ran\-go\-DB}, or the language of \texttt{Ori\-ent\-DB}). 
Naturally, Oracle's Spatial and Graph provides an \texttt{SQL}-based language \texttt{PGQL} (\texttt{Pro\-per\-ty\- Gra\-ph\- Que\-ry\- Lan\-gua\-ge}). 
But even a native graph database, such as  \texttt{Ti\-ger\-Gra\-ph}, queries its data with \texttt{GSQL}, a proprietary language based on \texttt{SQL}.

Some databases develop their graph languages more independently of the existing main approaches, such as \texttt{DQL} (\texttt{D\-gra\-ph\- Que\-ry\- Lan\-gua\-ge}) of  \texttt{D\-gra\-ph}, or \texttt{Ty\-pe\-QL} of  \texttt{Ty\-pe\-DB}. 
Development inspirations can also find their foundations in other known languages or concepts with \texttt{FQL} (\texttt{Fau\-na\- Que\-ry\- Lan\-gua\-ge}) of \texttt{Fau\-na\-DB} being an example of a \texttt{Ty\-pe\-Scri\-pt}-like approach, or a \texttt{JSON}-based \texttt{Flu\-ree\-QL} of the \texttt{Flu\-ree} database.

There are even projects focused on analytics which offer the ability to explore datasets using graph query languages without actually running 
a graph database, such as \gpeGRADOOP~\citep{DBLP:journals/pvldb/JunghannsKTGPR18} and \gpeGraphFrames~\citep{mishra2019graphframes}.

\subsection{Storage Architectures}~\label{chapter:publications:survey:sec:data_rep_and_indexing}

%TODO: consider subsections in this section: data representation, data distribution, indexing, query distribution

%TODO: mention computational representation of vertices/edges and the indexing (how is graph data and its properties accessed during processing?)

%%%%%%%%%%%%%%%%%%%%%%%%%%%%%%%%%%5
The combined efforts of academia and industry players 
have created a vibrant ecosystem of graph databases, the dimensions of which
we address in this survey. 
Among these dimensions, the matter of the internal representation 
of graphs becomes especially relevant and directly related to 
storage architecture design. 
%
%\textcolor{purple}{\sout{However, not all graph databases have readily available information on their storage architecture and graph representation designs (nor their choices).}}

In this section, we focus on selected unique storage architecture choices of the graph database solutions we analyzed, if provided.
We organize them into four main categories, according to the data structures they employ to architect the data storage and manage data internally: from more loosely to more strictly prescribed and advanced.

%We focus next on the storage architecture choices of some that are employed by some of the most popular graph database solutions we analyzed, as well as their uniqueness. 
% Although, not all graph databases had readily-available information on their internal storage architecture and graph representation designs (nor their choices). 

%########################  UNSTRUCT %%%%%%%%%%%%%%%%%%%%
\subsubsection{Unstructured or Loosely Structured Architectures}
In one extreme, a graph database may not adhere to  any specific external data structure and store the data just as binary (large) objects with no limitation to the internal structure.\\

\begin{itemize}

% BLOB
\item \textbf{BLOB-based Unstructured Data: }\dbPandaDB~\citep{zhao2021pandadb} is based on \texttt{Neo4j} and therefore, the data is stored in the form of 
a byte array, which is not useful by itself in understanding the metadata structure before deserializing it.
This database adds a semantic layer to the graph representation by manipulating unstructured data, which include 
metadata values such as MIME, length and version, among others. 
\dbPandaDB\ changes \texttt{Neo4j}'s property format by introducing binary large objects (BLOBs) as a data type 
to store unstructured data. 
Specifically, \texttt{Neo4j}'s \texttt{PropertyStore} byte array is adapted so that its last $28.5$ bytes store the metadata 
of unstructured data.
%For BLOBs over 10 kB, another file is used to store the binary content, such as a long string and array storage.
%If they are over 10 kB, they are not stored in native files but instead stored in \texttt{HBase}~\citep{DBLP:books/daglib/0027893}.
In summary, for unstructured data, the metadata is stored in the property store file (extended from \texttt{Neo4j}'s version), and 
the literal content is stored as BLOBs, either as separate files or in \texttt{HBase}.

The semantic information of the unstructured data can be computed and stored lazily or eagerly, with potential 
indexing in the latter case. 
Attention must be paid to the eager case, which demands a caching mechanism to maintain semantic information to 
avoid constant recomputation. 
It has a caching mechanism that assigns the values of a semantic space to a specific AI model with a serial number. 
When the AI model is updated, so is its serial number. 
The semantic values associated with that AI model are valid in the cache only if their associated serial number matches 
the latest serial number of the AI model.
\dbPandaDB\ chooses a suitable index for different semantic information, taking into account that, for example, 
facial features are represented as vectors or that the text content of the audio is in string format, among other examples. 
Numerical semantic data is indexed using the \texttt{B-\-Tree} structure~\citep{comer1979ubiquitous,graefe2011modern}, with an inverted 
index~\citep{witten1999managing,zobel2006inverted} being used for semantic information 
comprised of texts and strings, and high-dimension vector data relies on inverted vector search~\citep{buckley1985optimization}.
Indices can be built in batch or dynamic mode (the latter to update an already existing model based on new data).
Additional details can be found in~\citep{zhao2021pandadb}.

\end{itemize}

%######################## LINEAR STRUCT %%%%%%%%%%%%%%%%%%%%
\subsubsection{Linear Structured Architectures}

The most typical design for storage architecture employed by graph databases is to resort to a pre-defined linear data structure. 
This approach introduces some restrictions on how graph data is stored but simplifies management, allowing some degree of improved efficiency (e.g., direct access to properties) and extended semantics (e.g., transactional).\\

\begin{itemize}

\item \textbf{Double-linked list: } \dbNeoFRj~\citep{miller2013graph} stores properties as a double-linked list of property records, each holding a key and a value pointing to the next property. 
A specific type of record used in this scheme stores the content of properties in binary format.\footnote{See: \url{https://neo4j.com/developer/kb/understanding-data-on-disk/}}
\dbNeoFRj\ makes use of pointers for storing graph structure and properties of nodes and edges (which can become challenging if the entire 
graph does not fit into the server's memory).

\item \textbf{Sortledton:}~\citep{fuchs2022sortledton} is a universal graph data structure using an adjacency list-like design, 
which has one adjacency index and an adjacency list for each neighborhood.
The neighborhoods are sets of destinations, and the index is a map.
The authors chose this design based on their insight that optimizing for 
sequential vertex access is less beneficial than for any other pattern.
They consider three advantages: \textit{1)} an adjacency list-like data 
structure is embarrassingly parallel at the granularity of vertices because 
its neighborhoods are independent; \textit{2)} the maintenance of the index 
is simple and cheap because it is independent of changes to the neighborhoods;
\textit{3)} the adjacency list-like design allows reusing
well-studied map and set data structures from prior research.

\item \textbf{Record-based: }\dbArcadeDB~\citep{arcadedb_github} defines records as vertices, 
edges, and documents with a record ID (RID) assigned on creation. 
These are stored in pages (default size of 64KB) loaded in RAM as 
\texttt{By\-te\-Buf\-fer} objects.
There is the concept of \textbf{buckets}, which are physical files 
made of pages with multiple buckets per type to foster parallel 
job execution. 
Each bucket has an associated \textbf{index}, which uses the 
Log-structured merge-tree (LSM-Tree) algorithm~\citep{o1996log} for 
efficient insertion speed and concurrent compaction in the background. 
\dbArcadeDB\ enables data partitioning based on properties, which 
allows using a specific index on a lookup by key.

\item \textbf{Key-value abstraction: }\dbJanusGraph~\citep{janus} maps the graph to a key-value abstraction applied to an underlying storage layer (e.g., \dbCassandra\ or \texttt{HBase}).
While this achieves scalability, this abstraction also limits graph store flexibility, as storing node properties or adjacency lists as a single opaque object impedes finer-grained access to elements.

\item \textbf{Transactional Log-based: }\texttt{Transactional Edge Log} (\texttt{TEL})~\citep{zhu2019livegraph} 
is a data structure introduced by \dbLiveGraph\, which was designed in 
tune with the database's concurrency control algorithm. 
\texttt{TEL} combines a sequential memory layout with multi-versioning, 
containing cache-aligned timestamps and counters, which are leveraged 
to preserve the sequential nature of scans even when processing 
concurrent transactions.
%Coupled with the regular execution flow of the concurrency control algorithm, \texttt{TEL} enables speculation and prefetching. 
Vertices are stored in an individual way in what are described 
as \textit{vertex blocks}, while edges are grouped in what the 
authors describe as \texttt{TELs}, which are stored 
in a single large memory-mapped file managed by a memory allocator 
in \dbLiveGraph. 
There is a vertex index and an edge index, which store pointers 
to appropriate blocks via vertex/edge ID. 
Vertices are used with a standard copy-on-write approach so that the newest version of the vertex can be found 
through the vertex index, with each version pointing to 
the previous one within the vertex block.
Adjacency lists are stored as multi-versioned logs to 
support snapshot isolation efficiently, allowing read 
operations to proceed while avoiding interference with each 
other and write operations. 
% \textcolor{purple}{\sout{The default isolation level of \dbNeoFRj\ and many other commercial 
% DBMSs is read-committed, and the snapshot isolation used in \dbLiveGraph\ 
% is stronger (see Sections 3 and 6 of~\citep{zhu2019livegraph}).}
While the default isolation level of many other commercial DBMSs, such as \dbNeoFRj, is read-committed, \dbLiveGraph\ provides stronger snapshot isolation (see~\citep{zhu2019livegraph}).

\end{itemize}

%########################  LAYERED  %%%%%%%%%%%%%%%%%%%%
%\subsubsection{DAG-based Structured Architectures}
%\textcolor{purple}{TODO COMMENT: What does "DAG" stand for here? As we are in the area of graphs, the first comes into mind Directed Acyclig Graph, which isn't be the case.}
\subsubsection{Non-Linear Structured Architectures}
Alternatively, graph databases can define architectures that use non-linear data structures, in order to organize data storage and management. 
This approach is typically taken to achieve  improved flexibility (e.g., tracking modifications) or performance (e.g., indexing) at the expense of simplicity.  \\

\begin{itemize}

\item \textbf{Tree-based: }\dbByteGraph's \texttt{edge-tree}~\citep{li2022bytegraph} 
is a data structure similar to the B-tree to store vertex 
adjacency lists.
It can be configured to fit different workloads with various 
read-write requirements in order to balance read and write 
amplifications.
Regarding the durable storage layer, a persistent key-value 
store is used for fine-grained storage management of the 
\texttt{edge tree}. 
This is due to most data not having a fixed size (e.g., 
the number of neighbors or the length of edge/vertex 
properties). 
Each adjacency list is divided according to edge types 
and directions, with each list being stored as what 
the authors name the \texttt{edge tree}. 
It is composed of three types of nodes: root node, 
meta node, and edge node. 
The first two serve the purpose of indexing, while 
the edge nodes store the physical edge data. 
An \texttt{edge tree} will begin only with the 
root node and edge nodes, but as it increases in 
edge number, the meta nodes are created as a 
middle layer to index edge nodes (see Section 4.1 of 
~\citep{li2022bytegraph}).

\item \textbf{Layered abstraction: }
\dbTerminusDB\ introduces its \textit{\texttt{ter\-mi\-nus-sto\-re}} data structure~\citep{van2020succinct}, that may be regarded actually as a meta architecture. 
% https://terminusdb.com/blog/succinct-data-structures-for-modern-databases/
In fact, it represents a graph database using layers that conceptually map 
to the scheme of objects used in the \texttt{git} version control system. 
Each layer is identified by a unique 20-byte name. 
The initial layer represents a simple graph by using of succinct 
data structures such as front-coded dictionaries,
bit sequences, and wavelet trees inspired by \texttt{HDT}~\citep{martinez2012exchange}.
New layers are added over the initial one to add new 
\texttt{RDF} subjects, objects, predicates, or values~\citep{van2020succinct}.

\end{itemize}

\subsubsection{Relational storage}
Many systems rely on classical relational architecture to store their data. 
Although this might be obvious for engines that aim to support SQL/PGQ, the practice is far more prevalent, particularly in the RDF setting, where the dataset is viewed as one giant table with three columns denoting the subject-predicate-object element of an RDF triple.
Similar modeling is also possible in the case of property graphs. 
Engines can roughly be separated into the two common relational storage categories: row-based and column-based.
\begin{itemize}
    \item \textbf{Row-based storage.} Traditionally many RDF/SPARQL engines deploy this approach and view data as one big table (commonly dubbed \textit{triplestore})~\citep{AliSYHN22}. 
    The table itself might contain three or four columns, depending on whether named graphs  are supported. In many engines the underlying architecture is based on B+trees indexes which contain dictionary encoded IRIs in multiple subject-predicate-object permutations to speed up query execution. 
    These can be clustered or unclustered, depending on the particular engine. 
    Notable examples include \texttt{Vir\-tuo\-so}~\citep{erling2012virtuoso}, \texttt{TDB} (an RDF storage and query component of \texttt{Je\-na})~\citep{JenaTDB}, \dbBlazegraph~\citep{blazegraph}, \dbMillenniumDB~\citep{VrgocRAAABHNRR23} and \texttt{RDF-3X}~\citep{NeumannW08}. 
    Some engines like \texttt{MillenniumDB} also support property graphs, storing them using the same clustered B+tree architecture used in their RDF engine.
    \item \textbf{Column-based storage.} Columnar engines are prevalent in data analytics workspace and are increasingly being used for managing graph data. 
    For example \dbKuzu~\citep{JinFCLS23} is a fully fledged graph database using columnar storage. 
    Another example is \texttt{Duck\-DB}'s extension \texttt{Duck\-PGQ}~\citep{duckPGQ}. 
    Both of these deploy some sort of CSR-based encoding of the graph, which can be naturally stored in a columnar format. 
    Interestingly \texttt{Vir\-tuo\-so} offers both row-based and columnar storage capabilities.
\end{itemize}

%######################## ADVANCED %%%%%%%%%%%%%%%%%%%%
\subsubsection{Advanced Storage Architectures}
Finally, we address graph databases that employ more advanced storage architectures. 
They aim to address two main objectives: i) space efficiency through some form of information summarization or compression, and ii) scalability (and also performance) by making use of (distributed) shared memory to manage data across multiprocessors or clusters of machines.\\

\begin{itemize}

\item \textbf{Succinct Data Structures: }\dbZipG~\citep{khandelwal2017zipg} incorporates 
summarization techniques from the \texttt{Succinct} 
data structure~\citep{agarwal2015succinct} to 
achieve a memory-efficient representation. 
It transforms the input graph into a flat unstructured file, strategically storing a small amount of metadata 
along with the original input graph data in a way that the primitives of \texttt{Succinct} can be extended 
to efficiently implement the interactive graph queries of workloads such as those of Facebook \texttt{TAO}.
Each server in \dbZipG\ stores a set of update pointers to ensure that during execution, only the necessary 
logs (and their relevant bytes) are accessed for query execution.
The layout of \dbZipG\ has two flat unstructured files. 
It has the \texttt{No\-de\-Fi\-le}, which stores IDs of all vertices with a small amount of metadata to trade off storage 
(in uncompressed representation) for efficient random access into node properties.
The other file is the \texttt{Ed\-ge\-Fi\-le}, which holds all edge records and uses metadata and conversion of variable-length 
data into fixed-length data (before compressing) to enable the trade-off between storage (uncompressed representation) 
and optimization of random access into the records and complex operations such as binary searches over timestamps. 
For more details see Sections 2 and 3 of~\citep{khandelwal2017zipg}.

\item{\textbf{Data Compression:}}
There are data compression schemes where the graph is compressed by different techniques in the literature. 
Among these techniques, we find examples like the compressed sparse 
row (CSR)~\citep{bulucc2009parallel}, also known as 
the \textit{Yale format}, the compressed sparse column 
variant, the 
\texttt{Web\-Gra\-ph}~\citep{DBLP:conf/dcc/BoldiV04} format, and 
the $k^2$-tree structure~\citep{DBLP:journals/is/BrisaboaCBN17,pk-graph}, 
among others, enabling the analysis of graphs over a compressed 
representation, effectively enabling graph analysis with a 
lower memory footprint while using commodity hardware.

These techniques take advantage of aspects such as sparsity and the application of ordering criteria to its elements, 
to name a few.
However, changing the graph typically requires a conversion to 
an uncompressed format, applying the graph-modifying operations 
and then converting back to the compressed representation.
To overcome this bottleneck, the literature also saw the 
creation of compact graph schemes, enabling the modification 
of the graph while still retaining the benefits of compression. 
Techniques may be used to enable modifications over sets of 
static structures~\citep{DBLP:conf/pods/MunroNV15}, with a recent 
example of a dynamic $k^2$-tree~\citep{k2tree_extended,pk-graph}.
Other techniques exist, harnessing the benefits of compression 
with very low-overhead decompression operations as seen in 
\texttt{Log\-Gra\-ph\-(Gra\-ph)}~\citep{besta2018log}.

\item \textbf{Shared Memory: }\dbGTran's graph-native data 
store~\citep{chen2022g} was designed to address the 
challenges of graph data irregularity and multi-hop 
traversal-based query costs. 
The memory space for each machine is split into two parts, 
one following an architecture where nothing is shared 
to store a local piece of the graph partition, while 
the other part obeys a shared-memory architecture, composing 
an RDMA-based distributed memory space to enable remote access 
of property data.
The data store is composed of a \texttt{sto\-ra\-ge} 
\texttt{lay\-er} and an \texttt{OLTP} layer. 
It offers a multi-version schema to support a consistent 
view for concurrent transactions.

The \texttt{sto\-ra\-ge} \texttt{lay\-er} represents a 
property graph in two parts: one part for graph topology, 
the vertices and edges as adjacency lists; the other 
part represents property data (the keys and values 
of vertices and edges). 
\dbGTran\ uses an edge-cut partitioning strategy with 
hashing to partition a graph into $N$ shards among $N$ 
server nodes, with each shard storing one shard of 
vertices together with their in/out edges. 
A global address space is built over all deployed nodes 
so that the location of any object in the 
\texttt{sto\-ra\-ge} \texttt{lay\-er} can be retrieved 
by ID.
The \texttt{OLTP} layer has a group of worker threads 
to process incoming transactions by interacting with 
the data store, with each server node registering 
a chunk of memory at NIC, dividing the memory space 
of the node into RDMA-allocated memory and local memory. 
This is done to allow remote data access and fast 
communication.
For more details, see Sections 4.1 and 4.2 
of~\citep{chen2022g}.\\

\end{itemize}

The nature of these contributions is important in the realm 
of space-efficiency, but additional dimensions must be 
considered when thinking of integrating them in full-fledged 
databases.
The desire to enable parallel processing within a machine 
and/or across machines in a cluster adds a layer of complexity 
in the design of storage architecture of graph databases.
Some graph databases use compression techniques with an explicit 
focus on single-machine setups, offering only the feature of 
high-availability to the cluster scenario 
(e.g., \texttt{Neo4j}~\citep{neo4j_graph_db}).
Others were designed with sharding in mind (distributing the 
graph across machines), demanding the integration of the internal 
graph representation with considerations pertaining to 
consistency, coordination and latency of the storage architecture 
when designing it (e.g works adressing consistency~\citep{graph-db-integrity-17, graph-db-inconsistency-24}). This is required not only to distribute a graph across machines 
but also to enable graph queries or global graph algorithms to 
effectively make use of all the computing resources. 

%Among the graph databases mentioned in this document, there are those with a graph-oriented storage design
%\textcolor{cyan}{~LV: a couple of cites here would be ok.} 
%\textcolor{green}{~AF: either adding a cite e.g., or removing JanusGraph at the end of this sentence, since above we have the same issue when we mention Neo4j}, 
%while others support the graph representation over a generic storage (e.g. \texttt{Ja\-nus\-Gra\-ph}~\citep{janus}).

% DO NOT DELETE THIS COMMENTED SECTION.
% \subsection{Graph Datasets and Resources}~\label{chapter:publications:survey:sec:datasets_and_resources}

% WDC - Hyperlink Graphs
% http://webdatacommons.org/hyperlinkgraph/index.html\#toc1

% Netdata
% http://www-personal.umich.edu/~mejn/netdata/

% CovidGraph
% https://covidgraph.org/

% Graph Challenge Champions | Graph Challenge
% https://graphchallenge.mit.edu/champions

% Laboratory for Web Algorithmics
% https://law.di.unimi.it/datasets.php

% Web Data Commons
% http://webdatacommons.org/

% House of Graphs
% https://hog.grinvin.org/

% LDBC Social Network Benchmark~\citep{erling2015ldbc}

% Beyond Macrobenchmarks: Microbenchmark-based Graph 
% Database Evaluation~\citep{lissandrini2018beyond}

% Facebook LinkBench TAO~\citep{armstrong2013linkbench}
% END inlined from sections/model_and_query_languages/storage_architectures.tex

% END inlined from sections/model_and_query_languages/model_and_query_languages.tex

%%%%%%%%%%%%%%%%%%%%%%%%%%%%%%%%%%%%%%%%%%%%%%%%
%%%%%%%%%%%%%%%%%%%%%%%%%%%%%%%%%%%%%%%%%%%%%%%% GRAPH DATABASES
%%%%%%%%%%%%%%%%%%%%%%%%%%%%%%%%%%%%%%%%%%%%%%%%

% BEGIN inlined from sections/graph_databases.tex
\section{Graph Databases}\label{chapter:publications:survey:sec:graph-databases}
%
%\textcolor{purple}{TODO intro of this section}
There are many dimensions to graph databases, both in the realm of the technological challenges their developers must address and also in the breadth of features/properties that different systems offer.
In this section we highlight relevant aspects which must be considered in the design and development of graph databases. 
This is followed by a comprehensive set of features which are crucial for evaluating graph databases as well as an analysis of the features across a plethora of databases from both industry and open-source origins. 
The purpose of this extensive  analysis is to provide guiding principles for the analysis and choice of graph database solutions, raising awareness of both existing graph database system as well as aspects that must be taken in consideration for the analysis and choice of graph databases.
This section thus sheds light on key-aspects for both those who would set out to design and construct graph databases as well as those who would develop with and use them.

\subsection{Challenges}\label{chapter:publications:survey:sec:graph-databases:challenges}

In the scope of distributed graph transaction processing, 
several challenges have been noted in the literature~\citep{chen2022g}:

\begin{enumerate}

\item The irregularity of graph data~\citep{sun2015sqlgraph} can lead 
to inadequate locality for reads and writes after continuous updates.

\item After multi-hops (fan-out~\citep{shalita2016social}), the cost 
of traversal-based queries can become very high. 
The reason lies in the common power-law distribution on vertex 
degrees and the small-world phenomenon (i.e., high clustering coefficient and low distances) of many real-world graphs.

\item As a result, very large read/write sets may be 
involved in graph transactions. 
Adding to this, the connectedness of graph data can increase 
the contention/abort rate and lead to lower throughput when processing 
transactions.

\item Traditional centralized system architectures (with a primary 
node coordinator) may limit OLTP scalability (with the already mentioned contention likelihood, CPU overheads, and network bandwidth being typical sources of bottlenecks~\citep{zamanian2016end}).

\end{enumerate}

\subsection{Databases and Storage Systems Review}\label{chapter:publications:survey:sec:graph-databases:dbs-and-systems}

In this section, we list and describe different graph databases, which we group according to their type of supported graph model (e.g., property graph model, \texttt{RDF}, hybrid). 

\textbf{How to read this section.} 
Section~\ref{chapter:publications:survey:sec:graph-databases:dbs-and-systems} can be read either top-down, or selectively by graph model: property graphs (Section~\ref{chapter:publications:survey:sec:graph-databases:dbs-and-systems:prop_graph}), RDF engines (Section~\ref{chapter:publications:survey:sec:graph-databases:dbs-and-systems:RDF}), multi-model systems (Section~\ref{chapter:publications:survey:sec:graph-databases:dbs-and-systems:multiple_models}), and alternative models (Section~\ref{chapter:publications:survey:sec:graph-databases:dbs-and-systems:alternative_models}).
For a quick overview and cross-system comparison, we recommend starting with Tables~\ref{new:table:graph_database_features_1_1_1} to~\ref{new:table:graph_database_features_3_2_2}, which summarize feature coverage across all systems, and then jumping to the summarized per-system descriptions for details.
The overall aggregate analysis and main takeaways from the feature distribution analysis are discussed in Section~\ref{chapter:publications:survey:sec:analysis}.

We address the graph query languages of the graph databases. 
They either implement their own custom language (neglecting interoperability even though there may be automatic converters developed by third parties), or adhere to those that have been widely used and adopted by many projects (e.g., \gqlCypher, \gqlGremlin, \gqlSPARQL). 
The list includes open-source and commercial products, as well as database systems implemented to explore novel techniques as part of research activities.

For further information on the evolution of graph databases, we point the reader to contributions~\citep{angles2008survey,Pokorny2015}, which cover, among other aspects, data models and query languages. 

In cooperation with industry, we gathered forty six features relevant for potential users of graph databases. 
We group these into three main dimensions (or areas) (\textit{Product}, \textit{Database} and \textit{Data}), each divided into a few classes consisting of up to seven features (or aspects) that we evaluated for every listed graph database. 

This is illustrated in Figure~\ref{fig:dim:class:feat:tree}. We address each dimension after explaining our scoring approach. 

% BEGIN inlined from tables/dimension-class-feature-tree-v2.tex
\begin{figure}[h!]
\centering
%\caption{Description of dimensions, classes and features of analysis.}

% Curly-braces tree with aligned hierarchy columns
\begingroup
\renewcommand{\arraystretch}{1.15}
\setlength{\arraycolsep}{3pt}

% --- choose widths based on the longest labels in your data ---
\newlength{\DimW}
\newlength{\ClassW}
\settowidth{\DimW}{\textbf{Database}}
\settowidth{\ClassW}{\textit{Software Development}}

\newcommand{\Dim}[1]{\text{\makebox[\DimW][l]{\textbf{#1}}}}
\newcommand{\Cls}[1]{\text{\makebox[\ClassW][l]{\textit{#1}}}}
\newcommand{\Feat}[1]{\text{#1}}

\begin{adjustbox}{center,max width=\linewidth,max totalheight=0.9\textheight}
$
\displaystyle
\Dim{Dimensions}\ \ \ \;=\;
\left\{
\begin{array}{@{}l@{}}
% -------------------- Product --------------------
\Dim{Product}\;=\;
\left\{
\begin{array}{@{}l@{}}
\Cls{Adoption}\;=\;
\left\{
\begin{array}{@{}l@{}}
\Feat{Active development}\\
\Feat{Commercial support}\\
\Feat{Live community}\\
\Feat{Open source}\\
\Feat{Pricing}\\
\Feat{Trendiness}
\end{array}
\right.\\[6pt]
\Cls{Deployment}\;=\;
\left\{
\begin{array}{@{}l@{}}
\Feat{Containerization}\\
\Feat{Work as dedicated instance}\\
\Feat{Work as embedded}\\
\Feat{Testing in-memory version}\\
\Feat{Operating on \texttt{Linux}}\\
\Feat{Operating on \texttt{Windows}}\\
\Feat{SaaS offering}
\end{array}
\right.
\end{array}
\right.\\[10pt]
% -------------------- Database --------------------
\Dim{Database}\;=\;
\left\{
\begin{array}{@{}l@{}}
\Cls{Convenience}\;=\;
\left\{
\begin{array}{@{}l@{}}
\Feat{Automatic updates}\\
\Feat{Client-side caching}\\
\Feat{Data versioning}\\
\Feat{Live backups}
\end{array}
\right.\\[6pt]
\Cls{Distribution}\;=\;
\left\{
\begin{array}{@{}l@{}}
\Feat{Cluster re-balancing}\\
\Feat{Data distribution}\\
\Feat{High-availability}\\
\Feat{Query distribution}\\
\Feat{Replication support}
\end{array}
\right.\\[6pt]
\Cls{Software Development}\;=\;
\left\{
\begin{array}{@{}l@{}}
\Feat{Data types defined}\\
\Feat{Logging/Auditing}\\
\Feat{Object-graph mapper}\\
\Feat{Reactive programming}\\
\Feat{Documentation up-to-date}
\end{array}
\right.
\end{array}
\right.\\[10pt]
% -------------------- Data --------------------
\Dim{Data}\;=\;
\left\{
\begin{array}{@{}l@{}}
\Cls{Access}\;=\;
\left\{
\begin{array}{@{}l@{}}
\Feat{Binary protocol}\\
\Feat{CLI}\\
\Feat{GUI}\\
\Feat{Multi-database}\\
\Feat{(graph-)Native}\\
\Feat{REST API}\\
\Feat{Query language}
\end{array}
\right.\\[6pt]
\Cls{Consistency}\;=\;
\left\{
\begin{array}{@{}l@{}}
\Feat{Granular locking}\\
\Feat{Multiple isolation levels}\\
\Feat{Read committed transaction}\\
\Feat{Transaction support}
\end{array}
\right.\\[6pt]
\Cls{Control}\;=\;
\left\{
\begin{array}{@{}l@{}}
\Feat{Constraints}\\
\Feat{Schema}\\
\Feat{Secondary indexes}\\
\Feat{Server-side procedures}\\
\Feat{Triggers}
\end{array}
\right.\\[6pt]
\Cls{Security}\;=\;
\left\{
\begin{array}{@{}l@{}}
\Feat{Authentication}\\
\Feat{Authorization}\\
\Feat{Data encryption}
\end{array}
\right.
\end{array}
\right.
\end{array}
\right.
$
\end{adjustbox}

\endgroup
\caption{Description of dimensions, classes and features of analysis.}
\label{fig:dim:class:feat:tree}
\end{figure}

\subsubsection{Scoring Methodology}\label{chapter:publications:survey:sec:graph-databases:dbs-and-systems:scoring_methodology}

We identified candidate systems from benchmark/industry lists, prior surveys, and curated community directories, and retained those providing persistent graph storage and a documented graph model and query/access interface. 
For each feature, we relied primarily on official vendor documentation, repositories/release notes, authoritative benchmark/academic sources and developer mailing lists. 
A feature was credited only when explicitly documented or evidenced in an official release; ambiguous cases were scored conservatively and the rationale recorded in the per-system appendix in Tables~\ref{new:table:feature_description_P1} and~\ref{new:table:feature_description_P2}. 
Intermediate scores (0.25/0.5/0.75) capture limited/partial support. 
Disagreements were resolved by re-checking primary sources and reaching consensus. 
%The full evidence trail and extracted data are provided in the companion repository to facilitate reproduction and future updates.

\subsubsection*{Product}

The Product dimension contains features specific for embracing a software product as well as its provided manipulation means. 
These features are, therefore, common for any software system but at the same time considered to be very important for potential graph-database users. 
We divide them into the following two classes with their respective features.
%, ordered alphabetically.
\begin{itemize}
    \item \textit{Adoption} - Active development / Commercial support / Live community / Open source / Pricing / Trendiness.
    %\item \textit{Deployment} - Containerization / Embedded option / In-memory option / Operates on \texttt{Li\-nux} / Operates on \texttt{Win\-dows} / Saas Offering / Standalone instance option.

    \item \textit{Deployment} - Containerization /
Work as dedicated instance /
Work as embedded /
Testing in-memory version /
Operating on \texttt{Li\-nux} /
Operating on \texttt{Win\-dows} /
SaaS offering.

\end{itemize}

For \textit{active development} and \textit{trendiness}, we consider a reference date of \FeaturesReferenceDate\ for scoring, and we check the relevant activity in the time preceding this date.
We note that \textit{active development} has been obtained from a mixture of automated and manual verifications. 

For databases with public repositories in GitHub/GitLab/SourceForge, automatic verification is deployed to check the date of the latest commit, release and details of alternative branches in the event that the master branch does not demonstrate relevant activity within the last six months before the reference date. 
If a repository has been expressively archived or the repository discontinued with no mention of a replacement, the corresponding database receives a score of 0 for \textit{active development}.
Tables~\ref{table:annex_fact_1}--\ref{table:annex_fact_5} in the paper appendix show the details of activity verification for all the databases considered in this survey.

In the event that databases do not have repositories available, manual verification is performed by checking the date of the latest release and the date of the latest update on the official website, as well as checking for any news or announcements regarding the database.

Regarding the \textit{trendiness} feature, it is based on downloaded CSV data from Google Trends. 
We note that due to the topic overlap for certain database names, it was necessary to adjust the query keywords for some databases to obtain more accurate results:

\begin{itemize}
  \item \dbChronoGraph\ used the expression \textit{ChronoGraph graph database} (colliding with the concept of watches).
  \item \dbGaffer\ used the expression \textit{Gaffer graph database} (colliding with a type of tape).
  \item \dbGraphflow\ used the expression \textit{Graphflow graph database}.
  \item \dbKuzu\ used the expression \textit{Kuzu graph} (colliding with the name of a fibrous vine native to East Asia).
  \item \dbLiveGraph\ used the expression \textit{LiveGraph storage}.
  \item \dbTAO\ used the expression \textit{TAO graph database} (colliding with unrelated Bittensor's \$TAO cryptocurrency).
  \item \dbVirtuoso\ used the expression \textit{Virtuoso graph} (colliding with the concept of a virtuoso).
  \item \dbWeaver\ used the expression \textit{Weaver transactional graph database} (colliding with a verb with many overlaps).
  \item \dbZipG\ used the expression \textit{ZipG graph} (colliding with the concept of a zip and the name of a file format).
\end{itemize}

\subsubsection*{Database}

Features specific for any database product, specifically focused on needs of graph database users, are grouped within the Database dimension. 
These features are categorized into three main classes covering distribution capabilities, specifics for software development on top of the graph database, and features enabling smooth experience.
\begin{itemize}
    \item \textit{Convenience} - Automatic updates / Client-side caching / Data versioning / Live backups.
%    \item \textit{Distribution} - Cluster re-balancing / Data distribution / High availability / Query distribution / Replication support.

\item \textit{Distribution} -     Cluster Re-balancing /
Data Distribution /
High-Availability /
Query Distribution /
Replication support.

    %\item \textit{Software Development} - Data types / Logging/Auditing / Object-Graph Mapper / Reactive programming / Up-to-date documentation.

 \item \textit{Software Development} - Data types defined /
Logging/Auditing /
Object-Graph Mapper /
Reactive programming  /
Documentation up-to-date.

\end{itemize}

\subsubsection*{Data}

In the Data dimension, we address operations with data organized into four main feature classes: \textit{Access}, \textit{Consistency}, \textit{Control} and \textit{Security}. 
Data access covers means of read/write operations such as API and query languages as well as architectural level of access - whether the database is graph-native or accesses the data through an extra graph layer. 
Other areas focus on ensuring proper locking and transactional support, schema and optimization capabilities, and user access, rights, and encryption.
\begin{itemize}
    \item \textit{Access} - Binary protocol / CLI / GUI / Multi-database / (graph-)Native / REST API / Query language.
    \item \textit{Consistency} - Granular locking / Multiple isolation levels / Read committed transaction / Transaction support.
    \item \textit{Control} - Constraints / Schema / Secondary indexes / Server-side procedures / Triggers.
    \item \textit{Security} - Authentication / Authorization / Data encryption.
\end{itemize}

For completeness, we include a short description of each feature in the paper appendix (Tables~\ref{new:table:feature_description_P1} and~\ref{new:table:feature_description_P2}). 
They can also be accessed in our GitHub repository devoted to this survey.\footnote{\url{https://github.com/svitaluc/survey-gdbs}}

%\subsubsection*{Overall Analysis}

\subsubsection*{Review Results}

All the systems were extensively reviewed and the result of their analysis is shown in Tables~\ref{new:table:graph_database_features_1_1_1} to~\ref{new:table:graph_database_features_3_2_2}. 
The tables show the assessment of all the considered features  on  the databases studied, with features lined up grouped by class (double line separators), and classes ordered by their area - or dimension (triple line separator), following the same sequence by which they were described earlier.

The scoring assigned to each feature always ranges in the interval from 0 to 1 (i.e. $0$, $0.25$, $0.50$, $0.75$, $1.00$) based on whether feature is absent ($0$), or completely fulfilled/present/provided/supported ($1.00$). 
When relevant and possible, we also differentiate when this is achieved only limitedly ($0.25$), partially ($0.50$), or significantly ($0.75$). 
For clarity and ease of reference, we resort to associating such values with a graphical representation (i.e., respectively \emph{empty \pie{0}, quarter-filled \pie{90}, half-filled \pie{180}, three quarter-filled \pie{270}, and full circle \pie{360}}).
%

%Since the description of every feature and their rules for evaluation would be too extensive for this paper, they can be accessed either in the Appendix TODO YES OR NO?, or in our github repository devoted to this survey\footnote{\url{https://github.com/svitaluc/survey-gdbs}}.

Next, we will further cover individual systems in detail analyzing how they cumulatively fulfill or address each of the dimensions (or areas), as a whole, along with a short description of more relevant aspects.
For every dimension (or area) (\emph{Product, Database} and \emph{Data}) we summarize the evaluation of its features to one index, counted as a simple average of all the underlying numerical values of feature scores (in the scale 0, 0.25, 0,50, 0,75, 1.00 presented earlier). 
%All the features are always assigned with a rating starting from zero signaling feature absence up to a maximum of one, meaning the feature is fully offered. 
%Therefore, the index also ranges in the interval from zero to one, denoting a better offer in a given field with a higher resulting value.

Note that the evaluation is based on information available to public and on scientific papers, if present. 

Regarding the open source feature, we note that the absence of evidence is not necessarily evidence of absence (especially regarding enterprise products): to the best of our ability, we have analyzed each system's information ecosystem for clues on licensing status.

Consequently, proprietary databases of private companies (such as \dbTAO\ of Meta or \dbByteGraph\ of ByteDance) 
may have more features than publicly reported that are understandably kept as a corporate secret.

Our evaluation is necessarily documentation-driven, relying on publicly available material and scientific papers when present.
When a feature is not publicly documented, we conservatively score it as absent or not verifiable according to the rubric (Tables~\ref{new:table:feature_description_P1} and \ref{new:table:feature_description_P2}).
Therefore, the reported scores should be interpreted as a lower bound on publicly documented capabilities, and may underestimate proprietary/enterprise systems whose implementations or operational features are not disclosed.

We mitigate this by prioritizing primary sources and applying consistent, conservative scoring in ambiguous cases.

%%%%%%%%%%%%%%%
%%%%%%%%%%%%%%% PART I
%%%%%%%%%%%%%%%

\setlength{\tabcolsep}{3pt}
\renewcommand{\arraystretch}{1.1}

\begin{table}[!t]
\centering
\caption{Summary of graph database distinctive features (Part I-a).}
\label{new:table:graph_database_features_1_1_1}

\small
\setlength{\tabcolsep}{4pt}
\renewcommand{\arraystretch}{1.15}

\begin{adjustbox}{max width=\textwidth}
\begin{tabular}{@{}>{\centering\arraybackslash}p{0.85cm}>{\centering\arraybackslash}p{1.15cm}l*{9}{c}@{}}
\toprule
\multicolumn{2}{@{}c}{} & \multicolumn{1}{@{}l}{Feature}
& \rot{Alibaba GDB}
& \rot{ChronoGraph}
& \rot{DataStax Enterprise}
& \rot{Dgraph}
& \rot{Graphflow}
& \rot{JanusGraph}
& \rot{NebulaGraph}
& \rot{Neo4j}
& \rot{RedisGraph/FalkorDB}
\\
\midrule

% BEGIN inlined from tables/pretty/new-table1_1_1.tex
\multirow{13}{*}{\rowcat{Product}} & \multirow{6}{*}{\rowsubcat{Adoption}} & Active development & \pie{0} & \pie{360} & \pie{0} & \pie{360} & \pie{0} & \pie{360} & \pie{360} & \pie{360} & \pie{180} \\
 &  & Commercial support & \pie{360} & \pie{0} & \pie{360} & \pie{360} & \pie{0} & \pie{0} & \pie{360} & \pie{360} & \pie{360} \\
 &  & Live community & \pie{0} & \pie{0} & \pie{90} & \pie{180} & \pie{0} & \pie{360} & \pie{360} & \pie{360} & \pie{360} \\
 &  & Open source & \pie{0} & \pie{360} & \pie{0} & \pie{360} & \pie{360} & \pie{360} & \pie{360} & \pie{360} & \pie{360} \\
 &  & Pricing & \pie{360} & \pie{0} & \pie{0} & \pie{0} & \pie{360} & \pie{360} & \pie{0} & \pie{180} & \pie{360} \\
 &  & Trendiness & \pie{0} & \pie{0} & \pie{0} & \pie{360} & \pie{0} & \pie{360} & \pie{360} & \pie{360} & \pie{360} \\
\addlinespace[2pt]
\cmidrule(lr){2-12}
\addlinespace[2pt]
 & \multirow{7}{*}{\rowsubcat{Deployment}} & Containerization & \pie{0} & \pie{0} & \pie{360} & \pie{360} & \pie{0} & \pie{360} & \pie{360} & \pie{360} & \pie{360} \\
 &  & Work as dedicated instance & \pie{360} & \pie{0} & \pie{360} & \pie{360} & \pie{360} & \pie{360} & \pie{360} & \pie{360} & \pie{360} \\
 &  & Work as embedded & \pie{0} & \pie{360} & \pie{0} & \pie{0} & \pie{0} & \pie{360} & \pie{0} & \pie{360} & \pie{0} \\
 &  & Testing in-memory version & \pie{0} & \pie{0} & \pie{0} & \pie{360} & \pie{180} & \pie{360} & \pie{0} & \pie{360} & \pie{360} \\
 &  & Operating on Linux & \pie{360} & \pie{360} & \pie{360} & \pie{360} & \pie{360} & \pie{360} & \pie{360} & \pie{360} & \pie{360} \\
 &  & Operating on Windows & \pie{0} & \pie{360} & \pie{180} & \pie{360} & \pie{360} & \pie{360} & \pie{180} & \pie{360} & \pie{270} \\
 &  & SaaS offering & \pie{360} & \pie{360} & \pie{360} & \pie{360} & \pie{360} & \pie{360} & \pie{180} & \pie{360} & \pie{0} \\
\addlinespace[2pt]
\midrule
\addlinespace[2pt]
\multirow{14}{*}{\rowcat{Database}} & \multirow{4}{*}{\rowsubcat{Convenience}} & Automatic updates & \pie{360} & \pie{0} & \pie{0} & \pie{0} & \pie{0} & \pie{0} & \pie{0} & \pie{0} & \pie{0} \\
 &  & Client side caching & \pie{0} & \pie{0} & \pie{0} & \pie{0} & \pie{0} & \pie{0} & \pie{0} & \pie{0} & \pie{180} \\
 &  & Data versioning support & \pie{0} & \pie{360} & \pie{0} & \pie{0} & \pie{0} & \pie{0} & \pie{0} & \pie{0} & \pie{0} \\
 &  & Live backups & \pie{360} & \pie{0} & \pie{360} & \pie{360} & \pie{0} & \pie{360} & \pie{360} & \pie{360} & \pie{180} \\
\addlinespace[2pt]
\cmidrule(lr){2-12}
\addlinespace[2pt]
 & \multirow{5}{*}{\rowsubcat{Distribution}} & Cluster re-balancing & \pie{0} & \pie{0} & \pie{360} & \pie{360} & \pie{0} & \pie{0} & \pie{360} & \pie{0} & \pie{0} \\
 &  & Data distribution & \pie{0} & \pie{0} & \pie{360} & \pie{360} & \pie{0} & \pie{360} & \pie{360} & \pie{360} & \pie{0} \\
 &  & High-availability & \pie{360} & \pie{0} & \pie{360} & \pie{360} & \pie{0} & \pie{360} & \pie{360} & \pie{360} & \pie{180} \\
 &  & Query distribution & \pie{180} & \pie{0} & \pie{360} & \pie{360} & \pie{0} & \pie{0} & \pie{360} & \pie{360} & \pie{0} \\
 &  & Replication support & \pie{360} & \pie{0} & \pie{360} & \pie{360} & \pie{0} & \pie{360} & \pie{360} & \pie{360} & \pie{360} \\
\addlinespace[2pt]
\cmidrule(lr){2-12}
\addlinespace[2pt]
 & \multirow{5}{*}{\rowsubcat{Software\\Development}} & Data types defined & \pie{0} & \pie{360} & \pie{360} & \pie{360} & \pie{360} & \pie{360} & \pie{360} & \pie{360} & \pie{360} \\
 &  & Logging/Auditing & \pie{360} & \pie{360} & \pie{360} & \pie{180} & \pie{180} & \pie{360} & \pie{360} & \pie{360} & \pie{180} \\
 &  & Object-graph mapper & \pie{0} & \pie{360} & \pie{360} & \pie{0} & \pie{0} & \pie{0} & \pie{0} & \pie{360} & \pie{360} \\
 &  & Reactive programming & \pie{0} & \pie{0} & \pie{0} & \pie{0} & \pie{0} & \pie{0} & \pie{0} & \pie{360} & \pie{180} \\
 &  & Documentation up-to-date & \pie{360} & \pie{90} & \pie{180} & \pie{180} & \pie{0} & \pie{270} & \pie{360} & \pie{360} & \pie{360}
% END inlined from tables/pretty/new-table1_1_1.tex

\tabularnewline
\bottomrule
\end{tabular}
\end{adjustbox}
\end{table}

\begin{table}[!t]
\centering
\caption{Summary of graph database distinctive features (Part I-b).}
\label{new:table:graph_database_features_1_1_2}

\small
\setlength{\tabcolsep}{4pt}
\renewcommand{\arraystretch}{1.15}

\begin{adjustbox}{max width=\textwidth}
\begin{tabular}{@{}>{\centering\arraybackslash}p{0.85cm}>{\centering\arraybackslash}p{1.15cm}l*{9}{c}@{}}
\toprule
\multicolumn{2}{@{}c}{} & \multicolumn{1}{@{}l}{Feature}
& \rot{Alibaba GDB}
& \rot{ChronoGraph}
& \rot{DataStax Enterprise}
& \rot{Dgraph}
& \rot{Graphflow}
& \rot{JanusGraph}
& \rot{NebulaGraph}
& \rot{Neo4j}
& \rot{RedisGraph/FalkorDB}
\\
\midrule

% BEGIN inlined from tables/pretty/new-table1_1_2.tex
\multirow{19}{*}{\rowcat{Data}} & \multirow{7}{*}{\rowsubcat{Access}} & Binary protocol & \pie{360} & \pie{270} & \pie{360} & \pie{360} & \pie{0} & \pie{360} & \pie{360} & \pie{360} & \pie{360} \\
 &  & CLI & \pie{0} & \pie{360} & \pie{360} & \pie{0} & \pie{360} & \pie{360} & \pie{360} & \pie{360} & \pie{360} \\
 &  & GUI & \pie{0} & \pie{0} & \pie{360} & \pie{360} & \pie{0} & \pie{0} & \pie{360} & \pie{360} & \pie{0} \\
 &  & Multi-database & \pie{0} & \pie{0} & \pie{0} & \pie{0} & \pie{0} & \pie{360} & \pie{0} & \pie{360} & \pie{0} \\
 &  & Graph-native data & \pie{360} & \pie{0} & \pie{0} & \pie{360} & \pie{360} & \pie{360} & \pie{360} & \pie{360} & \pie{360} \\
 &  & REST API & \pie{0} & \pie{360} & \pie{0} & \pie{360} & \pie{0} & \pie{360} & \pie{180} & \pie{360} & \pie{360} \\
 &  & Query language & \pie{360} & \pie{360} & \pie{360} & \pie{360} & \pie{270} & \pie{360} & \pie{0} & \pie{360} & \pie{360} \\
\addlinespace[2pt]
\cmidrule(lr){2-12}
\addlinespace[2pt]
 & \multirow{4}{*}{\rowsubcat{Consistency}} & Granular locking & \pie{360} & \pie{0} & \pie{0} & \pie{0} & \pie{0} & \pie{180} & \pie{0} & \pie{360} & \pie{0} \\
 &  & Multiple isolation levels & \pie{360} & \pie{0} & \pie{0} & \pie{0} & \pie{0} & \pie{180} & \pie{0} & \pie{0} & \pie{0} \\
 &  & Read committed transaction & \pie{360} & \pie{360} & \pie{0} & \pie{0} & \pie{0} & \pie{180} & \pie{360} & \pie{360} & \pie{0} \\
 &  & Transaction support & \pie{360} & \pie{360} & \pie{0} & \pie{360} & \pie{0} & \pie{360} & \pie{360} & \pie{360} & \pie{360} \\
\addlinespace[2pt]
\cmidrule(lr){2-12}
\addlinespace[2pt]
 & \multirow{5}{*}{\rowsubcat{Control}} & Constraints & \pie{180} & \pie{0} & \pie{360} & \pie{0} & \pie{0} & \pie{360} & \pie{0} & \pie{360} & \pie{360} \\
 &  & Schema support & \pie{180} & \pie{180} & \pie{360} & \pie{360} & \pie{0} & \pie{360} & \pie{360} & \pie{360} & \pie{0} \\
 &  & Secondary indexes & \pie{360} & \pie{360} & \pie{360} & \pie{360} & \pie{0} & \pie{360} & \pie{360} & \pie{360} & \pie{360} \\
 &  & Server side procedures & \pie{0} & \pie{0} & \pie{0} & \pie{0} & \pie{0} & \pie{0} & \pie{360} & \pie{360} & \pie{360} \\
 &  & Triggers & \pie{0} & \pie{0} & \pie{360} & \pie{0} & \pie{0} & \pie{180} & \pie{360} & \pie{360} & \pie{180} \\
\addlinespace[2pt]
\cmidrule(lr){2-12}
\addlinespace[2pt]
 & \multirow{3}{*}{\rowsubcat{Security}} & Authentication & \pie{360} & \pie{0} & \pie{360} & \pie{360} & \pie{0} & \pie{360} & \pie{360} & \pie{360} & \pie{180} \\
 &  & Authorization & \pie{0} & \pie{0} & \pie{360} & \pie{360} & \pie{0} & \pie{360} & \pie{360} & \pie{360} & \pie{180} \\
 &  & Data encryption & \pie{0} & \pie{0} & \pie{360} & \pie{360} & \pie{0} & \pie{0} & \pie{0} & \pie{0} & \pie{180}
% END inlined from tables/pretty/new-table1_1_2.tex

\tabularnewline
\bottomrule
\end{tabular}
\end{adjustbox}
\end{table}

%%%%%%%%%%%%%%%
%%%%%%%%%%%%%%% PART II
%%%%%%%%%%%%%%%

\begin{table}[!t]
\centering
\caption{Summary of graph database distinctive features (Part II-a).}
\label{new:table:graph_database_features_1_2_1}

\small
\setlength{\tabcolsep}{4pt}
\renewcommand{\arraystretch}{1.15}

\begin{adjustbox}{max width=\textwidth}
\begin{tabular}{@{}>{\centering\arraybackslash}p{0.85cm}>{\centering\arraybackslash}p{1.15cm}l*{9}{c}@{}}
\toprule
\multicolumn{2}{@{}c}{} & \multicolumn{1}{@{}l}{Feature}
& \rot{SAP Hana}
& \rot{Sparksee}
& \rot{TigerGraph}
& \rot{Weaver}
& \rot{AllegroGraph}
& \rot{BlazeGraph}
& \rot{BrightstarDB}
& \rot{Cray Graph Engine}
& \rot{Ontotext GraphDB}
\\
\midrule

% BEGIN inlined from tables/pretty/new-table1_2_1.tex
\multirow{13}{*}{\rowcat{Product}} & \multirow{6}{*}{\rowsubcat{Adoption}} & Active development & \pie{0} & \pie{0} & \pie{360} & \pie{0} & \pie{0} & \pie{0} & \pie{0} & \pie{0} & \pie{360} \\
 &  & Commercial support & \pie{360} & \pie{360} & \pie{360} & \pie{0} & \pie{360} & \pie{0} & \pie{0} & \pie{360} & \pie{360} \\
 &  & Live community & \pie{0} & \pie{0} & \pie{180} & \pie{0} & \pie{0} & \pie{90} & \pie{360} & \pie{0} & \pie{360} \\
 &  & Open source & \pie{0} & \pie{0} & \pie{0} & \pie{360} & \pie{0} & \pie{360} & \pie{360} & \pie{0} & \pie{0} \\
 &  & Pricing & \pie{360} & \pie{0} & \pie{0} & \pie{0} & \pie{360} & \pie{0} & \pie{0} & \pie{0} & \pie{0} \\
 &  & Trendiness & \pie{360} & \pie{0} & \pie{360} & \pie{0} & \pie{360} & \pie{360} & \pie{180} & \pie{0} & \pie{360} \\
\addlinespace[2pt]
\cmidrule(lr){2-12}
\addlinespace[2pt]
 & \multirow{7}{*}{\rowsubcat{Deployment}} & Containerization & \pie{360} & \pie{0} & \pie{360} & \pie{180} & \pie{360} & \pie{0} & \pie{0} & \pie{0} & \pie{360} \\
 &  & Work as dedicated instance & \pie{360} & \pie{360} & \pie{360} & \pie{360} & \pie{360} & \pie{360} & \pie{360} & \pie{360} & \pie{360} \\
 &  & Work as embedded & \pie{0} & \pie{360} & \pie{360} & \pie{0} & \pie{0} & \pie{360} & \pie{360} & \pie{0} & \pie{360} \\
 &  & Testing in-memory version & \pie{0} & \pie{0} & \pie{180} & \pie{0} & \pie{0} & \pie{0} & \pie{0} & \pie{0} & \pie{0} \\
 &  & Operating on Linux & \pie{360} & \pie{360} & \pie{360} & \pie{360} & \pie{360} & \pie{270} & \pie{270} & \pie{360} & \pie{360} \\
 &  & Operating on Windows & \pie{0} & \pie{360} & \pie{180} & \pie{270} & \pie{0} & \pie{270} & \pie{270} & \pie{0} & \pie{360} \\
 &  & SaaS offering & \pie{360} & \pie{0} & \pie{0} & \pie{0} & \pie{360} & \pie{0} & \pie{0} & \pie{0} & \pie{0} \\
\addlinespace[2pt]
\midrule
\addlinespace[2pt]
\multirow{14}{*}{\rowcat{Database}} & \multirow{4}{*}{\rowsubcat{Convenience}} & Automatic updates & \pie{0} & \pie{0} & \pie{0} & \pie{0} & \pie{0} & \pie{0} & \pie{0} & \pie{0} & \pie{0} \\
 &  & Client side caching & \pie{360} & \pie{0} & \pie{0} & \pie{0} & \pie{360} & \pie{0} & \pie{0} & \pie{0} & \pie{0} \\
 &  & Data versioning support & \pie{360} & \pie{0} & \pie{0} & \pie{0} & \pie{360} & \pie{0} & \pie{0} & \pie{0} & \pie{360} \\
 &  & Live backups & \pie{360} & \pie{360} & \pie{360} & \pie{0} & \pie{360} & \pie{0} & \pie{0} & \pie{360} & \pie{180} \\
\addlinespace[2pt]
\cmidrule(lr){2-12}
\addlinespace[2pt]
 & \multirow{5}{*}{\rowsubcat{Distribution}} & Cluster Re-balancing & \pie{360} & \pie{0} & \pie{0} & \pie{0} & \pie{360} & \pie{0} & \pie{0} & \pie{0} & \pie{0} \\
 &  & Data distribution & \pie{360} & \pie{0} & \pie{360} & \pie{360} & \pie{360} & \pie{360} & \pie{360} & \pie{360} & \pie{270} \\
 &  & High-availability & \pie{360} & \pie{0} & \pie{360} & \pie{360} & \pie{360} & \pie{360} & \pie{360} & \pie{180} & \pie{360} \\
 &  & Query distribution & \pie{360} & \pie{0} & \pie{360} & \pie{360} & \pie{360} & \pie{360} & \pie{360} & \pie{360} & \pie{270} \\
 &  & Replication support & \pie{360} & \pie{0} & \pie{180} & \pie{180} & \pie{360} & \pie{360} & \pie{360} & \pie{0} & \pie{360} \\
\addlinespace[2pt]
\cmidrule(lr){2-12}
\addlinespace[2pt]
 & \multirow{5}{*}{\rowsubcat{Software\\Development}} & Data types defined & \pie{360} & \pie{360} & \pie{360} & \pie{0} & \pie{360} & \pie{180} & \pie{180} & \pie{180} & \pie{360} \\
 &  & Logging/Auditing & \pie{360} & \pie{0} & \pie{180} & \pie{180} & \pie{360} & \pie{360} & \pie{360} & \pie{360} & \pie{360} \\
 &  & Object-graph mapper & \pie{360} & \pie{0} & \pie{0} & \pie{0} & \pie{360} & \pie{0} & \pie{0} & \pie{360} & \pie{360} \\
 &  & Reactive programming & \pie{360} & \pie{360} & \pie{0} & \pie{0} & \pie{360} & \pie{0} & \pie{0} & \pie{0} & \pie{0} \\
 &  & Documentation up-to-date & \pie{360} & \pie{360} & \pie{270} & \pie{0} & \pie{360} & \pie{360} & \pie{360} & \pie{360} & \pie{360}
% END inlined from tables/pretty/new-table1_2_1.tex

\tabularnewline
\bottomrule
\end{tabular}
\end{adjustbox}
\end{table}

\begin{table}[!t]
\centering
\caption{Summary of graph database distinctive features (Part II-b).}
\label{new:table:graph_database_features_1_2_2}

\small
\setlength{\tabcolsep}{4pt}
\renewcommand{\arraystretch}{1.15}

\begin{adjustbox}{max width=\textwidth}
\begin{tabular}{@{}>{\centering\arraybackslash}p{0.85cm}>{\centering\arraybackslash}p{1.15cm}l*{9}{c}@{}}
\toprule
\multicolumn{2}{@{}c}{} & \multicolumn{1}{@{}l}{Feature}
& \rot{SAP Hana}
& \rot{Sparksee}
& \rot{TigerGraph}
& \rot{Weaver}
& \rot{AllegroGraph}
& \rot{BlazeGraph}
& \rot{BrightstarDB}
& \rot{Cray Graph Engine}
& \rot{Ontotext GraphDB}
\\
\midrule

% BEGIN inlined from tables/pretty/new-table1_2_2.tex
\multirow{19}{*}{\rowcat{Data}} & \multirow{7}{*}{\rowsubcat{Access}} & Binary protocol & \pie{360} & \pie{0} & \pie{0} & \pie{360} & \pie{0} & \pie{360} & \pie{360} & \pie{360} & \pie{0} \\
 &  & CLI & \pie{360} & \pie{0} & \pie{360} & \pie{0} & \pie{360} & \pie{360} & \pie{360} & \pie{360} & \pie{360} \\
 &  & GUI & \pie{360} & \pie{0} & \pie{360} & \pie{0} & \pie{360} & \pie{360} & \pie{360} & \pie{360} & \pie{360} \\
 &  & Multi-database & \pie{360} & \pie{0} & \pie{360} & \pie{0} & \pie{360} & \pie{0} & \pie{0} & \pie{0} & \pie{0} \\
 &  & Graph-native data & \pie{0} & \pie{360} & \pie{360} & \pie{0} & \pie{360} & \pie{360} & \pie{360} & \pie{360} & \pie{360} \\
 &  & REST API & \pie{360} & \pie{0} & \pie{360} & \pie{0} & \pie{360} & \pie{360} & \pie{360} & \pie{360} & \pie{360} \\
 &  & Query language & \pie{0} & \pie{360} & \pie{360} & \pie{0} & \pie{360} & \pie{360} & \pie{360} & \pie{360} & \pie{360} \\
\addlinespace[2pt]
\cmidrule(lr){2-12}
\addlinespace[2pt]
 & \multirow{4}{*}{\rowsubcat{Consistency}} & Granular locking & \pie{360} & \pie{0} & \pie{0} & \pie{0} & \pie{0} & \pie{0} & \pie{0} & \pie{0} & \pie{0} \\
 &  & Multiple isolation levels & \pie{360} & \pie{360} & \pie{0} & \pie{0} & \pie{360} & \pie{360} & \pie{360} & \pie{0} & \pie{0} \\
 &  & Read committed transaction & \pie{360} & \pie{0} & \pie{180} & \pie{180} & \pie{0} & \pie{360} & \pie{360} & \pie{0} & \pie{360} \\
 &  & Transaction support & \pie{360} & \pie{360} & \pie{360} & \pie{360} & \pie{360} & \pie{360} & \pie{360} & \pie{0} & \pie{360} \\
\addlinespace[2pt]
\cmidrule(lr){2-12}
\addlinespace[2pt]
 & \multirow{5}{*}{\rowsubcat{Control}} & Constraints & \pie{360} & \pie{360} & \pie{0} & \pie{0} & \pie{360} & \pie{0} & \pie{0} & \pie{180} & \pie{360} \\
 &  & Schema support & \pie{360} & \pie{360} & \pie{360} & \pie{0} & \pie{360} & \pie{0} & \pie{0} & \pie{180} & \pie{180} \\
 &  & Secondary indexes & \pie{360} & \pie{360} & \pie{0} & \pie{0} & \pie{360} & \pie{360} & \pie{360} & \pie{360} & \pie{360} \\
 &  & Server side procedures & \pie{360} & \pie{0} & \pie{360} & \pie{0} & \pie{360} & \pie{360} & \pie{360} & \pie{360} & \pie{360} \\
 &  & Triggers & \pie{360} & \pie{0} & \pie{0} & \pie{0} & \pie{360} & \pie{0} & \pie{0} & \pie{0} & \pie{0} \\
\addlinespace[2pt]
\cmidrule(lr){2-12}
\addlinespace[2pt]
 & \multirow{3}{*}{\rowsubcat{Security}} & Authentication & \pie{360} & \pie{0} & \pie{360} & \pie{0} & \pie{360} & \pie{0} & \pie{0} & \pie{360} & \pie{360} \\
 &  & Authorization & \pie{360} & \pie{0} & \pie{360} & \pie{0} & \pie{360} & \pie{0} & \pie{0} & \pie{360} & \pie{360} \\
 &  & Data encryption & \pie{360} & \pie{360} & \pie{360} & \pie{0} & \pie{0} & \pie{0} & \pie{0} & \pie{0} & \pie{0}
% END inlined from tables/pretty/new-table1_2_2.tex

\tabularnewline
\bottomrule
\end{tabular}
\end{adjustbox}
\end{table}

%%%%%%%%%%%%%%%
%%%%%%%%%%%%%%% PART III
%%%%%%%%%%%%%%%

\begin{table}[!t]
\centering
\caption{Summary of graph database distinctive features (Part III-a).}
\label{new:table:graph_database_features_2_1_1}

\small
\setlength{\tabcolsep}{4pt}
\renewcommand{\arraystretch}{1.15}

\begin{adjustbox}{max width=\textwidth}
\begin{tabular}{@{}>{\centering\arraybackslash}p{0.85cm}>{\centering\arraybackslash}p{1.15cm}l*{8}{c}@{}}
\toprule
\multicolumn{2}{@{}c}{} & \multicolumn{1}{@{}l}{Feature}
& \rot{Neptune}
& \rot{Altair Graph Lakehouse}
& \rot{ArangoDB}
& \rot{IBM System G}
& \rot{OrientDB}
& \rot{OSG}
& \rot{Stardog}
& \rot{Virtuoso}
\\
\midrule

% BEGIN inlined from tables/pretty/new-table2_1_1.tex
\multirow{13}{*}{\rowcat{Product}} & \multirow{6}{*}{\rowsubcat{Adoption}} & Active development & \pie{360} & \pie{180} & \pie{360} & \pie{0} & \pie{360} & \pie{360} & \pie{360} & \pie{360} \\
 &  & Commercial support & \pie{360} & \pie{360} & \pie{360} & \pie{0} & \pie{360} & \pie{360} & \pie{360} & \pie{360} \\
 &  & Live community & \pie{360} & \pie{0} & \pie{180} & \pie{0} & \pie{180} & \pie{180} & \pie{360} & \pie{360} \\
 &  & Open source & \pie{0} & \pie{0} & \pie{360} & \pie{0} & \pie{360} & \pie{0} & \pie{0} & \pie{360} \\
 &  & Pricing & \pie{360} & \pie{360} & \pie{0} & \pie{0} & \pie{0} & \pie{360} & \pie{0} & \pie{360} \\
 &  & Trendiness & \pie{360} & \pie{0} & \pie{360} & \pie{360} & \pie{360} & \pie{0} & \pie{360} & \pie{360} \\
\addlinespace[2pt]
\cmidrule(lr){2-11}
\addlinespace[2pt]
 & \multirow{7}{*}{\rowsubcat{Deployment}} & Containerization & \pie{0} & \pie{360} & \pie{360} & \pie{0} & \pie{360} & \pie{180} & \pie{360} & \pie{360} \\
 &  & Work as dedicated instance & \pie{360} & \pie{360} & \pie{360} & \pie{360} & \pie{360} & \pie{360} & \pie{360} & \pie{360} \\
 &  & Work as embedded & \pie{0} & \pie{0} & \pie{0} & \pie{0} & \pie{360} & \pie{0} & \pie{0} & \pie{0} \\
 &  & Testing in-memory version & \pie{0} & \pie{360} & \pie{360} & \pie{360} & \pie{360} & \pie{0} & \pie{0} & \pie{0} \\
 &  & Operating on Linux & \pie{360} & \pie{360} & \pie{360} & \pie{360} & \pie{360} & \pie{360} & \pie{360} & \pie{360} \\
 &  & Operating on Windows & \pie{0} & \pie{0} & \pie{360} & \pie{0} & \pie{360} & \pie{360} & \pie{0} & \pie{360} \\
 &  & SaaS offering & \pie{360} & \pie{0} & \pie{0} & \pie{0} & \pie{0} & \pie{360} & \pie{360} & \pie{360} \\
\addlinespace[2pt]
\midrule
\addlinespace[2pt]
\multirow{11}{*}{\rowcat{Database}} & \multirow{4}{*}{\rowsubcat{Convenience}} & Automatic updates & \pie{360} & \pie{0} & \pie{0} & \pie{0} & \pie{0} & \pie{360} & \pie{0} & \pie{0} \\
 &  & Client side caching & \pie{360} & \pie{0} & \pie{0} & \pie{0} & \pie{0} & \pie{360} & \pie{0} & \pie{0} \\
 &  & Data versioning support & \pie{0} & \pie{360} & \pie{0} & \pie{0} & \pie{0} & \pie{360} & \pie{0} & \pie{0} \\
 &  & Live backups & \pie{360} & \pie{360} & \pie{360} & \pie{0} & \pie{360} & \pie{360} & \pie{360} & \pie{360} \\
\addlinespace[2pt]
\cmidrule(lr){2-11}
\addlinespace[2pt]
 & \multirow{5}{*}{\rowsubcat{Distribution}} & Cluster Re-balancing & \pie{0} & \pie{0} & \pie{360} & \pie{0} & \pie{0} & \pie{180} & \pie{0} & \pie{0} \\
 &  & Data distribution & \pie{0} & \pie{360} & \pie{360} & \pie{360} & \pie{360} & \pie{360} & \pie{0} & \pie{360} \\
 &  & High-availability & \pie{360} & \pie{360} & \pie{360} & \pie{0} & \pie{360} & \pie{360} & \pie{360} & \pie{0} \\
 &  & Query distribution & \pie{180} & \pie{360} & \pie{360} & \pie{360} & \pie{180} & \pie{360} & \pie{360} & \pie{360} \\
 &  & Replication support & \pie{360} & \pie{360} & \pie{360} & \pie{0} & \pie{360} & \pie{360} & \pie{360} & \pie{360} \\
\addlinespace[2pt]
\cmidrule(lr){2-11}
\addlinespace[2pt]
& \multirow{5}{*}{\rowsubcat{Software\\Development}} & Data types defined & \pie{180} & \pie{180} & \pie{360} & \pie{180} & \pie{360} & \pie{180} & \pie{180} & \pie{0} \\
 &  & Logging/Auditing & \pie{360} & \pie{360} & \pie{360} & \pie{0} & \pie{360} & \pie{360} & \pie{360} & \pie{360} \\
 &  & Object-graph mapper & \pie{0} & \pie{0} & \pie{360} & \pie{0} & \pie{0} & \pie{360} & \pie{0} & \pie{0} \\
 &  & Reactive programming & \pie{360} & \pie{0} & \pie{0} & \pie{0} & \pie{0} & \pie{360} & \pie{360} & \pie{0} \\
 &  & Documentation up-to-date & \pie{360} & \pie{360} & \pie{360} & \pie{0} & \pie{180} & \pie{360} & \pie{360} & \pie{180}
% END inlined from tables/pretty/new-table2_1_1.tex

\tabularnewline
\bottomrule
\end{tabular}
\end{adjustbox}
\end{table}

\begin{table}[!t]
\centering
\caption{Summary of graph database distinctive features (Part III-b).}
\label{new:table:graph_database_features_2_1_2}

\small
\setlength{\tabcolsep}{4pt}
\renewcommand{\arraystretch}{1.15}

\begin{adjustbox}{max width=\textwidth}
\begin{tabular}{@{}>{\centering\arraybackslash}p{0.85cm}>{\centering\arraybackslash}p{1.15cm}l*{8}{c}@{}}
\toprule
\multicolumn{2}{@{}c}{} & \multicolumn{1}{@{}l}{Feature}
& \rot{Neptune}
& \rot{Altair Graph Lakehouse}
& \rot{ArangoDB}
& \rot{IBM System G}
& \rot{OrientDB}
& \rot{OSG}
& \rot{Stardog}
& \rot{Virtuoso}
\\
\midrule

% BEGIN inlined from tables/pretty/new-table2_1_2.tex
\multirow{19}{*}{\rowcat{Data}} & \multirow{7}{*}{\rowsubcat{Access}} & Binary protocol & \pie{360} & \pie{0} & \pie{360} & \pie{360} & \pie{360} & \pie{360} & \pie{0} & \pie{0} \\
 &  & CLI & \pie{360} & \pie{360} & \pie{360} & \pie{360} & \pie{360} & \pie{360} & \pie{360} & \pie{360} \\
 &  & GUI & \pie{360} & \pie{360} & \pie{360} & \pie{0} & \pie{360} & \pie{360} & \pie{360} & \pie{360} \\
 &  & Multi-database & \pie{0} & \pie{0} & \pie{360} & \pie{0} & \pie{360} & \pie{360} & \pie{360} & \pie{360} \\
 &  & Graph-native data & \pie{360} & \pie{360} & \pie{0} & \pie{360} & \pie{360} & \pie{360} & \pie{0} & \pie{0} \\
 &  & REST API & \pie{360} & \pie{360} & \pie{360} & \pie{360} & \pie{360} & \pie{360} & \pie{360} & \pie{360} \\
 &  & Query language & \pie{360} & \pie{360} & \pie{180} & \pie{360} & \pie{360} & \pie{360} & \pie{360} & \pie{360} \\
\addlinespace[2pt]
\cmidrule(lr){2-11}
\addlinespace[2pt]
 & \multirow{4}{*}{\rowsubcat{Consistency}} & Granular locking & \pie{360} & \pie{0} & \pie{0} & \pie{0} & \pie{0} & \pie{360} & \pie{360} & \pie{360} \\
 &  & Multiple isolation levels & \pie{360} & \pie{0} & \pie{0} & \pie{0} & \pie{360} & \pie{360} & \pie{360} & \pie{360} \\
 &  & Read committed transaction & \pie{360} & \pie{0} & \pie{0} & \pie{0} & \pie{360} & \pie{360} & \pie{0} & \pie{360} \\
 &  & Transaction support & \pie{360} & \pie{360} & \pie{360} & \pie{360} & \pie{360} & \pie{360} & \pie{360} & \pie{360} \\
\addlinespace[2pt]
\cmidrule(lr){2-11}
\addlinespace[2pt]
 & \multirow{5}{*}{\rowsubcat{Control}} & Constraints & \pie{360} & \pie{0} & \pie{360} & \pie{0} & \pie{0} & \pie{360} & \pie{360} & \pie{360} \\
 &  & Schema support & \pie{180} & \pie{360} & \pie{180} & \pie{0} & \pie{360} & \pie{360} & \pie{360} & \pie{0} \\
 &  & Secondary indexes & \pie{360} & \pie{360} & \pie{360} & \pie{0} & \pie{360} & \pie{360} & \pie{360} & \pie{360} \\
 &  & Server side procedures & \pie{0} & \pie{0} & \pie{180} & \pie{0} & \pie{360} & \pie{180} & \pie{360} & \pie{360} \\
 &  & Triggers & \pie{0} & \pie{0} & \pie{0} & \pie{0} & \pie{360} & \pie{360} & \pie{0} & \pie{360} \\
\addlinespace[2pt]
\cmidrule(lr){2-11}
\addlinespace[2pt]
 & \multirow{3}{*}{\rowsubcat{Security}} & Authentication & \pie{360} & \pie{360} & \pie{360} & \pie{0} & \pie{360} & \pie{360} & \pie{360} & \pie{360} \\
 &  & Authorization & \pie{360} & \pie{360} & \pie{180} & \pie{0} & \pie{360} & \pie{360} & \pie{360} & \pie{360} \\
 &  & Data encryption & \pie{360} & \pie{360} & \pie{360} & \pie{0} & \pie{360} & \pie{360} & \pie{0} & \pie{0}
% END inlined from tables/pretty/new-table2_1_2.tex

\tabularnewline
\bottomrule
\end{tabular}
\end{adjustbox}
\end{table}

%%%%%%%%%%%%%%%
%%%%%%%%%%%%%%% PART IV
%%%%%%%%%%%%%%%

\begin{table}[!t]
\centering
\caption{Summary of graph database distinctive features (Part IV-a).}
\label{new:table:graph_database_features_2_2_1}

\small
\setlength{\tabcolsep}{4pt}
\renewcommand{\arraystretch}{1.15}

\begin{adjustbox}{max width=\textwidth}
\begin{tabular}{@{}>{\centering\arraybackslash}p{0.85cm}>{\centering\arraybackslash}p{1.15cm}l*{8}{c}@{}}
\toprule
\multicolumn{2}{@{}c}{} & \multicolumn{1}{@{}l}{Feature}
& \rot{Cosmos DB}
& \rot{FaunaDB}
& \rot{Google Cayley}
& \rot{HyperGraphDB}
& \rot{Objectivity/DB}
& \rot{KatanaGraph}
& \rot{TerminusDB}
& \rot{MemGraph}
\\
\midrule

% BEGIN inlined from tables/pretty/new-table2_2_1.tex
\multirow{13}{*}{\rowcat{Product}} & \multirow{6}{*}{\rowsubcat{Adoption}} & Active development & \pie{0} & \pie{0} & \pie{360} & \pie{0} & \pie{0} & \pie{0} & \pie{360} & \pie{360} \\
 &  & Commercial support & \pie{360} & \pie{360} & \pie{0} & \pie{0} & \pie{360} & \pie{360} & \pie{360} & \pie{360} \\
 &  & Live community & \pie{180} & \pie{0} & \pie{0} & \pie{180} & \pie{0} & \pie{180} & \pie{270} & \pie{180} \\
 &  & Open source & \pie{0} & \pie{180} & \pie{360} & \pie{360} & \pie{0} & \pie{360} & \pie{360} & \pie{0} \\
 &  & Pricing & \pie{360} & \pie{360} & \pie{0} & \pie{0} & \pie{0} & \pie{360} & \pie{360} & \pie{0} \\
 &  & Trendiness & \pie{360} & \pie{0} & \pie{360} & \pie{0} & \pie{0} & \pie{0} & \pie{360} & \pie{360} \\
\addlinespace[2pt]
\cmidrule(lr){2-11}
\addlinespace[2pt]
 & \multirow{7}{*}{\rowsubcat{Deployment}} & Containerization & \pie{180} & \pie{360} & \pie{360} & \pie{0} & \pie{0} & \pie{0} & \pie{360} & \pie{360} \\
 &  & Work as dedicated instance & \pie{360} & \pie{360} & \pie{360} & \pie{360} & \pie{360} & \pie{0} & \pie{360} & \pie{360} \\
 &  & Work as embedded & \pie{0} & \pie{0} & \pie{360} & \pie{360} & \pie{0} & \pie{0} & \pie{0} & \pie{0} \\
 &  & Testing in-memory version & \pie{360} & \pie{0} & \pie{360} & \pie{0} & \pie{0} & \pie{0} & \pie{0} & \pie{270} \\
 &  & Operating on Linux & \pie{0} & \pie{360} & \pie{360} & \pie{360} & \pie{360} & \pie{360} & \pie{360} & \pie{360} \\
 &  & Operating on Windows & \pie{360} & \pie{0} & \pie{0} & \pie{180} & \pie{360} & \pie{0} & \pie{270} & \pie{180} \\
 &  & SaaS offering & \pie{360} & \pie{360} & \pie{0} & \pie{0} & \pie{360} & \pie{0} & \pie{360} & \pie{360} \\
\addlinespace[2pt]
\midrule
\addlinespace[2pt]
\multirow{14}{*}{\rowcat{Database}} & \multirow{4}{*}{\rowsubcat{Convenience}} & Automatic updates & \pie{360} & \pie{0} & \pie{0} & \pie{0} & \pie{0} & \pie{0} & \pie{0} & \pie{0} \\
 &  & Client side caching & \pie{0} & \pie{0} & \pie{180} & \pie{360} & \pie{0} & \pie{0} & \pie{0} & \pie{0} \\
 &  & Data versioning support & \pie{0} & \pie{360} & \pie{0} & \pie{360} & \pie{0} & \pie{0} & \pie{360} & \pie{0} \\
 &  & Live backups & \pie{360} & \pie{360} & \pie{0} & \pie{0} & \pie{0} & \pie{0} & \pie{360} & \pie{360} \\
\addlinespace[2pt]
\cmidrule(lr){2-11}
\addlinespace[2pt]
 & \multirow{5}{*}{\rowsubcat{Distribution}} & Cluster Re-balancing & \pie{180} & \pie{0} & \pie{0} & \pie{0} & \pie{0} & \pie{360} & \pie{0} & \pie{360} \\
 &  & Data distribution & \pie{360} & \pie{360} & \pie{360} & \pie{90} & \pie{360} & \pie{360} & \pie{0} & \pie{360} \\
 &  & High-availability & \pie{360} & \pie{360} & \pie{180} & \pie{0} & \pie{360} & \pie{0} & \pie{0} & \pie{360} \\
 &  & Query distribution & \pie{360} & \pie{360} & \pie{360} & \pie{90} & \pie{0} & \pie{0} & \pie{0} & \pie{360} \\
 &  & Replication support & \pie{360} & \pie{360} & \pie{360} & \pie{0} & \pie{360} & \pie{0} & \pie{0} & \pie{360} \\
\addlinespace[2pt]
\cmidrule(lr){2-11}
\addlinespace[2pt]
 & \multirow{5}{*}{\rowsubcat{Software\\Development}} & Data types defined & \pie{0} & \pie{180} & \pie{360} & \pie{360} & \pie{180} & \pie{360} & \pie{180} & \pie{0} \\
 &  & Logging/Auditing & \pie{360} & \pie{360} & \pie{360} & \pie{0} & \pie{360} & \pie{0} & \pie{360} & \pie{360} \\
 &  & Object-graph mapper & \pie{360} & \pie{0} & \pie{360} & \pie{360} & \pie{0} & \pie{0} & \pie{0} & \pie{360} \\
 &  & Reactive programming & \pie{360} & \pie{0} & \pie{0} & \pie{0} & \pie{0} & \pie{0} & \pie{0} & \pie{0} \\
 &  & Documentation up-to-date & \pie{360} & \pie{360} & \pie{360} & \pie{360} & \pie{0} & \pie{0} & \pie{360} & \pie{270}
% END inlined from tables/pretty/new-table2_2_1.tex

\tabularnewline
\bottomrule
\end{tabular}
\end{adjustbox}
\end{table}

\begin{table}[!t]
\centering
\caption{Summary of graph database distinctive features (Part IV-b).}
\label{new:table:graph_database_features_2_2_2}

\small
\setlength{\tabcolsep}{4pt}
\renewcommand{\arraystretch}{1.15}

\begin{adjustbox}{max width=\textwidth}
\begin{tabular}{@{}>{\centering\arraybackslash}p{0.85cm}>{\centering\arraybackslash}p{1.15cm}l*{8}{c}@{}}
\toprule
\multicolumn{2}{@{}c}{} & \multicolumn{1}{@{}l}{Feature}
& \rot{Cosmos DB}
& \rot{FaunaDB}
& \rot{Google Cayley}
& \rot{HyperGraphDB}
& \rot{Objectivity/DB}
& \rot{KatanaGraph}
& \rot{TerminusDB}
& \rot{MemGraph}
\\
\midrule

% BEGIN inlined from tables/pretty/new-table2_2_2.tex
\multirow{19}{*}{\rowcat{Data}} & \multirow{7}{*}{\rowsubcat{Access}} & Binary protocol & \pie{0} & \pie{0} & \pie{0} & \pie{360} & \pie{360} & \pie{0} & \pie{0} & \pie{360} \\
 &  & CLI & \pie{360} & \pie{360} & \pie{360} & \pie{0} & \pie{360} & \pie{0} & \pie{0} & \pie{360} \\
 &  & GUI & \pie{360} & \pie{0} & \pie{360} & \pie{0} & \pie{360} & \pie{0} & \pie{0} & \pie{360} \\
 &  & Multi-database & \pie{360} & \pie{0} & \pie{0} & \pie{360} & \pie{0} & \pie{0} & \pie{0} & \pie{0} \\
 &  & Graph-native data & \pie{0} & \pie{360} & \pie{360} & \pie{0} & \pie{360} & \pie{360} & \pie{0} & \pie{360} \\
 &  & REST API & \pie{360} & \pie{0} & \pie{360} & \pie{0} & \pie{360} & \pie{0} & \pie{360} & \pie{0} \\
 &  & Query language & \pie{360} & \pie{360} & \pie{360} & \pie{180} & \pie{180} & \pie{0} & \pie{180} & \pie{360} \\
\addlinespace[2pt]
\cmidrule(lr){2-11}
\addlinespace[2pt]
 & \multirow{4}{*}{\rowsubcat{Consistency}} & Granular locking & \pie{0} & \pie{0} & \pie{0} & \pie{0} & \pie{0} & \pie{0} & \pie{360} & \pie{0} \\
 &  & Multiple isolation levels & \pie{360} & \pie{0} & \pie{0} & \pie{180} & \pie{0} & \pie{0} & \pie{0} & \pie{0} \\
 &  & Read committed transaction & \pie{360} & \pie{0} & \pie{0} & \pie{90} & \pie{0} & \pie{0} & \pie{0} & \pie{360} \\
 &  & Transaction support & \pie{180} & \pie{360} & \pie{360} & \pie{360} & \pie{360} & \pie{360} & \pie{360} & \pie{360} \\
\addlinespace[2pt]
\cmidrule(lr){2-11}
\addlinespace[2pt]
 & \multirow{5}{*}{\rowsubcat{Control}} & Constraints & \pie{0} & \pie{360} & \pie{180} & \pie{360} & \pie{0} & \pie{0} & \pie{360} & \pie{360} \\
 &  & Schema support & \pie{0} & \pie{360} & \pie{180} & \pie{180} & \pie{360} & \pie{0} & \pie{0} & \pie{270} \\
 &  & Secondary indexes & \pie{360} & \pie{360} & \pie{360} & \pie{360} & \pie{0} & \pie{0} & \pie{0} & \pie{360} \\
 &  & Server side procedures & \pie{360} & \pie{360} & \pie{0} & \pie{0} & \pie{0} & \pie{0} & \pie{0} & \pie{360} \\
 &  & Triggers & \pie{360} & \pie{0} & \pie{0} & \pie{180} & \pie{0} & \pie{0} & \pie{0} & \pie{360} \\
\addlinespace[2pt]
\cmidrule(lr){2-11}
\addlinespace[2pt]
 & \multirow{3}{*}{\rowsubcat{Security}} & Authentication & \pie{360} & \pie{360} & \pie{0} & \pie{0} & \pie{360} & \pie{0} & \pie{360} & \pie{360} \\
 &  & Authorization & \pie{360} & \pie{360} & \pie{0} & \pie{0} & \pie{360} & \pie{0} & \pie{360} & \pie{360} \\
 &  & Data encryption & \pie{360} & \pie{0} & \pie{0} & \pie{0} & \pie{0} & \pie{0} & \pie{0} & \pie{0}
% END inlined from tables/pretty/new-table2_2_2.tex

\tabularnewline
\bottomrule
\end{tabular}
\end{adjustbox}
\end{table}

%%%%%%%%%%%%%%%
%%%%%%%%%%%%%%% PART V
%%%%%%%%%%%%%%%

\begin{table}[!t]
\centering
\caption{Summary of graph database distinctive features (Part V-a).}
\label{new:table:graph_database_features_3_1_1}

\small
\setlength{\tabcolsep}{4pt}
\renewcommand{\arraystretch}{1.15}

\begin{adjustbox}{max width=\textwidth}
\begin{tabular}{@{}>{\centering\arraybackslash}p{0.85cm}>{\centering\arraybackslash}p{1.15cm}l*{9}{c}@{}}
\toprule
\multicolumn{2}{@{}c}{} & \multicolumn{1}{@{}l}{Feature}
& \rot{PandaDB}
& \rot{Gaffer}
& \rot{ZipG}
& \rot{LiveGraph}
& \rot{G-Tran}
& \rot{ByteGraph}
& \rot{Apache HugeGraph}
& \rot{TypeDB}
& \rot{SurrealDB}
\\
\midrule

% BEGIN inlined from tables/pretty/new-table3_1_1.tex
\multirow{13}{*}{\rowcat{Product}} & \multirow{6}{*}{\rowsubcat{Adoption}} & Active development & \pie{0} & \pie{0} & \pie{0} & \pie{0} & \pie{0} & \pie{0} & \pie{360} & \pie{360} & \pie{360} \\
 &  & Commercial support & \pie{0} & \pie{0} & \pie{0} & \pie{0} & \pie{0} & \pie{0} & \pie{0} & \pie{360} & \pie{360} \\
 &  & Live community & \pie{0} & \pie{0} & \pie{0} & \pie{0} & \pie{0} & \pie{0} & \pie{360} & \pie{360} & \pie{180} \\
 &  & Open source & \pie{360} & \pie{360} & \pie{360} & \pie{360} & \pie{360} & \pie{0} & \pie{360} & \pie{360} & \pie{180} \\
 &  & Pricing & \pie{0} & \pie{0} & \pie{0} & \pie{0} & \pie{0} & \pie{0} & \pie{0} & \pie{360} & \pie{0} \\
 &  & Trendiness & \pie{180} & \pie{0} & \pie{0} & \pie{0} & \pie{360} & \pie{180} & \pie{0} & \pie{360} & \pie{360} \\
\addlinespace[2pt]
\cmidrule(lr){2-12}
\addlinespace[2pt]
 & \multirow{7}{*}{\rowsubcat{Deployment}} & Containerization & \pie{360} & \pie{0} & \pie{0} & \pie{0} & \pie{0} & \pie{0} & \pie{360} & \pie{360} & \pie{360} \\
 &  & Work as dedicated instance & \pie{360} & \pie{0} & \pie{360} & \pie{360} & \pie{360} & \pie{360} & \pie{360} & \pie{360} & \pie{360} \\
 &  & Work as embedded & \pie{360} & \pie{0} & \pie{0} & \pie{0} & \pie{0} & \pie{0} & \pie{0} & \pie{0} & \pie{360} \\
 &  & Testing in-memory version & \pie{0} & \pie{0} & \pie{0} & \pie{360} & \pie{360} & \pie{0} & \pie{360} & \pie{360} & \pie{360} \\
 &  & Operating on Linux & \pie{0} & \pie{360} & \pie{360} & \pie{360} & \pie{360} & \pie{0} & \pie{360} & \pie{360} & \pie{360} \\
 &  & Operating on Windows & \pie{360} & \pie{360} & \pie{270} & \pie{270} & \pie{360} & \pie{0} & \pie{360} & \pie{360} & \pie{360} \\
 &  & SaaS offering & \pie{0} & \pie{0} & \pie{0} & \pie{0} & \pie{0} & \pie{0} & \pie{0} & \pie{360} & \pie{360} \\
\addlinespace[2pt]
\midrule
\addlinespace[2pt]
\multirow{14}{*}{\rowcat{Database}} & \multirow{4}{*}{\rowsubcat{Convenience}} & Automatic updates & \pie{0} & \pie{0} & \pie{0} & \pie{0} & \pie{0} & \pie{0} & \pie{360} & \pie{0} & \pie{0} \\
 &  & Client side caching & \pie{360} & \pie{0} & \pie{0} & \pie{0} & \pie{0} & \pie{0} & \pie{0} & \pie{0} & \pie{0} \\
 &  & Data versioning support & \pie{0} & \pie{0} & \pie{0} & \pie{0} & \pie{0} & \pie{360} & \pie{0} & \pie{0} & \pie{0} \\
 &  & Live backups & \pie{360} & \pie{0} & \pie{0} & \pie{0} & \pie{0} & \pie{0} & \pie{360} & \pie{0} & \pie{0} \\
\addlinespace[2pt]
\cmidrule(lr){2-12}
\addlinespace[2pt]
 & \multirow{5}{*}{\rowsubcat{Distribution}} & Cluster Re-balancing & \pie{0} & \pie{0} & \pie{0} & \pie{0} & \pie{0} & \pie{0} & \pie{360} & \pie{0} & \pie{360} \\
 &  & Data distribution & \pie{0} & \pie{360} & \pie{360} & \pie{0} & \pie{360} & \pie{360} & \pie{360} & \pie{0} & \pie{360} \\
 &  & High-availability & \pie{360} & \pie{0} & \pie{0} & \pie{0} & \pie{0} & \pie{360} & \pie{360} & \pie{360} & \pie{360} \\
 &  & Query distribution & \pie{180} & \pie{360} & \pie{360} & \pie{0} & \pie{360} & \pie{360} & \pie{360} & \pie{360} & \pie{360} \\
 &  & Replication support & \pie{360} & \pie{360} & \pie{0} & \pie{0} & \pie{0} & \pie{360} & \pie{360} & \pie{360} & \pie{360} \\
\addlinespace[2pt]
\cmidrule(lr){2-12}
\addlinespace[2pt]
 & \multirow{5}{*}{\rowsubcat{Software\\Development}} & Data types defined & \pie{360} & \pie{180} & \pie{180} & \pie{180} & \pie{180} & \pie{180} & \pie{180} & \pie{180} & \pie{180} \\
 &  & Logging/Auditing & \pie{360} & \pie{360} & \pie{0} & \pie{0} & \pie{0} & \pie{360} & \pie{360} & \pie{360} & \pie{360} \\
 &  & Object-graph mapper & \pie{360} & \pie{360} & \pie{360} & \pie{0} & \pie{360} & \pie{0} & \pie{360} & \pie{0} & \pie{360} \\
 &  & Reactive programming & \pie{360} & \pie{0} & \pie{0} & \pie{0} & \pie{0} & \pie{0} & \pie{0} & \pie{360} & \pie{0} \\
 &  & Documentation up-to-date & \pie{360} & \pie{360} & \pie{360} & \pie{360} & \pie{360} & \pie{0} & \pie{360} & \pie{360} & \pie{360}
% END inlined from tables/pretty/new-table3_1_1.tex

\tabularnewline
\bottomrule
\end{tabular}
\end{adjustbox}
\end{table}

\begin{table}[!t]
\centering
\caption{Summary of graph database distinctive features (Part V-b).}
\label{new:table:graph_database_features_3_1_2}

\small
\setlength{\tabcolsep}{4pt}
\renewcommand{\arraystretch}{1.15}

\begin{adjustbox}{max width=\textwidth}
\begin{tabular}{@{}>{\centering\arraybackslash}p{0.85cm}>{\centering\arraybackslash}p{1.15cm}l*{9}{c}@{}}
\toprule
\multicolumn{2}{@{}c}{} & \multicolumn{1}{@{}l}{Feature}
& \rot{PandaDB}
& \rot{Gaffer}
& \rot{ZipG}
& \rot{LiveGraph}
& \rot{G-Tran}
& \rot{ByteGraph}
& \rot{Apache HugeGraph}
& \rot{TypeDB}
& \rot{SurrealDB}
\\
\midrule

% BEGIN inlined from tables/pretty/new-table3_1_2.tex
\multirow{19}{*}{\rowcat{Data}} & \multirow{7}{*}{\rowsubcat{Access}} & Binary protocol & \pie{360} & \pie{0} & \pie{0} & \pie{0} & \pie{0} & \pie{0} & \pie{360} & \pie{360} & \pie{360} \\
 &  & CLI & \pie{360} & \pie{0} & \pie{0} & \pie{0} & \pie{0} & \pie{0} & \pie{360} & \pie{360} & \pie{360} \\
 &  & GUI & \pie{360} & \pie{0} & \pie{0} & \pie{0} & \pie{0} & \pie{0} & \pie{360} & \pie{360} & \pie{360} \\
 &  & Multi-database & \pie{360} & \pie{0} & \pie{0} & \pie{0} & \pie{0} & \pie{0} & \pie{0} & \pie{0} & \pie{360} \\
 &  & Graph-native data & \pie{360} & \pie{360} & \pie{360} & \pie{360} & \pie{360} & \pie{360} & \pie{360} & \pie{0} & \pie{0} \\
 &  & REST API & \pie{360} & \pie{360} & \pie{0} & \pie{0} & \pie{0} & \pie{0} & \pie{360} & \pie{0} & \pie{360} \\
 &  & Query language & \pie{360} & \pie{0} & \pie{0} & \pie{0} & \pie{360} & \pie{360} & \pie{360} & \pie{180} & \pie{180} \\
\addlinespace[2pt]
\cmidrule(lr){2-12}
\addlinespace[2pt]
 & \multirow{4}{*}{\rowsubcat{Consistency}} & Granular locking & \pie{360} & \pie{0} & \pie{0} & \pie{360} & \pie{360} & \pie{360} & \pie{360} & \pie{0} & \pie{0} \\
 &  & Multiple isolation levels & \pie{0} & \pie{0} & \pie{0} & \pie{0} & \pie{0} & \pie{0} & \pie{180} & \pie{180} & \pie{180} \\
 &  & Read committed transaction & \pie{360} & \pie{0} & \pie{0} & \pie{0} & \pie{0} & \pie{360} & \pie{360} & \pie{360} & \pie{360} \\
 &  & Transaction support & \pie{360} & \pie{360} & \pie{0} & \pie{360} & \pie{360} & \pie{360} & \pie{360} & \pie{360} & \pie{360} \\
\addlinespace[2pt]
\cmidrule(lr){2-12}
\addlinespace[2pt]
 & \multirow{5}{*}{\rowsubcat{Control}} & Constraints & \pie{360} & \pie{0} & \pie{0} & \pie{0} & \pie{0} & \pie{0} & \pie{0} & \pie{360} & \pie{360} \\
 &  & Schema support & \pie{360} & \pie{360} & \pie{0} & \pie{0} & \pie{0} & \pie{360} & \pie{360} & \pie{360} & \pie{360} \\
 &  & Secondary indexes & \pie{360} & \pie{0} & \pie{0} & \pie{0} & \pie{0} & \pie{0} & \pie{360} & \pie{360} & \pie{360} \\
 &  & Server side procedures & \pie{360} & \pie{0} & \pie{0} & \pie{0} & \pie{0} & \pie{0} & \pie{0} & \pie{0} & \pie{360} \\
 &  & Triggers & \pie{360} & \pie{0} & \pie{0} & \pie{0} & \pie{0} & \pie{0} & \pie{0} & \pie{360} & \pie{360} \\
\addlinespace[2pt]
\cmidrule(lr){2-12}
\addlinespace[2pt]
 & \multirow{3}{*}{\rowsubcat{Security}} & Authentication & \pie{360} & \pie{360} & \pie{0} & \pie{0} & \pie{0} & \pie{0} & \pie{360} & \pie{360} & \pie{360} \\
 &  & Authorization & \pie{360} & \pie{360} & \pie{0} & \pie{0} & \pie{0} & \pie{0} & \pie{360} & \pie{360} & \pie{360} \\
 &  & Data encryption & \pie{0} & \pie{0} & \pie{0} & \pie{0} & \pie{0} & \pie{0} & \pie{0} & \pie{0} & \pie{0}
% END inlined from tables/pretty/new-table3_1_2.tex

\tabularnewline
\bottomrule
\end{tabular}
\end{adjustbox}
\end{table}

%%%%%%%%%%%%%%%
%%%%%%%%%%%%%%% PART VI
%%%%%%%%%%%%%%%

\begin{table}[!t]
\centering
\caption{Summary of graph database distinctive features (Part VI-a).}
\label{new:table:graph_database_features_3_2_1}

\small
\setlength{\tabcolsep}{3pt}
\renewcommand{\arraystretch}{1.12}

\begin{adjustbox}{max width=\textwidth}
\begin{tabular}{@{}>{\centering\arraybackslash}p{0.75cm}>{\centering\arraybackslash}p{1.00cm}l*{11}{c}@{}}
\toprule
\multicolumn{2}{@{}c}{} & \multicolumn{1}{@{}l}{Feature}
& \rot{Ultipa}
& \rot{BangDB}
& \rot{ArcadeDB}
& \rot{Fluree}
& \rot{HGraphDB}
& \rot{StellarDB}
& \rot{AgensGraph}
& \rot{GalaxyBase}
& \rot{MillenniumDB}
& \rot{Kuzu}
& \rot{TAO}
\\
\midrule

% BEGIN inlined from tables/pretty/new-table3_2_1.tex
\multirow{13}{*}{\rowcat{Product}} & \multirow{6}{*}{\rowsubcat{Adoption}} & Active development & \pie{180} & \pie{360} & \pie{360} & \pie{360} & \pie{360} & \pie{360} & \pie{360} & \pie{0} & \pie{360} & \pie{0} & \pie{0} \\
 &  & Commercial support & \pie{360} & \pie{360} & \pie{0} & \pie{360} & \pie{0} & \pie{360} & \pie{360} & \pie{360} & \pie{0} & \pie{0} & \pie{0} \\
 &  & Live community & \pie{360} & \pie{0} & \pie{360} & \pie{180} & \pie{360} & \pie{360} & \pie{270} & \pie{0} & \pie{270} & \pie{360} & \pie{0} \\
 &  & Open source & \pie{0} & \pie{360} & \pie{360} & \pie{360} & \pie{360} & \pie{0} & \pie{360} & \pie{0} & \pie{180} & \pie{360} & \pie{0} \\
 &  & Pricing & \pie{360} & \pie{360} & \pie{360} & \pie{0} & \pie{0} & \pie{0} & \pie{0} & \pie{0} & \pie{0} & \pie{0} & \pie{0} \\
 &  & Trendiness & \pie{180} & \pie{0} & \pie{180} & \pie{0} & \pie{0} & \pie{0} & \pie{0} & \pie{0} & \pie{0} & \pie{360} & \pie{0} \\
\addlinespace[2pt]
\cmidrule(lr){2-14}
\addlinespace[2pt]
 & \multirow{7}{*}{\rowsubcat{Deployment}} & Containerization & \pie{360} & \pie{360} & \pie{360} & \pie{360} & \pie{0} & \pie{0} & \pie{0} & \pie{360} & \pie{360} & \pie{360} & \pie{0} \\
 &  & Work as dedicated instance & \pie{360} & \pie{360} & \pie{360} & \pie{360} & \pie{0} & \pie{360} & \pie{360} & \pie{360} & \pie{360} & \pie{0} & \pie{360} \\
 &  & Work as embedded & \pie{0} & \pie{360} & \pie{360} & \pie{360} & \pie{0} & \pie{0} & \pie{360} & \pie{360} & \pie{0} & \pie{360} & \pie{0} \\
 &  & Testing in-memory version & \pie{0} & \pie{360} & \pie{0} & \pie{0} & \pie{0} & \pie{0} & \pie{0} & \pie{0} & \pie{0} & \pie{360} & \pie{0} \\
 &  & Operating on Linux & \pie{360} & \pie{360} & \pie{360} & \pie{360} & \pie{360} & \pie{360} & \pie{360} & \pie{180} & \pie{360} & \pie{360} & \pie{0} \\
 &  & Operating on Windows & \pie{360} & \pie{0} & \pie{360} & \pie{360} & \pie{360} & \pie{0} & \pie{360} & \pie{180} & \pie{270} & \pie{360} & \pie{0} \\
 &  & SaaS offering & \pie{360} & \pie{360} & \pie{0} & \pie{360} & \pie{0} & \pie{360} & \pie{360} & \pie{360} & \pie{0} & \pie{0} & \pie{0} \\
\addlinespace[2pt]
\midrule
\addlinespace[2pt]
\multirow{14}{*}{\rowcat{Database}} & \multirow{4}{*}{\rowsubcat{Convenience}} & Automatic updates & \pie{360} & \pie{0} & \pie{360} & \pie{360} & \pie{0} & \pie{0} & \pie{0} & \pie{0} & \pie{0} & \pie{0} & \pie{0} \\
 &  & Client side caching & \pie{0} & \pie{0} & \pie{0} & \pie{0} & \pie{360} & \pie{0} & \pie{0} & \pie{0} & \pie{0} & \pie{360} & \pie{360} \\
 &  & Data versioning support & \pie{0} & \pie{0} & \pie{0} & \pie{360} & \pie{180} & \pie{0} & \pie{0} & \pie{0} & \pie{0} & \pie{0} & \pie{0} \\
 &  & Live backups & \pie{360} & \pie{0} & \pie{360} & \pie{360} & \pie{0} & \pie{0} & \pie{360} & \pie{360} & \pie{0} & \pie{0} & \pie{360} \\
\addlinespace[2pt]
\cmidrule(lr){2-14}
\addlinespace[2pt]
 & \multirow{5}{*}{\rowsubcat{Distribution}} & Cluster Re-balancing & \pie{0} & \pie{0} & \pie{0} & \pie{0} & \pie{360} & \pie{0} & \pie{0} & \pie{0} & \pie{0} & \pie{0} & \pie{0} \\
 &  & Data distribution & \pie{360} & \pie{360} & \pie{0} & \pie{0} & \pie{360} & \pie{360} & \pie{0} & \pie{360} & \pie{0} & \pie{0} & \pie{360} \\
 &  & High-availability & \pie{360} & \pie{360} & \pie{360} & \pie{0} & \pie{360} & \pie{0} & \pie{360} & \pie{360} & \pie{0} & \pie{0} & \pie{360} \\
 &  & Query distribution & \pie{360} & \pie{360} & \pie{0} & \pie{0} & \pie{360} & \pie{0} & \pie{0} & \pie{360} & \pie{0} & \pie{0} & \pie{0} \\
 &  & Replication support & \pie{360} & \pie{360} & \pie{360} & \pie{0} & \pie{360} & \pie{0} & \pie{360} & \pie{360} & \pie{0} & \pie{0} & \pie{360} \\
\addlinespace[2pt]
\cmidrule(lr){2-14}
\addlinespace[2pt]
 & \multirow{5}{*}{\rowsubcat{Software\\Development}} & Data types defined & \pie{360} & \pie{180} & \pie{360} & \pie{360} & \pie{360} & \pie{180} & \pie{180} & \pie{180} & \pie{360} & \pie{360} & \pie{360} \\
 &  & Logging/Auditing & \pie{360} & \pie{360} & \pie{360} & \pie{360} & \pie{360} & \pie{0} & \pie{360} & \pie{360} & \pie{0} & \pie{360} & \pie{0} \\
 &  & Object-graph mapper & \pie{0} & \pie{0} & \pie{0} & \pie{0} & \pie{360} & \pie{0} & \pie{0} & \pie{0} & \pie{0} & \pie{0} & \pie{0} \\
 &  & Reactive programming & \pie{0} & \pie{0} & \pie{360} & \pie{0} & \pie{0} & \pie{0} & \pie{0} & \pie{0} & \pie{0} & \pie{0} & \pie{0} \\
 &  & Documentation up-to-date & \pie{360} & \pie{360} & \pie{360} & \pie{360} & \pie{0} & \pie{360} & \pie{360} & \pie{360} & \pie{180} & \pie{360} & \pie{0}
% END inlined from tables/pretty/new-table3_2_1.tex

\tabularnewline
\bottomrule
\end{tabular}
\end{adjustbox}
\end{table}

\begin{table}[!t]
\centering
\caption{Summary of graph database distinctive features (Part VI-b).}
\label{new:table:graph_database_features_3_2_2}

\small
\setlength{\tabcolsep}{3pt}
\renewcommand{\arraystretch}{1.12}

\begin{adjustbox}{max width=\textwidth}
\begin{tabular}{@{}>{\centering\arraybackslash}p{0.75cm}>{\centering\arraybackslash}p{1.00cm}l*{11}{c}@{}}
\toprule
\multicolumn{2}{@{}c}{} & \multicolumn{1}{@{}l}{Feature}
& \rot{Ultipa}
& \rot{BangDB}
& \rot{ArcadeDB}
& \rot{Fluree}
& \rot{HGraphDB}
& \rot{StellarDB}
& \rot{AgensGraph}
& \rot{GalaxyBase}
& \rot{MillenniumDB}
& \rot{Kuzu}
& \rot{TAO}
\\
\midrule

% BEGIN inlined from tables/pretty/new-table3_2_2.tex
\multirow{19}{*}{\rowcat{Data}} & \multirow{7}{*}{\rowsubcat{Access}} & Binary protocol & \pie{0} & \pie{0} & \pie{360} & \pie{0} & \pie{360} & \pie{0} & \pie{0} & \pie{360} & \pie{360} & \pie{360} & \pie{0} \\
 &  & CLI & \pie{360} & \pie{360} & \pie{360} & \pie{0} & \pie{0} & \pie{360} & \pie{360} & \pie{360} & \pie{360} & \pie{360} & \pie{0} \\
 &  & GUI & \pie{0} & \pie{360} & \pie{360} & \pie{0} & \pie{180} & \pie{360} & \pie{360} & \pie{360} & \pie{360} & \pie{360} & \pie{0} \\
 &  & Multi-database & \pie{0} & \pie{360} & \pie{360} & \pie{0} & \pie{0} & \pie{360} & \pie{360} & \pie{0} & \pie{0} & \pie{0} & \pie{0} \\
 &  & Graph-native data & \pie{360} & \pie{0} & \pie{360} & \pie{360} & \pie{0} & \pie{0} & \pie{0} & \pie{360} & \pie{360} & \pie{360} & \pie{0} \\
 &  & REST API & \pie{360} & \pie{360} & \pie{360} & \pie{360} & \pie{0} & \pie{0} & \pie{0} & \pie{360} & \pie{360} & \pie{360} & \pie{360} \\
 &  & Query language & \pie{180} & \pie{360} & \pie{360} & \pie{360} & \pie{360} & \pie{360} & \pie{360} & \pie{360} & \pie{360} & \pie{360} & \pie{0} \\
\addlinespace[2pt]
\cmidrule(lr){2-14}
\addlinespace[2pt]
 & \multirow{4}{*}{\rowsubcat{Consistency}} & Granular locking & \pie{0} & \pie{0} & \pie{0} & \pie{0} & \pie{0} & \pie{0} & \pie{0} & \pie{0} & \pie{0} & \pie{360} & \pie{0} \\
 &  & Multiple isolation levels & \pie{0} & \pie{0} & \pie{360} & \pie{0} & \pie{0} & \pie{0} & \pie{180} & \pie{0} & \pie{180} & \pie{360} & \pie{0} \\
 &  & Read committed transaction & \pie{0} & \pie{0} & \pie{360} & \pie{360} & \pie{360} & \pie{0} & \pie{360} & \pie{0} & \pie{360} & \pie{360} & \pie{0} \\
 &  & Transaction support & \pie{0} & \pie{360} & \pie{360} & \pie{360} & \pie{180} & \pie{0} & \pie{360} & \pie{360} & \pie{360} & \pie{360} & \pie{360} \\
\addlinespace[2pt]
\cmidrule(lr){2-14}
\addlinespace[2pt]
 & \multirow{5}{*}{\rowsubcat{Control}} & Constraints & \pie{180} & \pie{180} & \pie{180} & \pie{270} & \pie{0} & \pie{0} & \pie{360} & \pie{0} & \pie{0} & \pie{0} & \pie{0} \\
 &  & Schema support & \pie{360} & \pie{360} & \pie{360} & \pie{180} & \pie{360} & \pie{0} & \pie{360} & \pie{360} & \pie{0} & \pie{360} & \pie{360} \\
 &  & Secondary indexes & \pie{360} & \pie{360} & \pie{360} & \pie{360} & \pie{360} & \pie{0} & \pie{360} & \pie{360} & \pie{360} & \pie{0} & \pie{360} \\
 &  & Server side procedures & \pie{0} & \pie{0} & \pie{360} & \pie{0} & \pie{0} & \pie{0} & \pie{360} & \pie{360} & \pie{0} & \pie{0} & \pie{360} \\
 &  & Triggers & \pie{360} & \pie{180} & \pie{360} & \pie{0} & \pie{0} & \pie{0} & \pie{360} & \pie{360} & \pie{0} & \pie{0} & \pie{360} \\
\addlinespace[2pt]
\cmidrule(lr){2-14}
\addlinespace[2pt]
 & \multirow{3}{*}{\rowsubcat{Security}} & Authentication & \pie{360} & \pie{360} & \pie{360} & \pie{360} & \pie{360} & \pie{360} & \pie{360} & \pie{360} & \pie{360} & \pie{0} & \pie{0} \\
 &  & Authorization & \pie{360} & \pie{360} & \pie{360} & \pie{360} & \pie{0} & \pie{360} & \pie{360} & \pie{360} & \pie{0} & \pie{0} & \pie{0} \\
 &  & Data encryption & \pie{360} & \pie{0} & \pie{0} & \pie{360} & \pie{0} & \pie{0} & \pie{0} & \pie{0} & \pie{0} & \pie{0} & \pie{0}
% END inlined from tables/pretty/new-table3_2_2.tex

\tabularnewline
\bottomrule
\end{tabular}
\end{adjustbox}
\end{table}

\subsubsection{Property Graph Model}\label{chapter:publications:survey:sec:graph-databases:dbs-and-systems:prop_graph}

The following graph databases we list are focused on the property graph model. 
%
%\textcolor{purple}{\sout{Their information describes the use of this model; we found no graph query languages for \texttt{RDF} in their features, even if these databases are described as multi-model:}
%
It means that these graph databases directly employ the property graph model, and they do not support any query languages for \texttt{RDF} in their features.

\paragraph{\textbf{Alibaba Graph Database / TuGraph}}
%License:Open source
\Chart{0.38}{Product}\hspace{5mm}
\Chart{0.46}{Database}\hspace{5mm}
\Chart{0.53}{Data}\\
The Ant Group, which is behind the Alibaba B2B marketplace, has a cloud-oriented graph database service, providing horizontal scalability,  ACID transactions and supporting the \frameworkTinkerPop\ stack.
It supports the property graph model, the \gqlGremlin\ query language and there are programming interfaces for \texttt{Go}, \langJava, \texttt{.NET}, \texttt{No\-de.js} and \langPython.
For visualization, a paid third-party tool \emph{G.V() - Gremlin Graph Database IDE and Visualization Tool} \citep{gdotv} can be utilized.
Among online resources, we found a page which refers to this offering as \texttt{Gra\-ph\- Da\-ta\-ba\-se}~\citep{alibaba_graphdb} and another page with the same graphics but naming the offering as \texttt{Tu\-Gra\-ph}~\citep{tugraph_graphdb}, the latter in a public repository under \licenseApacheTwo.

\paragraph{\textbf{Apache HugeGraph}}
%License:Open source
\Chart{0.62}{Product}\hspace{5mm}
\Chart{0.75}{Database}\hspace{5mm}
\Chart{0.71}{Data}\\
\dbHugeGraph~\citep{apache_hugegraph_github} 
is a graph database written in \langJava\ and compliant with the \texttt{Tin\-ker\-Pop\- 3} framework.
% HugeGraph offers a binary protocol via TinkerPop and Gremlin.
% https://tinkerpop.apache.org/docs/current/reference/
It adopts the edge-cut partition scheme and offers a decoupling between 
the actual \texttt{sto\-ra\-ge\- lay\-er}, supporting technologies 
such as \dbRocksDB, \dbCassandra\ and \dbHBase\- among others. 
It integrates with \texttt{Flink}, \texttt{Spark} and other data processing 
engines.
\dbHugeGraph\ also has a \texttt{gra\-ph} \texttt{en\-gi\-ne} 
\texttt{lay\-er} which intermediates with the \texttt{sto\-ra\-ge} \texttt{lay\-er}, 
supports both \texttt{OLTP} and \texttt{OLAP} graph computation types and 
offers a \texttt{REST} API to query the graph as well as APIs for service monitoring 
and operations. 
Above the \texttt{gra\-ph} \texttt{en\-gi\-ne} \texttt{lay\-er} there is an 
\texttt{ap\-pli\-ca\-tion} \texttt{lay\-er} which contains the following 
relevant components: \textit{a)} \texttt{Hub\-ble}, a visual analytics 
platform for the process of data modelling, online/offline analysis and 
guided workflow for operating graph applications; \textit{b)} the \texttt{Loa\-der} 
data importing tool to convert from different formats and bulk import into 
the database; \textit{c)} \texttt{To\-ols}, command-line utilities to 
deploy, manage and restore data; \textit{d)} \texttt{Com\-pu\-ter}, a distributed 
graph processing system which implements \texttt{Pre\-gel} and can run on 
\texttt{Ku\-ber\-ne\-tes}; \textit{e)} \texttt{Cli\-ent}, a \texttt{Ja\-va}-based 
client to operate \texttt{Hu\-ge\-Gra\-ph}, with the mention of other languages 
being a possibility for the future.
Its source code is available online under the \licenseApacheTwo, 
having an active mailing list and development community.

\paragraph{\textbf{ByteGraph}}
%License:Proprietary
\Chart{0.12}{Product}\hspace{5mm}
\Chart{0.46}{Database}\hspace{5mm}
\Chart{0.32}{Data}\\
\dbByteGraph~\citep{li2022bytegraph} is a distributed graph database, 
motivated by the use cases and massive amounts of data produced 
by platforms such as TikTok, Douyin and Toutiao (products of ByteDance). 
The authors of \dbByteGraph\ define three categories of graph workloads: 
online analytics processing (OLAP) such as knowledge graphs and 
risk management; transaction processing (OLTP) such as what is 
used in recommendation and GNN sampling; and serving of fresh data 
to applications in real time (OLSP) such as e-commerce.
This graph database makes use of a two-tier architecture consisting 
of a durable storage layer and a cache layer for low-latency data access. 
\dbByteGraph\ optimizes the caching layer by introducing the 
\texttt{edge-tree}, a data structure similar to the B-tree
in order to store the adjacency lists of vertices. 
This provides high parallelism when accessing the neighbourhood 
of high-degree vertices and reduces the overall data loaded into 
memory for queries such as edge searching. 
The authors propose two adaptive optimizations on thread pools and 
indexes to adapt to sudden changes such as rapid workload increases. 
Scenarios of machine failure are also considered, as \dbByteGraph\ 
employs several techniques to guarantee availability such as 
geographic replications and weighted consistent 
hashing~\citep{mirrokni2018consistent}. 
Computation and storage are decoupled to enhance scalability.
It supports the \gqlGremlin\ \texttt{GQL} (the authors provide both 
a rule-based and a cost-based optimizer) and the full 
set of queries used in the evaluation of \dbByteGraph\ 
are displayed online on GitHub~\citep{bytegraph_queries_github}. 
We did not find its source code online.
It is an in-house solution for which we did not find release notes or changelog, but it has been continuously updated based on recent publications~\citep{bian2026discard}.

\paragraph{\textbf{ChronoGraph}}
%License:Mixed
\Chart{0.45}{Product}\hspace{5mm}
\Chart{0.30}{Database}\hspace{5mm}
\Chart{0.38}{Data}\\
\dbChronoGraph~\citep{haeusler2017chronograph} is a \frameworkTinkerPop\-compliant (offering \gqlGremlin\ to query the data in property graph model) graph database supporting ACID transactions, system-time content versioning and analysis. 
It is implemented as a key-value store enhanced with temporal information, using a B-tree data structure.
It currently exists as part of the Chronos Project, an ecosystem of projects which has \texttt{Chro\-no\-DB} (versioned key-value store), 
\dbChronoGraph\ (the database) and \texttt{Chro\-no\-Sphe\-re} (versioned Eclipse Modeling Framework repository).
The project is available online~\citep{chronograph_github} under the \texttt{aGPL\- v3} (open-source and academic purposes) and commercial licenses are available on demand. 
It is to be deployed as a process-embedded library.
No visualization functionalities accompany the database.

\paragraph{\textbf{DataStax Enterprise Graph (DSE)}}
%License:Mixed
\Chart{0.43}{Product}\hspace{5mm}
\Chart{0.68}{Database}\hspace{5mm}
\Chart{0.58}{Data}\\
\dbDataStaxEnterprise~\citep{datastax} is a proprietary fork of \texttt{Ti\-tan} and supporting \langJava.
It integrates with the \dbCassandra~\citep{Lakshman:2010:CDS:1773912.1773922} distributed database (over which it provides graph data models) and supports \frameworkTinkerPop\ (property graph with \gqlGremlin).
It is complemented by the \texttt{Da\-ta\-Stax\- Stu\-dio}, which allows for interactive querying and visualization of graph data similarly to \dbNeoFRj\ and \dbSAPHanaGraph.
There are several repositories belonging to the DataStax company, many under \licenseApacheTwo, among which a fork of the \texttt{Cas\-san\-dra} database with code pertaining to the company's product.\footnote{For a list of repositories: \url{https://github.com/orgs/datastax/repositories}}
There is a repository with release notes of their products~\citep{datastax_github}.

% https://github.com/dgraph-io/dgraph
% https://dgraph.io
\paragraph{\textbf{Dgraph}}
%License:Mixed
\Chart{0.86}{Product}\hspace{5mm}
\Chart{0.57}{Database}\hspace{5mm}
\Chart{0.58}{Data}\\
\dbDgraph~\citep{dgraph_graphql,dgraph_github} was written in \texttt{Go} and it is a distributed graph database offering horizontal scaling and ACID properties.
It is built to reduce disk seeks and minimize network usage footprint in cluster scenarios.
\dbDgraph\ is licensed under two licenses: the \licenseApacheTwo\ and a \texttt{D\-gra\-ph\- Com\-mu\-ni\-ty\- Li\-cen\-se}.
It automatically moves data to rebalance cluster shards.
It uses a simplified version of the \texttt{Gra\-ph\-QL} query language.
%
%\textcolor{purple}{\sout{Support for \gqlGremlin\ or \gqlCypher\ has been mentioned for the future but will depend on community efforts. %\footnote{\url{https://docs.dgraph.io/faq/}} 
%\dbDgraph\ has a scalability advantage over \dbNeoFRj\ as the latter may have multiple servers but they are merely replicas, while the former can grow horizontally (vertical scaling is expensive).}}
%
There is a proprietary enterprise version (conditions specified under a custom \texttt{D\-gra\-ph\- Com\-mu\-ni\-ty\- Li\-cen\-se}) with advanced features for backups and encryption.
Official clients include \langJava, \langJavaScript\ and \langPython.
\dbDgraph\ does not offer native visualization and global analytics functionalities (but third-party tools supporting JSON format can be applied).

\paragraph{\textbf{Galaxybase}}
%License:Proprietary
\Chart{0.50}{Product}\hspace{5mm}
\Chart{0.54}{Database}\hspace{5mm}
\Chart{0.65}{Data}\\
\dbGalaxybase~\citep{tong2024galaxybase} is a commercial distributed graph database, offering 
a \textit{Pioneer} version (with limited storage capacity\footnote{See: \url{https://createlink.com/download}}) and two superior tiers: \textit{Enterprise} and \textit{Ultimate}. 
The website states it offers compressed data storage, a native graph storage engine and partitioned 
sharded storage.\footnote{See: \url{https://createlink.com/galaxyProduct}}
It supports the property graph model and the enterprise line offers properties types such as lists and collections, 
as well as cleaning of invalid data. 
Indices may be created for properties and read-committed transactions are supported.
Queries may be written with \gqlCypher\ and there are APIs for \texttt{Ja\-va}, \texttt{Go}, \texttt{Py\-thon} 
and \texttt{Ja\-va\-Scri\-pt}. 
With respect to operational management, it offers data backups/recovery, load balancing, role-based access 
control at the graph level, with fine-grained permissions for edges and attributes.
It has a visual interface to explore the graph.

\paragraph{\textbf{Gaffer}}
%License:Open source
\Chart{0.23}{Product}\hspace{5mm}
\Chart{0.46}{Database}\hspace{5mm}
\Chart{0.32}{Data}\\
\dbGaffer\ is a graph database framework, providing modular storage for very large graphs with properties on 
vertices and edges.~\citep{gaffer_github}.
It allows the use of storage options such as \texttt{Accumulo}~\citep{halldorsson2013apache}, \texttt{HBase}~\citep{DBLP:books/daglib/0027893} 
and \texttt{Parquet}~\citep{vohra2016apache} (though documentation in the current version claims the last two will be deprecated\footnote{See: \url{https://github.com/gchq/Gaffer/issues/2556} and \url{https://github.com/gchq/Gaffer/issues/2590}} in version $2$ of \dbGaffer).
It was created by UK's intelligence, security, and cyber agency.
\dbGaffer\ is licensed under the \licenseApacheTwo\ and is covered by Crown Copyright.
Its source code is available online~\citep{gaffer_github} with development instructions.\footnote{See: \url{https://gchq.github.io/gaffer-doc/v1docs/summaries/getting-started.html}}
\dbGaffer\ contains a \texttt{time-library} enabling timestamps to be set on stored entities and edges.
It does this by offering classes such as \texttt{RBMBackedTimestampSet} which stores timestamps truncated to a certain resolution (e.g., minutes, hours and so on)  
and \texttt{BoundedTimestampSet} which allows for the accumulation of timestamps up to a maximum of $N$ (if it goes over this number, $N$ are sampled from the total).
It is written in \langJava\ and \langJavaScript\ and is able to interoperate with \texttt{Spark} \texttt{RDD}, \texttt{DataFrame} and \texttt{GraphFrame} representations. 
It also offers a \texttt{REST} API. 
Any distributed processing afforded by \dbGaffer\ is achieved by virtue of the underlying storage technology and use of \texttt{Spark}.

\paragraph{\textbf{Graphflow}}
%License:Open source
\Chart{0.46}{Product}\hspace{5mm}
\Chart{0.11}{Database}\hspace{5mm}
\Chart{0.14}{Data}\\
\dbGraphflow~\citep{kankanamge2017graphflow,10.14778/3342263.3342643,mhedhbi2020a+} was released as a prototype active graph database.
It is an in-memory graph store employing the property graph model and supporting both one-time and continuous sub-graph queries.
\dbGraphflow\ uses a one-time query processor called \textit{Generic Join} and a \textit{Delta Join} which enables the continuous sub-graph queries. 
It extends the \texttt{open\-Cy\-pher} language with triggers to perform actions upon certain conditions. 
We did not find information regarding its direct use beyond academic purposes nor about supporting ACID transactions.
Its code was available online~\citep{graphflow_github} under the \licenseApacheTwo, but now returns a 404 error as of March 2026.

\paragraph{\textbf{G-Tran}}
%License:Open source
\Chart{0.46}{Product}\hspace{5mm}
\Chart{0.32}{Database}\hspace{5mm}
\Chart{0.21}{Data}\\
\dbGTran~\citep{chen2022g} is available online~\citep{gtran_github} 
under \licenseApacheTwo\ and written in \texttt{C++}.
It is a remote direct memory access (RDMA)-enabled distributed 
in-memory graph database, offering support for serialization 
and snapshot isolation. 
The authors introduce a graph-native data store to enable 
desirable data locality and fast data access for transactional 
updates and queries.
\dbGTran\ offers a fully-decentralized architecture, relying on 
RDMA to process distributed transactions with the massively 
parallel processing (MPP) model, aiming to maximize the usage 
of computer resources. 
The authors propose a novel multi-version optimistic concurrency control
(\texttt{MV-OCC}) protocol to handle the challenge of large 
read/write sets in graph transactions. 
It was benchmarked against \dbJanusGraph, \dbArangoDB, \dbNeoFRj\ 
and \dbTigerGraph\ (Developer Edition). 
It offers snapshot isolation and strong consistency, 
aiming to provide low latency and high throughput using: \textit{a)} 
graph-native data store with efficient data memory layouts 
to support data locality and fast access for read/write graph transactions 
under frequent updates; \textit{b)} a decentralized architecture 
based on RDMA to avoid the bottlenecks of centralized transaction 
coordination.
\dbGTran\ uses \gqlGremlin\ for querying and supports the property 
graph model.

\paragraph{\textbf{HGraphDB}}
%License:Open source
\Chart{0.38}{Product}\hspace{5mm}
\Chart{0.68}{Database}\hspace{5mm}
\Chart{0.37}{Data}\\
\dbHGraphDB~\citep{hgraphdb_github} is an implementation of \frameworkTinkerPop 3, 
functioning as a client layer to use \dbHBase\ as a graph database and open-sourced under \licenseApacheTwo. 
It enables the creation of schemas for vertices and edges. 
Graph analytics can be achieved by providing integration with \texttt{Gi\-ra\-ph}, \texttt{Spark} \gpeGraphFrames\ 
and \texttt{Flink} \texttt{Gel\-ly}, which are able to access the data stored in \texttt{HBase} for computation. 
Because \frameworkTinkerPop\ is used as a layer over \texttt{HBa\-se}, \gqlGremlin-based tools are available.
The distribution, scalability and storage mechanisms are those of the underlying \texttt{HBa\-se}.
%\item \texttt{H\-Gra\-ph\-DB}~\citep{hgraphdb_github} is an implementation of \frameworkTinkerPop\ 3, 
%functioning as a client layer to use \dbHBase\ as a graph database and open-sourced under \licenseApacheTwo. 
%It enables the creation of schemas for vertices and edges. 
%Graph analytics can be achieved by providing integration with \texttt{Gi\-ra\-ph}, \texttt{Spark} \gpeGraphFrames\ 
%and \texttt{Flink} \texttt{Gel\-ly}. 
%Because \frameworkTinkerPop\ is used as a layer over \texttt{HBa\-se}, \gqlGremlin-based tools are available.
%The distribution, scalability and storage mechanisms are those of the underlying \texttt{HBa\-se}.

\paragraph{\textbf{JanusGraph}}
%License:Open source
\Chart{0.92}{Product}\hspace{5mm}
\Chart{0.52}{Database}\hspace{5mm}
\Chart{0.74}{Data}\\
\dbJanusGraph~\citep{janus} is an open-source project licensed~\citep{janus_github} under the \licenseApacheTwo.
A database optimized for storing (in adjacency list format) and querying large graphs with (billions of) edges and vertices distributed across a multi-machine cluster with ACID transactions.
\dbJanusGraph, which debuted in 2017, is based on the \langJava\ code base of the \texttt{Ti\-tan} graph database project and is supported by the likes of Google, IBM and the \texttt{Li\-nux} Foundation, to name a few.
Like \texttt{Ti\-tan}, it supports \dbCassandra, \texttt{HBase} and \texttt{Ber\-ke\-ley\-DB} (among others) as an underlying storage.
\dbJanusGraph\ can integrate platforms such as \texttt{Spark}, \texttt{Gi\-ra\-ph} and \texttt{Ha\-doop}.
It also natively integrates with the \frameworkTinkerPop\ graph stack, supporting \gqlGremlin\ applications, the query language and its graph server, with graphs in the property graph model. 
Due to supporting \frameworkTinkerPop, one may use one of its drivers to use \gqlGremlin\ from \texttt{E\-li\-xir}, \texttt{Go}, \langJava, \texttt{.NET}, \texttt{PHP}, \langPython, \langRuby\ and \langScala.
It supports global analytics using \texttt{Spark} integration as well.

\paragraph{\textbf{KatanaGraph}}
%License:Proprietary
\Chart{0.38}{Product}\hspace{5mm}
\Chart{0.21}{Database}\hspace{5mm}
\Chart{0.11}{Data}\\
\dbKatanaGraph\ is a library and set of 
applications to work efficiently with large graphs. 
It offers parallel graph analytics and machine learning 
algorithms, multithreaded execution with load balancing and 
a scalability focus on multi-socket systems. 
It provides scalable concurrent containers (e.g. bag, vector 
and list) and an interface to implement algorithms.
It is licensed under the \texttt{BSD-3-Clause} 
license.
\dbKatanaGraph's enterprise offering is known 
as \texttt{Graph Intelligence Platform}, providing solutions 
for scalable purposes such as querying (contextual searches), 
analytics (path-finding, centrality and community detection), 
mining (for pattern discovery) and prediction.
It has been tested with clusters of over 256 machines, 
supercomputers and public clusters on Amazon Web Services 
(AWS), Google Cloud Platform (GCP) and Microsoft Azure, and 
offers \texttt{C/C++} and \langPython\ programming bindings.\footnote{See: \url{https://katanagraph.com/resources/datasheets/katana-graph-intelligence-platform}}
\texttt{Ka\-ta\-na\-Gra\-ph} provides different partitioning 
policies (1D, 2D and hybrid vertex-cut) as well as 
user-specified policies and an optimized partition-aware 
communication layer.
Its distributed parallel computing framework is based on previous 
works such as the \texttt{Gluon}~\citep{dathathri2018gluon} 
communication engine and the \texttt{Cusp}~\citep{hoang2021cusp} 
streaming partitioner.
\dbKatanaGraph's value proposition is the integration 
of a graph computing subsystem of a graph database, analytics and 
machine learning. 
Although it is not a graph database in itself, it is worth 
mentioning as a scalable distributed system allowing for the 
ingestion and processing of data from many different sources 
while harnessing many types of computational resources. 
Its last release was in 2022, and we did not find information about its future development plans.

\paragraph{\textbf{Kuzu}}
%License:Open source
\Chart{0.62}{Product}\hspace{5mm}
\Chart{0.29}{Database}\hspace{5mm}
\Chart{0.61}{Data}\\
\dbKuzu~\citep{JinFCLS23} is an open-source, embedded property graph database intended to be integrated directly into applications rather than deployed as a standalone server. 
Its main focus is on high-performance single-node graph analytics. 
It adopts  strongly typed graph data model and provides a \gqlCypher-based (\texttt{GQL}-inspired) query language for expressive graph pattern matching and analytical workloads. 
It is implemented in \texttt{C++} with a vectorized execution engine that achieves high query throughput on local machines while supporting ACID transactions under read-committed isolation. 
The system is cross-platform, running on \texttt{Li\-nux} and \texttt{Win\-dows}, and integrates smoothly with developer environments such as \langPython. 
It favours simplicity and development ease through an embedded API and command-line interface, while deliberately omitting server-oriented and enterprise features including clustering, replication, authentication, and high availability; thus, more suited for embedded analytics, data science, and local graph processing rather than large-scale distributed deployments. 
It is under the \licenseMIT, and the project has been archived as of October 2025 (as per its GitHub repository).\citep{kuzu_github}

\paragraph{\textbf{LiveGraph}}
%License:Open source
\Chart{0.37}{Product}\hspace{5mm}
\Chart{0.11}{Database}\hspace{5mm}
\Chart{0.16}{Data}\\
\dbLiveGraph~\citep{zhu2019livegraph} is a graph storage system which 
optimizes graph workloads by ensuring that adjacency list scans 
are \textit{purely sequential}, that is, random accesses are never 
required, even in the face of concurrent transactions. 
It implements in-memory and out-of-core graph storage on a 
single server and offers the property graph model.
The authors present the \texttt{Transactional Edge Log} (\texttt{TEL}), 
a graph-aware data structure with a concurrency control mechanism.
\dbLiveGraph\ proposes to be a graph storage system supporting 
both transactional and real-time analytical workloads. 
The authors employ single-threaded micro-benchmarks and
micro-architectural analysis for comparison of different 
commonly-used data structures and assess the advantage of a 
sequential memory layout. 
This system guarantees the following properties: \textit{1)} 
that different versions of edges in the same adjacency list 
have to be stored in contiguous memory locations; \textit{2)} 
locating the right version of an edge should not depend 
on auxiliary data structures which could lead to the occurrence of 
random accesses; \textit{3)} the concurrency control 
algorithm does not require random access during scans.
Its source code is available online under \licenseApacheTwo\ and written in \texttt{C++}~\citep{livegraph_github}. 
It is a relevant approach to graph storage although the evaluated techniques are presented as the \dbLiveGraph\ system without being part of a database system (thus lacking all the functionalities typically found in them).

\paragraph{\textbf{Memgraph}}
%License:Proprietary
\Chart{0.70}{Product}\hspace{5mm}
\Chart{0.63}{Database}\hspace{5mm}
\Chart{0.72}{Data}\\
\dbMemgraph\ is a commercial graph database with both an on-premise version and a cloud service~\citep{memgraph_github}. 
It was developed for real-time streaming and is compatible with \dbNeoFRj. 
It supports ingestion of data from \texttt{Kafka}, \texttt{SQL} and plain \texttt{CSV} files and allows users 
to query data using \gqlCypher.
It is implemented in \texttt{C/C++}, offering an in-memory architecture and ACID compliance.
The source code is available online under a custom license called \texttt{Bu\-si\-nes\-s} \texttt{Sour\-ce}
\texttt{Li\-cen\-se\- 1.1\- (BSL)}, the \dbMemgraph\ \texttt{En\-ter\-pri\-se} \texttt{Li\-cen\-se\- (MEL)}.
It supports dynamic graph algorithms such as dynamic PageRank with a stream of data, where rankings 
are updated as soon as new graph objects are created in the database. 
%It was benchmarked against the \langPython\ \texttt{networkx} package, coming out five times faster for PageRank on a small graph ($|V| = 78.181, |E| = 310.227$).
\dbMemgraph\ provides graph functionality divided by modules. 
It has a module for traditional graph algorithms, covering classes of graph problems such as centrality measures, 
path-finding and community detection. 
Another module contains streaming graph algorithms for centrality and community detection 
algorithms. 
It also provides graph machine learning algorithms, such as link prediction and node classification based on 
graph neural networks (GNNs), as well as computing node embeddings on static graphs (\texttt{node2vec}) and a 
graph neural network algorithm to predict new edges and node labels from graph structure and available 
node and edge features.
There is also a module that supports integration with NVIDIA \texttt{CUDA} and \texttt{networkx}.
The commercial product also provides \texttt{Mem\-gra\-ph\- Lab}, a visual user interface to connect and explore 
data in \dbMemgraph\ instance and \texttt{MA\-GE}, an open-source repository of query modules and graph algorithms.

% https://github.com/vesoft-inc/nebula
% https://nebula-graph.io/
% https://docs.nebula-graph.io/manual-EN/1.overview/3.design-and-architecture/1.design-and-architecture/
\paragraph{\textbf{NebulaGraph}}
%License:Open source
\Chart{0.69}{Product}\hspace{5mm}
\Chart{0.64}{Database}\hspace{5mm}
\Chart{0.66}{Data}\\
\dbNebulaGraph\ is an open-source graph database (available online~\citep{nebulagraph_github}) licensed under \licenseApacheTwo, provides a custom \texttt{Ne\-bu\-la\- Gra\-ph\- Que\-ry\- Lan\-gu\-age\- (nGQL)} with syntax close to \texttt{SQL} and \gqlCypher\ support is planned.
It supports the property graph model, ACID transactions and is implemented with a separation of storage and computation, being able to scale horizontally.
It supports multiple storage engines like \texttt{HBase}~\citep{DBLP:books/daglib/0027893} (implementing the graph logic over these key-value stores) and \texttt{Ro\-cks\-DB}~\citep{rocksdb_github} and has clients in \texttt{Go}, \langJava\ and \langPython.
It also has the complementing \texttt{Ne\-bu\-la\- Gra\-ph\- Stu\-dio} for interactive visual querying and analytics. 

\paragraph{\textbf{Neo4j}}
%License:Mixed
\Chart{0.96}{Product}\hspace{5mm}
\Chart{0.71}{Database}\hspace{5mm}
\Chart{0.89}{Data}\\
\dbNeoFRj~\citep{Webber:2012:PIN:2384716.2384777} is a graph database with multiple editions~\citep{neo4j_github}: a community edition licensed under the free \texttt{GNU\- Ge\-ne\-ral\- Pu\-blic\- Li\-cen\-se\- (GPL)\- v3}, and a commercial one. 
We note that its public GitHub repository has last been updated two years ago (and therefore consider it is not in active development).
However, the company has been continuously updating the product/service offering.
It supports different programming languages \texttt{C/C++}, \langClojure, \texttt{Go}, \langHaskell, \langJava, \langJavaScript, \texttt{.NET}, \texttt{Perl}, \texttt{PHP}, \langPython, \texttt{R} and \langRuby. 
\dbNeoFRj\ is optimized for highly-connected data.
It relies on methods of data access for graphs without considering data locality. 
\dbNeoFRj's graph processing consists of mostly random data access.
For large graphs which require out-of-memory processing, the major performance bottleneck becomes the random access to secondary storage.
The authors created a system which supports ACID transactions, high availability, with operations that modify data occurring within transactions to guarantee consistency.
It uses the \gqlCypher\ query language and data is stored on disk as fixed-size records in linked lists.
\dbNeoFRj\ has a library offering many different graph algorithms. %\footnote{\url{https://neo4j.com/developer/graph-algorithms/}}
%\textcolor{purple}{\sout{As far as we know, \dbNeoFRj's scale-out capabilities are only true for read operations.}}
All writes are directed to the \dbNeoFRj\ cluster master, an architecture which has its limitations. 
Among other uses, \dbNeoFRj\ has also been employed for building applications using the \texttt{GRAND\-sta\-ck} framework~\citep{grandstack_farmework}. 
\dbNeoFRj\ also has an interactive graph explorer to query and update specific elements of the graph.
It also offers the \texttt{Aura} automated cloud platform database as a service.

\paragraph{\textbf{PandaDB}}
%License:Open source
\Chart{0.46}{Product}\hspace{5mm}
\Chart{0.68}{Database}\hspace{5mm}
\Chart{0.89}{Data}\\
\dbPandaDB~\citep{zhao2021pandadb} is an open-source distributed graph database for the management 
and querying of both structured and unstructured data in one framework. 
\dbPandaDB\ is aimed at applications like recommendation systems, financial technology (e.g., fraudulent cash-out 
detection) and knowledge graphs (e.g., new data operators to process the information of unstructured data to accelerate 
graph neural networks).
It consists of: \textit{1}; a graph data model to manage the data; \textit{2}; a \gqlCypher\ query language extension to understand the 
semantic information of the unstructured graph data; \textit{3} a cost model and query optimization techniques to speed up 
unstructured data processing.
It adopts the native graph database \dbNeoFRj\ as its foundation and estimates the cost of unstructured data operations 
(AI model inference is used to understand the unstructured data) with query plan optimization.
% https://arxiv.org/pdf/2107.01963.pdf
It is an academic software with multiple versions available.\footnote{See: \url{https://github.com/grapheco}} 
It is available under the \licenseApacheTwo\ and written mostly in \langScala~\citep{pandadb_github}.

\paragraph{\textbf{RedisGraph}}
%License:Mixed
\Chart{0.79}{Product}\hspace{5mm}
\Chart{0.46}{Database}\hspace{5mm}
\Chart{0.58}{Data}\\
\dbRedisGraph~\citep{cailliau2019redisgraph} is a property graph database which uses sparse matrices to represent a graph's adjacency matrix and uses linear algebra for graph queries.
It uses custom memory-efficient data structures stored in RAM, having on-disk persistence and tabular result sets.
Queries may be written in a subset of \gqlCypher\ and are internally translated into linear algebra expressions. 
It has a custom license and client libraries for \dbRedisGraph\ have been developed in \texttt{E\-li\-xir}, \texttt{Go}, \langJava, \langJavaScript, \texttt{PHP}, \langPython, \langRuby\ and \texttt{Rust}, complementing existing accesses that \texttt{Re\-dis} already supports. 
As far as we know, it only works in single-server mode (bounded by the machine's RAM) and it does not support ACID properties.
It has a Community Edition under a custom license, the latter 
with  source code available online~\citep{redis_github}.

\paragraph{\textbf{SAP Hana Graph}}
%License:Proprietary
\Chart{0.54}{Product}\hspace{5mm}
\Chart{0.93}{Database}\hspace{5mm}
\Chart{0.89}{Data}\\
\dbSAPHanaGraph~\citep{rudolf2013graph,hwang2018graph,hana_github} is a column-oriented, in-memory relational database management system. 
It performs different type of data analysis, among which, graph data processing with the property graph model and ACID transactions.
This graph functionality includes interpretation of \gqlCypher\ and a visual graph manipulation tool.
Its graph processing capabilities have served use cases like fraud detection and route planning.
No source code is available online as the product is purely commercial.

\paragraph{\textbf{Sparksee}}
%License:Proprietary
\Chart{0.38}{Product}\hspace{5mm}
\Chart{0.29}{Database}\hspace{5mm}
\Chart{0.42}{Data}\\
\dbSparksee~\citep{martinez2011dex,sparksee_former_dex} (formerly \texttt{DEX}) is a property graph database offering ACID transactions and representing the graph using bitmap data structures with high compression rates (with each bitmap partitioned into chunks that fit disk pages).
The graphs in \dbSparksee\ are labelled multigraphs and it has multiple licenses depending on the purpose, with free licenses for evaluation, research and development.
It offers APIs in \texttt{C++}, \texttt{.NET}, \langJava, \texttt{Ob\-jec\-ti\-ve-C}, \langPython\ and mobile devices.
We found no capabilities for data visualization in \dbSparksee, though it is able to export data to formats supported by third-party software.

\paragraph{\textbf{StellarDB}}
%License:Proprietary
\Chart{0.46}{Product}\hspace{5mm}
\Chart{0.18}{Database}\hspace{5mm}
\Chart{0.32}{Data}\\
\dbStellarDB~\citep{stellardb} is a distributed graph database with a custom graph storage 
structure, employing compression algorithms and partitioning strategies to reduce the storage footprint 
and to focus on even data distribution in the cluster. 
It offers a \gqlCypher-based query language as well as \texttt{SQL}, and has a library of graph algorithms 
that are parallel-capable. 
Graph visualization is offered through an interface which supports 2D and 3D visualization display. 
It runs on a family of \texttt{Li\-nux} operating systems.\footnote{See: \url{https://www.transwarp.cn/doc/tdh/9.3.3/Installation-guide--install-preparation-chapter}}
\texttt{Stel\-lar\-DB} has database management features such as user permission authentication, 
cluster status monitoring, logging, data encryption and backup recoveries. 
It is subject to commercial licensing and it is not open-source. 
\texttt{Stel\-lar\-DB} depends on \texttt{HDFS}, \texttt{Yarn} and \texttt{Zoo\-Kee\-per}.
More information is available at the website of \texttt{Stel\-lar\-DB}.\footnote{See: \url{https://www.transwarp.cn/en/product/stellardb}}  

\paragraph{\textbf{TAO}}
%License:Proprietary
\Chart{0.13}{Product}\hspace{5mm}
\Chart{1.00}{Database}\hspace{5mm}
\Chart{0.67}{Data}\\
\dbTAO~\citep{bronson2013tao} debuted as Facebook's distributed graph store.
It is a geographically-distributed data store, providing efficient and timely access to the social graph using a fixed set of queries. 
It replaced \texttt{memcache} for many data types and runs on thousands of machines in a distributed fashion. 
It divides the data into logical shards, with each shard assigned to a  \texttt{MySQL} instance, and  being contained in a logical database. 
Database servers manage one or more shards, requiring tuning shard-to-server mapping to balance load across hosts. 
All object types are stored in one table, and all association types in another table.
\dbTAO\ has a caching layer which implements a complete API for clients and handles all communication with databases.
This layer has multiple cache servers (which together form a tier), and the cache is filled on demand with an LRU eviction policy.
It was designed in a way to limit maximum size of tiers by creating a hierarchy, with a \textit{leader} tier and \textit{follower} tiers.
Later, \texttt{RAMP-TAO} was presented~\citep{cheng2021ramp} as a protocol over \dbTAO\ which ensures atomic visibility 
for an environment that is geographically-distributed and read-optimized.
We chose to include \dbTAO\ due to its research and engineering relevance to enable 
operation on a massive graph use-case.

\paragraph{\textbf{TigerGraph}}
%License:Proprietary
\Chart{0.77}{Product}\hspace{5mm}
\Chart{0.48}{Database}\hspace{5mm}
\Chart{0.66}{Data}\\
\dbTigerGraph~\citep{deutsch2019tigergraph,tigergraph} is a commercial graph database (formerly \texttt{GraphSQL}\footnote{See: \url{https://www.zdnet.com/article/tigergraph-a-graph-database-born-to-roar/}}) implemented in \texttt{C++} and comes in three versions: developer edition (supporting only single-machine, no distribution and is only for non-production, research or educational purposes), cloud edition (as a managed service) and enterprise edition (allowing for horizontal scalability - distributed graphs).
It supports ACID consistency, access through a \texttt{REST} API, has a custom \texttt{SQL}-like query language (\texttt{GSQL}) and features a graphical user interface named \texttt{Gra\-ph\-Stu\-dio} to perform interactive graph data analytics. 
The \dbTigerGraph\ model was designed with its graph vertices, edges and their attributes to support an engine that performs massively-parallel processing to compute queries and analytics.

%\textcolor{purple}{\sout{Each vertex and edge acts as both a unit of storage and computation, integrating and extending both TLAV and TLEV paradigms.
%It supports the property graph model plus extensions to enable the massively-parallel processing.}}

\paragraph{\textbf{Ultipa}}
%License:Proprietary
\Chart{0.69}{Product}\hspace{5mm}
\Chart{0.64}{Database}\hspace{5mm}
\Chart{0.53}{Data}\\
\dbUltipa~\citep{ultipa_graph} focuses on achieving high scalability 
by employing an \texttt{HTAP} (\texttt{Hy\-brid} \texttt{Tran\-sac\-tio\-nal} and 
\texttt{A\-nal\-y\-ti\-cal} \texttt{Pro\-ce\-s\-sing}) architecture. 
This architecture (a practical implementation was published in recent years~\citep{huang2020tidb}) 
aims to remove the need of having multiple copies of data 
and the data offloading from databases to warehouses.
It has its own \dbUltipa\ \texttt{Que\-ry} \texttt{Lan\-gu\-age} (\texttt{UQL}), 
and the authors intend to evolve it in order to keep up with compatibility and 
functionality of the developing international \texttt{GQL} standard of 
the LDBC (Linked Data Benchmark Council).
There is the \dbUltipa\ \texttt{Ma\-na\-ger} which unifies the management 
of the database, cluster, datasets, schemas and account management, enabling 
users to configure and visualize graph data.
\dbUltipa\ \texttt{Cloud} offers a pay-as-you-go model and is a solution 
integrating the \dbUltipa\ \texttt{Ma\-na\-ger} and \dbUltipa\ \texttt{Ser\-ver}. 
It offers shared, standard and enterprise versions, the last one's price defined on a per-contact 
basis.
On the website, drivers are mentioned for \texttt{Ja\-va}, \texttt{Py\-thon}, \texttt{C++}, 
\texttt{Go}, \texttt{No\-de} and a \texttt{REST} API.
Its licensing is commercial and there is no open-source free version.

\paragraph{\textbf{Weaver}}
%License:Open source
\Chart{0.33}{Product}\hspace{5mm}
\Chart{0.29}{Database}\hspace{5mm}
\Chart{0.13}{Data}\\
\dbWeaver~\citep{DBLP:journals/pvldb/DubeyHES16} is an open-source~\citep{weaver_github} graph database (custom permissive license) for efficient, transactional graph analytics.
It introduced the concept of refinable timestamps.
It is a mechanism to obtain a rough ordering of distributed operations if that is sufficient, but also fine-grained orderings when they become necessary.
It is capable of distributing a graph across multiple shards while supporting concurrency.
Refinable timestamps allow for the existence of a multi-version graph: write operations use their timestamps as a mark for vertices and edges.
This allows for the existence of consistent versions of the graph so that long-running analysis queries can operate on a consistent version of the graph, as well as historical queries.
\dbWeaver\ is written in \texttt{C++}, offering binding options for \langPython. 
We did not find any support for popular graph query languages.

\paragraph{\textbf{ZipG}}
%License:Proprietary
\Chart{0.29}{Product}\hspace{5mm}
\Chart{0.32}{Database}\hspace{5mm}
\Chart{0.05}{Data}\\
\dbZipG\ was designed as a memory-efficient graph store for interactive querying~\citep{khandelwal2017zipg}.
It employs a compressed graph representation supporting a range of graph queries which are applied directly on the 
compressed representation. 
It was evaluated using queries from different graph query workloads (e.g., such as the Facebook \texttt{TAO} use-cases), 
achieving an order of magnitude higher in throughput compared to \dbNeoFRj\ and \texttt{Ti\-tan}.
\dbZipG\ was implemented in \texttt{C++} and has a package running on \texttt{Spark} and written in \langScala. 
It supports distributed execution and its source-code is available online~\citep{zipg_github}, albeit missing licensing information.
It warrants mentioning for its use of a compressed graph representation and support for graph query operations, but it does not offer traits of a graph database such as transaction support, replication, graphical user interfaces or graph query languages.

\subsubsection{RDF Data Model}\label{chapter:publications:survey:sec:graph-databases:dbs-and-systems:RDF}

The following graph databases we address are focused on the \texttt{RDF} data model and variations (including support for the property graph model by representing them as \texttt{RDF}~\citep{hartig2014reconciliation}):

\paragraph{\textbf{AllegroGraph}}
%License:Proprietary
\Chart{0.54}{Product}\hspace{5mm}
\Chart{0.93}{Database}\hspace{5mm}
\Chart{0.79}{Data}\\
\dbAllegroGraph~\citep{allegrograph} is a proprietary commercial graph database with clients under \texttt{E\-cli\-pse\- Pu\-blic\- Li\-cen\-se\- v1\- (EPL\- v1)} which supports several programming languages (\texttt{C\#}, \texttt{C}, \texttt{Com\-mon\- Lisp}, \langClojure, \langJava, \texttt{Perl}, \langPython, \langScala) that was purpose-built for \texttt{RDF} (triple-store).
It supports an array of mechanisms to access the information it stores, namely reasoning with an ontology (\texttt{RDFS++\- Rea\-so\-ning}), materialized reasoning (generating new triples based on inference rules - \texttt{OWL2\- RL\- Ma\-te\-ria\-li\-zed\- Rea\-so\-ner}), \gqlSPARQL\ queries, \texttt{Pro\-log} and also low-level APIs.

\paragraph{\textbf{Blazegraph}}
%License:Proprietary
\Chart{0.44}{Product}\hspace{5mm}
\Chart{0.46}{Database}\hspace{5mm}
\Chart{0.58}{Data}\\
\dbBlazegraph~\citep{blazegraph} (formerly \texttt{Big\-da\-ta}) is an \texttt{RDF} database able to support up to billions of edges in a single machine and available under the \texttt{GNU\- Ge\-ne\-ral\- Pu\-blic\- Li\-cen\-se\- v2.0}, supporting \gqlSPARQL\ and the \texttt{Tin\-ker\-Pop\- Blue\-prin\-ts} API.
It has \texttt{.NET} and \langPython\ clients.
One of its associated internal projects is \texttt{bla\-ze\-gra\-ph-grem\-lin}, which allows the storage of property graphs internally in \texttt{RDF} format, which can then be queried with \gqlSPARQL. 
It also offers \dbBlazegraph\ \texttt{Work\-Ben\-ch} as a GUI to visualize and query data.
It essentially has an alternative approach to \texttt{RDF} reification, enabling labelled property graph capabilities on \texttt{RDF} graphs, with the ability to query the graphs in \gqlGremlin\ as well. 
Stored procedures are supported via \gqlSPARQL\ 
extensions and transactions are supported through 
Multi-Version Optimistic Concurrency Control (OCC).
Snapshot isolation is offered and authentication may be enabled with mechanisms outside the database itself. 
It uses a federated architecture to distribute the data across the cluster resources (with queries also functioning across the cluster).
\dbBlazegraph\ has been acquired by Amazon.\footnote{See: \url{https://www.trademarkia.com/blazegraph-86498414}}

\paragraph{\textbf{BrightstarDB}}
%License:Open source
\Chart{0.46}{Product}\hspace{5mm}
\Chart{0.46}{Database}\hspace{5mm}
\Chart{0.58}{Data}\\
\dbBrightstarDB\ is an open-source~\citep{brightstardb_github} multi-threaded multi-platform (including mobile) \texttt{.NET} \texttt{RDF} store, supporting \gqlSPARQL\ and binding of \texttt{RDF} resources to \texttt{.NET} dynamic objects (it has tools to use \texttt{.NET} interfaces and generate concrete classes to persist their data in \dbBrightstarDB). 
It is licensed under the \texttt{MIT\- Li\-cen\-se} and there is also an Enterprise version.
\dbBrightstarDB\ supports single-threaded writes and multi-threaded reads, with ACID transactions. 
It does not support horizontal scaling.

\paragraph{\textbf{Cray Graph Engine}}
%License:Proprietary
\Chart{0.23}{Product}\hspace{5mm}
\Chart{0.50}{Database}\hspace{5mm}
\Chart{0.58}{Data}\\
\dbCrayGraphEngine~\citep{rickett2018loading} is an \texttt{RDF} triple database offering the \gqlSPARQL\ query language. 
As a commercial product, it was designed while considering different architectures of proprietary systems (containing the \texttt{Cray\- A\-ries\- in\-ter\-con\-nect}~\citep{alverson2012cray}) of the company behind \texttt{CGE}.
It offers APIs for \langJava, \langPython\ and \texttt{Spark}, having a number of pre-built graph algorithms.
It is not open-source and its back-end relies on internal queries written in \texttt{C++} to work with a global address space using multiple processes, across multiple compute nodes, to share data and synchronize operations.
This product being both proprietary and reliant on custom hardware has the consequence of not being so widespread.
However, its results and special-purpose architecture make it a competitive platform which harnessed innovation in design as a graph database.

\paragraph{\textbf{Fluree}}
%License:Open source
\Chart{0.73}{Product}\hspace{5mm}
\Chart{0.43}{Database}\hspace{5mm}
\Chart{0.54}{Data}\\
\dbFluree~\citep{fluree} is an open-source (\licenseEclipseTwo)~\citep{fluree_github} immutable (i.e., append-only) 
graph database with a cloud-native architecture, supporting the \texttt{JSON} linked data (\texttt{JSON-LD}) 
format\footnote{\texttt{JSON-LD} is a standard for providing data context and interoperability.}. 
It targets knowledge graphs and may be run in containers, embedded in applications (\langClojure\ and 
\texttt{No\-de}) or as a stand-alone JVM service. 
\texttt{Flu\-ree} can run both as a local instance or as a cloud-managed service.
For graph querying it supports \texttt{SQL}, \gqlSPARQL, \texttt{Gra\-ph\-QL} and the custom 
\texttt{Flu\-ree\-QL}.
Data access control policies are expressed in \texttt{JSON-LD} as well.
The documentation states that to begin storing data, a ledger (in the sense of blockchain) must be 
created in \texttt{Flu\-ree} (analogous to the creation of a database within traditional database management 
systems). 
It records data operations and the transactions/queries will run against it and a ledger can 
enable the operation of more than one application.

\paragraph{\textbf{Ontotext GraphDB}}
%License:Proprietary
\Chart{0.82}{Product}\hspace{5mm}
\Chart{0.64}{Database}\hspace{5mm}
\Chart{0.66}{Data}\\
\dbOntotextGraphDB~\citep{ontotext_graphdb} is a graph database focused on \texttt{RDF} data and offering ACID transaction properties.  
It comes in three editions: free which is used for smaller projects and for testing and is only able to execute at most two concurrent queries; standard which can load and query statements at scale; enterprise edition which offers horizontal scalability and other features.
It supports \gqlSPARQL\ and offers a \langJava\ programming API. 
No source code for the free version is available.

\paragraph{\textbf{TerminusDB}}
%License:Open source
\Chart{0.81}{Product}\hspace{5mm}
\Chart{0.32}{Database}\hspace{5mm}
\Chart{0.34}{Data}\\
\dbTerminusDB\ is an open-source graph database 
and document store~\citep{van2020succinct}, described also as a 
toolkit for building collaborative applications.
It offers a document API to build knowledge graphs from 
\texttt{JSON} documents.
It is available under \licenseApacheTwo, offering bindings 
for \texttt{JavaScript} and \langPython~\citep{terminusdb_github} and 
employing delta encoding schemes similar to those of 
\texttt{git}\footnote{See: \url{https://git-scm.com/}}. 
It debuts its own \texttt{Web Object Query Language} (\texttt{WOQL}) and 
relies on an internal data structure called \texttt{\textit{ter\-mi\-nus\--sto\-re}} written 
in \texttt{Rust}. 
It is focused on the OWL model (not property graph), extending \texttt{RDF} with 
classes, restrictions and strongly-typed properties. 
\dbTerminusDB\ represents a graph as a series of layers, similarly to \texttt{git}.
Vertices and edges are enriched with classes and restrictions to model structure of data 
to enable finer granularity in how changes are expressed. 
Rather than different sequential versions of data, the storage is organized as base data and modifications that are 
stored as new layers over a base version of the graph.
%
%Additionally, the authors note that while there has been an explosion of database designs including graph, key-value stores, document and multi-model offerings, most of them are still heavily influenced by the philosophies of the relational database management systems of the 1970s~\citep{codd1970relational}.

\subsubsection{Multiple Data Models}\label{chapter:publications:survey:sec:graph-databases:dbs-and-systems:multiple_models}

The following graph databases support at least both the property graph model and/or \texttt{RDF} explicitly plus other data models (e.g., at least any of: document collections, relational model, object model):

\paragraph{\textbf{AgensGraph}}
%License:Mixed
\Chart{0.67}{Product}\hspace{5mm}
\Chart{0.39}{Database}\hspace{5mm}
\Chart{0.71}{Data}\\
\dbAgensGraph~\citep{agensgraph} is a multi-model database that supports  
graph manipulation (property graph model with \texttt{JSON} to represent properties) using 
\gqlCypher\ and \texttt{SQL} in a hybrid query processing model. 
It is based on the \texttt{Post\-gre\-SQL} relational database management system, which 
offers horizontal and vertical scalability.
It includes ACID transactions, multi-version concurrency control, stored procedures, triggers, 
constraints and monitoring.\footnote{See: \url{https://www.skaiworldwide.com/resource}}
There is an enterprise edition which offers monitoring, high availability and memory optimizations. 
A community edition is also available~\citep{agensgraph_github} under the \licenseApacheTwo\ (mostly implemented in \texttt{C}). 
SKAI Worldwide Co., Ltd. (formerly Bitnine Global Inc.) is the company behind \texttt{A\-gens\-Gra\-ph}.%, also offers a cloud solution called \texttt{AGCloud}.\footnote{See: \url{https://agcloud.bitnine.net/}}
Visualization capability is offered through the complementary product \texttt{A\-gens\-Brow\-ser}.\footnote{See: \url{https://worldwide-homepage-bucket.s3.ap-northeast-2.amazonaws.com/manual/agenbrowser+web+manual+pdf_quick+guide_(EN).pdf}}
\dbAgensGraph\ is based on \texttt{Post\-gre\-SQL\- RDB\-MS}.\footnote{See: \url{https://www.postgresql.org/about/news/announcing-the-release-of-agensgraph-v2160-3149/}}

\paragraph{\textbf{Altair Graph Lakehouse}}
%License:Proprietary
\Chart{0.50}{Product}\hspace{5mm}
\Chart{0.61}{Database}\hspace{5mm}
\Chart{0.58}{Data}\\
\dbAltairGraphLakehouse~\citep{anzodb} is a proprietary database built to enable \texttt{RDF} with \gqlSPARQL\ and the property graph with \gqlCypher\ queries to analyse big graphs (trillions of relationships). It has \langJava\ and \texttt{C++} APIs to create functions, aggregates and services.
It supports ACID transactions and also supports an \texttt{RDF+} inference engine following W3C standards and uses compressed in-memory and on disk storage of data.
This database is described as beyond a transaction-oriented database and as a \texttt{Gra\-ph\- On\-li\-ne\- Ana\-ly\-tics\- Pro\-ce\-ssing\- (GO\-LAP)} database, enabling interactive view, analysis and update of graph data, in a way similar to the interactive capabilities of the \dbNeoFRj\ database.
It comes as a single-machine (and memory usage limitations) free edition and an enterprise edition which supports unlimited cluster size.
It supports third-party visualization tools.
We did not find details on its internal data structures nor source code online.
\dbAltairGraphLakehouse\footnote{\url{https://2025.help.altair.com/2025.0/graphlakehouse/userdoc/Home.htm}} was rebranded from \dbAnzoGraphDB\footnote{See: \url{https://hub.docker.com/r/cambridgesemantics/anzograph-db}} (Cambridge Semantics was acquired by Altair), which was previously \texttt{SPAR\-QL\-Ver\-se}.
The company behind this product, Altair Engineering Inc., has been acquired by Siemens AG.\footnote{See: \url{https://press.siemens.com/global/en/pressrelease/siemens-acquires-altair-create-most-complete-ai-powered-portfolio-industrial-software}}

\paragraph{\textbf{Amazon Neptune}}
%License:Proprietary
\Chart{0.62}{Product}\hspace{5mm}
\Chart{0.64}{Database}\hspace{5mm}
\Chart{0.82}{Data}\\
\dbAmazonNeptune~\citep{bebee2018amazon} is a managed proprietary service (freeing the user from having to focus on management tasks, provisioning, patching, etc.) that is ACID-compliant and focused on highly-connected datasets. Among its use cases there are recommendation engines, fraud detection and drug discovery, among others.
Its implementation language and internal graph representation have not been disclosed and it supports both the property graph and \texttt{RDF} models, offering \gqlGremlin\ and \gqlSPARQL\ to query them.
As a full-fledged commercial product, it also has many features related to backup, replicas, security and management tasks, using product features such as Amazon's \texttt{S3}, \texttt{EC2} and \texttt{Cloud\-Wa\-tch} to offer scalability.
Usage samples are available online~\citep{neptune_github,neptune_github_updated}.

\paragraph{\textbf{ArangoDB}}
%License:Mixed
\Chart{0.86}{Product}\hspace{5mm}
\Chart{0.71}{Database}\hspace{5mm}
\Chart{0.63}{Data}\\
\dbArangoDB\ is an open-source~\citep{arangodb_github} multi-threaded database with support for graph storage (as well as key/value pairs and documents) that is available in both a proprietary license and the \licenseApacheTwo.
It is written in \langJavaScript\ from the browser to the back-end and all data is stored as \texttt{JSON} documents.
\dbArangoDB\ provides a storage engine for mostly in-memory operations and an alternative storage engine based on \texttt{Ro\-cks\-DB}, enabling datasets that are much bigger than RAM.
It guarantees ACID transactions for multi-document and multi-collection queries in a single instance and for single-document operations in cluster mode.
Replication and sharding are offered, allowing users to set up the database in a master-slave configuration or to spread bigger datasets across multiple servers.
It exposes a \texttt{Pre\-gel}-like API to express graph algorithms (implying access to the stored data in the database), has a custom \texttt{SQL}-like query language called \texttt{AQL (A\-ran\-go\-DB\- Que\-ry\- Lan\-gua\-ge)} and includes a built-in graph explorer.

\paragraph{\textbf{ArcadeDB}}
%License:Open source
\Chart{0.73}{Product}\hspace{5mm}
\Chart{0.57}{Database}\hspace{5mm}
\Chart{0.87}{Data}\\
\dbArcadeDB~\citep{arcadedb_github} is a multi-model database which shares the same 
model of \dbOrientDB\ and is licensed under \licenseApacheTwo. 
It comes with \texttt{Do\-cker} and \texttt{Ku\-ber\-ne\-tes} support and offers the \texttt{Ar\-ca\-de\-DB} 
\texttt{Stu\-dio} as a visual tool to facilitate query writing and graph exploration.
It can be used through \texttt{JVM}-compatible languages and also provides drivers (only a subset of 
operations are implemented); \texttt{Re\-dis} (transient entries in the server, persistent entries in the database) 
and \texttt{Mon\-go\-DB} (CRUD operations and queries). 
There is a high-availability module which enables replication, and when a server joins a cluster, an election 
process takes place (RAFT protocol). 
It supports \gqlGremlin\ and partially supports \gqlCypher\ through the \texttt{cy\-pher-for-grem\-lin} 
transpiler.
\texttt{Gra\-ph\-QL} is partially supported, enabling commands through \langJava\ and HTTP APIs and a 
\texttt{Post\-gres} driver.
The \texttt{Post\-gres} protocol may be used, enabling the execution of queries through different 
programming language drivers.

\paragraph{\textbf{BangDB}}
%License:Open source
\Chart{0.77}{Product}\hspace{5mm}
\Chart{0.46}{Database}\hspace{5mm}
\Chart{0.58}{Data}\\
\dbBangDB~\citep{bangdb_github} is a multi-model database under a \texttt{BSD} license. 
It supports representations such as key-value, documents, columns, large files, temporal and graphs.
Regarding graph storage, it supports \gqlCypher\ queries and provides data science libraries. 
A community version is offered, as well as priced enterprise versions (basic and pro) which offer 
a dashboard and an array of domain-specific applications.
Its cloud data platform is called \texttt{Am\-pe\-re}, which provides a set of composable workflows 
to enable a no-code approach to implementing functionality and offers a \texttt{REST} API.
\dbBangDB\ integrates its graph functionality with a streaming component, enabling the creation 
of graphs as data streams into the system.
It also offers CEP (complex event processing) on the data to find patterns/anomalies in continuous manner.

\paragraph{\textbf{IBM System G}}
%License:Proprietary
\Chart{0.31}{Product}\hspace{5mm}
\Chart{0.18}{Database}\hspace{5mm}
\Chart{0.32}{Data}\\
\dbIBMSystemG~\citep{ibm_suite_g} is a suite of functionalities, able to support the property graph (with \gqlGremlin) as well as \texttt{RDF} (though we did not find comments on \texttt{RDF}-specific query languages).
It is comprised of proprietary components as well as open-source (\texttt{Tin\-ker\-Pop} and \texttt{HBa\-se}) and comes with visual query capabilities, providing visual feedback into query building and result analysis to ease the debugging process.
ACID transactions are supported and the graph is represented in its native store with a data structure similar to compressed sparse vectors, using offsets to delimit, for each graph element, the latest and earliest temporal information of the element. 
Its \texttt{Que\-ry\- Ma\-na\-ger} is responsible for the communication with the distributed shards to enable query execution over the distributed graph.
It offers \texttt{g\-She\-ll}, a \texttt{ba\-sh}-like client to manage graph stores. 
The \texttt{IBM\- Sys\-tem\- G} product has been discontinued\footnote{Now-invalid page: \url{http://systemg.research.ibm.com/download.html}}.

%%%%%%%  NEW DOMAGOJ %%%%%%%
\paragraph{\textbf{MillenniumDB}}
%License:Open source
\Chart{0.46}{Product}\hspace{5mm}
\Chart{0.11}{Database}\hspace{5mm}
\Chart{0.55}{Data}\\
\dbMillenniumDB\ in an open-source graph system~\citep{millenniumdb_github} supporting property graphs, \texttt{RDF} and multilayered graph format. 
The system is actively being developed at IMFD Chile, a research institute focusing on fundamental data problems. \dbMillenniumDB\ comes equipped with three query languages: \texttt{GQL} for property graph querying, \texttt{SPAR\-QL} 1.1 for \texttt{RDF}, and its own proprietary \gqlCypher-like language for the extended model they propose. 
The system itself uses classical relational architecture and index-organized storage with four permutations of edge data being stored. 
One novel aspect of the system is that it deploys worst-case optimal~\citep{wco} join algorithms whenever possible, showing some perfomance benefits in join-intensive queries~\citep{VrgocRAAABHNRR23}. 
The system provides several access APIs, a console client and drivers for \langPython\ and \langJavaScript. It is under the \licenseGPLTwo.

\paragraph{\textbf{Oracle Spatial and Graph}}
%License:Proprietary
\Chart{0.62}{Product}\hspace{3mm}
\Chart{0.93}{Database}\hspace{3mm}
\Chart{0.97}{Data}\\
% https://docs.oracle.com/en/database/oracle/oracle-database/18/spgdg/using-property-graphs-oracle-database.html#GUID-85A4E0F7-0C3C-411D-912A-FD4744FADE44
\dbOracleSpatialAndGraph~\citep{oracledb_graph_features} is a long-standing commercial product which has spatial and graph capabilities, among which the property graph model using \texttt{PGQL} and the \texttt{RDF} model with \gqlSPARQL. 
It also supports a feature-rich studio with notebook interpreters, shell user-interface and graph visualization.
There are different \langJava\ APIs, one for the \texttt{O\-ra\-cle\- Spa\-tial\- and\- Gra\-ph\- Pro\-per\-ty\- Gra\-ph}, another for \texttt{Tin\-ker\-Pop\- Blue\-prin\-ts} and \texttt{Da\-ta\-ba\-se\- Pro\-per\-ty\- Gra\-ph}.

\paragraph{\textbf{OrientDB}}
%License:Mixed
\Chart{0.81}{Product}\hspace{5mm}
\Chart{0.50}{Database}\hspace{5mm}
\Chart{0.84}{Data}\\
\dbOrientDB~\citep{DBLP:conf/bncod/0001DLT21} is a distributed (property model) graph database that supports \frameworkTinkerPop\ and functions both as a graph (native) database and \texttt{No\-SQL} document database as well, with a Community Edition licensed~\citep{orientdb_github} under \licenseApacheTwo\ and a commercial edition.
There are drivers supporting \dbOrientDB\ at least in the following languages: \langClojure, \texttt{Go}, \langJava, \langJavaScript, \texttt{.NET}, \texttt{No\-de.js}, \texttt{PHP}, \langPython, \texttt{R}, \langRuby\ and \langScala.
It supports sharding (horizontal scaling), has ACID support and offers an adapted \texttt{SQL} for querying.

% https://docs.stardog.com/
\paragraph{\textbf{Stardog}}
%License:Proprietary
\Chart{0.62}{Product}\hspace{5mm}
\Chart{0.54}{Database}\hspace{5mm}
\Chart{0.74}{Data}\\
\dbStardog~\citep{stardog} is a proprietary (with a limited-time free trial version and a paid Enterprise license) knowledge graph database with a graph model based on \texttt{RDF} and extensions to support the property graph model.
It is horizontally-scalable, supports ACID operations, \texttt{Gra\-ph\-QL}, \gqlGremlin\ and \gqlSPARQL\ for querying and introspection and may be programmed in \langClojure, \texttt{Groovy}, \langJava, \langJavaScript, \texttt{.NET} and \texttt{Spring}.
For exploration, it features \texttt{Star\-dog\- Stu\-dio}. 
We did not find its pricing scheme on its website. 
The commercial offering is based on cloud deployments  relying on the \texttt{Li\-nux} (Ubuntu) ecosystem.

\paragraph{\textbf{SurrealDB}}
%License:Mixed
\Chart{0.85}{Product}\hspace{5mm}
\Chart{0.65}{Database}\hspace{5mm}
\Chart{0.79}{Data}\\
\dbSurrealDB~\citep{surrealdb_github} was built in 
\texttt{Rust} as a single library.
% Snapshot isolation in SurrealDB: https://surrealdb.com/blog/vart-a-persistent-data-structure-for-snapshot-isolation#understanding-isolation-levels-in-acid
It is a multi-model database to be used as both an embedded 
database library and as a database server which may be deployed 
in a distributed cluster. 
It uses tables with the added functionality of advanced 
nested fields and arrays, using inter-document record 
links to enable high-performance related queries 
without the overhead of multiple join operations. 
Its authors claim this storage design choice enables 
full graph database functionality using simple \texttt{SQL} 
extensions, with vertices represented as records.
A \texttt{Do\-cker} image is offered, with developers having 
access to different \texttt{Ja\-va\-Scri\-pt} client-side app 
tutorials as well as server-side language bindings (e.g. \texttt{Go\-lang} 
and \texttt{Rust}). 
It supports ACID transactions, full-text indexing, graph querying, 
\texttt{GraphQL} and both structured and unstructured data.
\texttt{Sur\-re\-al\-DB} offers row-level permissions-based access 
control. 
Versioned data features are in development\footnote{See: \url{https://surrealdb.com/features}}.
It is licensed under a custom \texttt{Bu\-si\-ne\-ss} \texttt{Sour\-ce} 
\texttt{License\- 1.1}, claiming that the \textit{Licensed Work} will 
be made available under the \licenseApacheTwo\ in April 2027.

% https://docs.openlinksw.com/virtuoso/
\paragraph{\textbf{Virtuoso}}
%License:Mixed
\Chart{0.83}{Product}\hspace{5mm}
\Chart{0.39}{Database}\hspace{5mm}
\Chart{0.79}{Data}\\
\dbVirtuoso~\citep{erling2012virtuoso} is a multi-model database management system that supports relational as well as property graphs. It has an open-source~\citep{virtuoso_github} edition under the \texttt{GPLv2} license as well as a proprietary enterprise edition.
At least the following programming languages are supported: \texttt{C/C++}, \texttt{C\#}, \langJava, \langJavaScript, \texttt{.NET}, \texttt{PHP}, \langPython, \langRuby\ and \texttt{Vi\-su\-al\- Ba\-sic}.
It supports horizontal scaling, has functionalities for interactive data exploration and supports \gqlSPARQL.

\subsubsection{Alternative Data Models}\label{chapter:publications:survey:sec:graph-databases:dbs-and-systems:alternative_models}

We lastly note the following databases that have been used to represent graphs, though not having an explicit description of supporting the property graph model or \texttt{RDF}:

\paragraph{\textbf{Azure Cosmos DB}}
%License:Proprietary
\Chart{0.62}{Product}\hspace{5mm}
\Chart{0.75}{Database}\hspace{5mm}
\Chart{0.71}{Data}\\
\dbAzureCosmosDB~\citep{paz2018introduction} is a commercial database solution that is multi-model, globally-distributed, schema-agnostic, horizontally-scalable and fully supports ACID.
It is classified as a \texttt{No\-SQL} database, but the multi-model API is a relevant offering, for it can expose stored data for example as table rows (\dbCassandra), collections (\texttt{Mon\-go\-DB}) and most importantly as graphs (\gqlGremlin).
It has connectors for \langJava, \texttt{.NET}, \langPython\ and \texttt{Xa\-ma\-rin}.
It is also a fully-managed service with scalability, freeing developer resources from topics such as data centre deployments, software upgrades and other operations.
An online repository of source code information (including SDKs, example applications, connectors and helper libraries) is available~\citep{cosmos_github}. 

\paragraph{\textbf{FaunaDB}}
%License:Open source
\Chart{0.50}{Product}\hspace{5mm}
\Chart{0.61}{Database}\hspace{5mm}
\Chart{0.53}{Data}\\
\dbFaunaDB~\citep{faunadb} is a distributed database platform targeting the modern cloud and container-centric environments. 
It has a custom \texttt{Fauna Query Language} (\texttt{FQL}) which operates on schema types such as documents, collections, indices, sets and databases.
This language can be accessed through drivers in languages such as \texttt{Android}, \texttt{C\#}, \texttt{Go}, \langJava, \langJavaScript, \langPython, \langRuby, \langScala\ and \texttt{Swi\-ft}.
\dbFaunaDB\ supports concurrency, ACID transactions and offers a RESTful HTTP API. 
As of March 2025, the Fauna Inc. services have been discontinued.\footnote{See: \url{https://fauna-livid.vercel.app/blog/the-future-of-fauna}}

\paragraph{\textbf{Google Cayley}}
%License:Open source
\Chart{0.62}{Product}\hspace{5mm}
\Chart{0.57}{Database}\hspace{5mm}
\Chart{0.42}{Data}\\
\dbGoogleCayley\ is an open-source~\citep{cayley_github} database behind Google's Knowledge Graph, having been tested at least since 2014 (and it is the spiritual successor to \texttt{graphd}~\citep{Meyer:2010:OST:1807167.1807283}).
It is a community-driven database written in \texttt{Go}, including a REPL, a RESTful API, a query editor and visualizer.
It supports \texttt{Giz\-mo} (query language inspired by \gqlGremlin) and \texttt{Gra\-ph\-QL}.
\texttt{Cay\-ley} supports multiple storage back-ends such as \texttt{LevelDB}, \texttt{Post\-gre\-SQL}, \texttt{Mon\-go\-DB} (distributed stores) and also an ephemeral in-memory storage.
The ability to support ACID transactions is delegated to the underlying storage back-end.
\texttt{Cay\-ley} being distributed depends on the underlying storage being distributed as well.
%Also in active development as of 2019.

\paragraph{\textbf{HyperGraphDB}}
%License:Open source
\Chart{0.38}{Product}\hspace{5mm}
\Chart{0.39}{Database}\hspace{5mm}
\Chart{0.38}{Data}\\
\dbHyperGraphDB~\citep{iordanov2010hypergraphdb} is as an open-source 
general purpose data storage mechanism.
It is used to store hypergraphs (a graph generalization where an edge can 
join any number of vertices). 
The low-level storage is based on \texttt{Ber\-ke\-ley\-DB} and is 
implemented in \langJava. 
Despite the description on the website mentioning distribution, the 
support for either distributed sharding or distributed replication is not 
provided in its current implementation~\citep{hypergraphdb_not_distributed} (only federating independent databases is possible). It has support for transactions with MVCC based conflict detection.  
%and we did not find mentions of ACID transaction support.

\paragraph{\textbf{ObjectivityDB}}
%License:Proprietary
\Chart{0.38}{Product}\hspace{5mm}
\Chart{0.32}{Database}\hspace{5mm}
\Chart{0.50}{Data}\\
% https://www.objectivity.com/products/objectivitydb/
\dbObjectivityDB~\citep{objectivitydb} is a database 
technology powering the massively scalable graph software platform 
\texttt{Thing\-Span} (formerly known as \texttt{In\-fi\-ni\-te\-Gra\-ph}).  
It is a fully-distributed database (able to scale horizontally) offering 
APIs in \texttt{C++}, \texttt{C\#}, \langJava\ and \langPython.
\dbObjectivityDB\ is described as a distributed 
object database, supporting many data models (among which highly complex 
and inter-related data).
%It is schema-based and uses a proprietary query language called \texttt{DO}. 
It is schema-based and licensing is defined on a case-by-case basis.
At the time of writing, the websites of the company are not accessible and their domain appears to have expired.\footnote{See: \url{http://www.objectivity.com/}} 
%Both the \dbObjectivityDB\footnote{See: \url{http://www.objectivity.com/}}  and \texttt{In\-fi\-ni\-te\-Gra\-ph}\footnote{See: \url{https://infinitegraph.com/}}  resolve to a web hosting entry page.

\paragraph{\textbf{TypeDB}}
%License:Open source
\Chart{0.92}{Product}\hspace{5mm}
\Chart{0.46}{Database}\hspace{5mm}
\Chart{0.63}{Data}\\
% On TypeDB access control: https://typedb.com/docs/typedb/managing/security
\dbTypeDB~\citep{typedb_github} is designed to 
be a database capable of natively expressing inheritance hierarchies 
and dependencies. 
This database is offered under \licenseAffero, with its 
own query language \texttt{Ty\-pe\-QL} and is written in 
\texttt{Ja\-va}.
It aims to implement conceptual data models without forcing 
them into predefined structures, which the authors claim to be a
direct focus not supported in other databases.
%The authors of \dbTypeDB\ note the limitations of  relational databases the continuous maintenance  and integrity control in the application layer when implementing  class-table inheritance and normalizing multi-valued attributes in relational databases.
%They further point out limitations of document databases in terms  of insertion/retrieval of highly-interconnected data, as  similarly-shaped data is assigned to the same collection without a schema, delegating integrity and structure maintenance to the developer.
\dbTypeDB\ also claims to focus on the challenges 
of polymorphism handling in graph databases due to a lack of 
a proper schema, leading to type hierarchies being implemented 
as a structure-less data labels.
This database offers a cloud-agnostic interface to manage 
instances from providers such as Amazon, GCP or Azure. 
It also enables the use of clusters with multiple database 
instances to improve scalability. 
Access requires authentication and data encryption is supported. 
%as well as user-defined policies for password strength and rotation.
%At the time of writing, a waiting list is offered on the \dbTypeDB\ website and we have found no mention of pricing.
Three versions are offered: \texttt{Ty\-pe\-DB\- Co\-re} which is 
open-source and available via \texttt{Do\-cker} container and 
builds for \texttt{Win\-dows}, \texttt{Li\-nux} and \texttt{mac\-OS}; 
\texttt{Ty\-pe\-DB\- En\-ter\-pri\-se} which extends the former 
with security and high-availability features for running in 
production (as well as authentication); \texttt{Ty\-pe\-DB\- Cl\-oud} 
which is built on \texttt{Ku\-ber\-ne\-tes} and enables easy 
deployment across different teams and projects in different 
cloud providers.\\

%We have found other graph databases (\textcolor{red}{TODO: place these in appropriate sections above!})

%\vspace{1cm}

Other notable mentions include the \texttt{OQGRAPH}~\citep{mariadb_oqgraph} graph storage engine of \texttt{Ma\-ria\-DB}, developed to handle hierarchies and vertices with many connections and intended for retrieving hierarchical information such as graphs, routes and social relationships in \texttt{SQL}. 
We also note the following resources for further deepening of the aspects involved in the evolution of the study of graph databases~\citep{jin2010gblender,buerli2012current,shimpi2012overview,kolomivcenko2013experimental,Vaikuntam2014Evaluation,de2014model,henderson2014system,robinson2015graph}, covering aspects such as the internal graph representations, experimental comparisons, principles for querying and extracting information from the graph and designing graph databases to make use of distributed infrastructures.

%\section{Data Representation and Indexing}~\label{chapter:publications:survey:sec:data_rep_and_indexing}

\subsection{Analysis and Discussion}~\label{chapter:publications:survey:sec:analysis}
Here, we draw the main findings about which features appear to be considered more and less relevant by the community and industry, and to what extent they are provided (or supported), either fully, or at least partially to some extent.

We perform this by looking into how each feature is addressed by the analyzed systems and categorize them according to how wide their support is, or lack thereof, across systems, as shown in Tables~\ref{new:table:graph_database_features_summary_P1} and~\ref{new:table:graph_database_features_summary_P2}.

There are multiple angles of interest regarding the relationship between features and systems. We do not intend to be exhaustive  as the table can be read across features over different support levels individually and cumulatively. 
Herein we refer the features in groups that we find to have more statistical impact  w.r.t. assessing  their relevance in the analyzed systems and how they can influence choice decisions:

\begin{itemize}

%%% 100% Support in >=50% of systems
\item \textbf{Features supported by majority of systems at a relevant level:}. 

%%% 100\% fully supported in >=50\% of the systems

A fair number of features are fully supported by a majority of systems (indicated by \emph{full circle}). 
We address them according to dimension and class. 
The Product Dimension includes features related to adoption and deployment. 
In the Adoption class, active development is a key factor, with 60.78\% of graph database systems demonstrating full engagement. 
Commercial support is fully provided by 64.71\% of studied systems. 
Open source availability is observed in 56.86\% of systems. 
In the Deployment class, work as a dedicated instance is a feature fully supported by 92\% of systems. 
Operating on \texttt{Li\-nux} is widely available, with 88.24\% of systems providing full compatibility.  

The Database Dimension focuses on features that enhance the database experience. 
In the Convenience class, live backups are supported by 58\% of systems. 
In the Distribution class, data distribution is fully implemented in 68\% of the studied graph database systems. 
High availability is ensured by 66\% of the systems. 
Query distribution is fully supported by 60.78\% of systems. 
Replication support is a widely available feature, with 70\% of systems providing full implementation. 
In the Software Development class, logging and auditing capabilities are fully implemented in 74\% of systems. 
Documentation up-to-date is a crucial software development feature, with 70.59\% of systems ensuring full maintenance.  

The Data Dimension addresses features related to data access, consistency, control, and security. 
In the Access class, CLI support is fully available in 72\% of systems. 
GUI availability is a relatively widely adopted feature, with 60\% of systems providing complete support. Graph-native data storage is fully integrated into 66\% of systems. 
REST API functionality is available in 64\% of systems. 
Query language support is a key feature, with 70\% of systems providing full implementation. 
In the Consistency class, transaction support is usually considered a fundamental requirement, fully available in 84\% of systems. 
In the Control class, schema support if fully provided in 54\% of systems, while secondary indexes are widely supported, with full implementation in 74\% of systems. 
In the Security class, authentication is a critical security feature, with 72\% of systems ensuring full implementation. 
Authorization mechanisms are fully supported in 66\% of the studied systems.

% additionl features where the number systems with significant (75%) or full support reach >=50%.

If we consider significant support (\emph{three quarter-filled circle}) as relevant, there is another feature that reaches support from a majority of systems: Operating on \texttt{Win\-dows}. 

If we further consider partial support (\emph{half-filled circle}) as relevant for a decision, then additional features become included in those that are supported by a majority of systems, all being: Live community, Operating on \texttt{Win\-dows} 
Data types defined, Read committed transaction, and
Constraints.\\

\item \textbf{Features  at most only partially supported  in  a majority of  systems}:
We can also analyze the table by checking what are the features that are lacking widespread relevant support, i.e. those where a majority of systems offer no support (\emph{empty  circle}), only limited support (\emph{quarter-filled circle}) or  partial support (\emph{half-filled circles}). 
Conversely, we could regard these as the  features where significant  or full support is only offered  by a minority of systems.

Such features are those that may be  considered not essential to provide, or more difficult to implement by most of the community. 
We identified them as  those features that in a majority of the systems analyzed do not go beyond being partially supported. 

Considering the Product dimension, these include
Live community (66.67\%), Pricing (56.89\%), Trendiness (64.71\%), Work as embedded (66\%), Testing in-memory version (64.7\%), and SaaS offering (51.22\%), 

Considering the Database dimension, they include
Automatic updates (84\%), Client side caching (85.71\%), Data versioning support (78\%), Cluster Re-balancing (78\%), Data types defined (54\%), Object-Graph Mapper (58\%), Reactive programming (78\%), 

In the Data dimension, we identify Multi-database (64\%), Granular locking (74\%), Multiple isolation levels (76\%), Read committed transaction (56\%), Constraints (56\%), Server side procedures (61.22\%), Triggers (67.34\%), Data encryption (74\%).

If instead we consider a more restrictive  threshold of only having limited supported (\emph{quarter-filled circle}) to identify features of less interest from the systems analyzed, this set is reduced to:

Considering the Product dimension, they still include 
Pricing (55.17\%), Trendiness (50.98\%), Work as embedded (66.0\%), Testing in-memory version (60.78\%).

In the Database dimension, they include Automatic updates (84.0\%), Client side caching (81.63\%), Data versioning support (76.0\%), Cluster Re-balancing (74.0\%), Object-Graph Mapper (58.0\%), and programming (76.0\%). 

Considering the Data dimension, those are Reactive Multi-database (64.0\%), Granular locking (72.0\%), Multiple isolation levels (64.0\%), Read committed transaction (50.0\%), Server side procedures (57.14\%), Triggers (59.18\%), and Data encryption (72.0\%).

Finally, if we only consider features  without any type of support whatsoever in  a majority of systems, as an indicator of very low interest in, or high difficulty of, systems providing them, we still have the following set: Pricing (55.17\%), Work as embedded (66.0\%), Testing in-memory version (60.78\%) in the Product Dimension; Automatic updates (84.0\%), Client side caching (81.63\%), Data versioning support (76.0\%), Cluster Re-balancing (74.0\%), Object-Graph Mapper (58.0\%), Reactive programming (76.0\%) in the Database dimension; and Multi-database (64.0\%), Granular locking (72.0\%), Multiple isolation levels (64.0\%), Server side procedures (57.14\%), Triggers (59.18\%), Data encryption (72.0\%) in the Data Dimension.

%Finally, if we only consider features  without any type of support whatsoever in  a majority of systems, as a indicator of very low interest in, or high difficulty of, systems providing them, we still have the following set: Pricing (55.17\%), Work as embedded (66.0\%), Testing in-memory version (60.78\%), Automatic updates (84.0\%), Client side caching (81.63\%), Data versioning support (76.0\%), Cluster Re-balancing (74.0\%), Object-Graph Mapper (58.0\%), Reactive programming (76.0\%), Multi-database (64.0\%), Granular locking (72.0\%), Multiple isolation levels (64.0\%), Server side procedures (57.14\%), Triggers (59.18\%), Data encryption (72.0\%).

\end{itemize}

%%%%% Summary table P1.

\begin{table}[!t]
\centering
\caption{Summary of feature level of adoption across graph databases and systems (Part I).}
\label{new:table:graph_database_features_summary_P1}

\small
\setlength{\tabcolsep}{4pt}
\renewcommand{\arraystretch}{1.15}

\begin{adjustbox}{max width=\textwidth}
\begin{tabular}{@{}
  >{\raggedright\arraybackslash}p{4.2cm}
  S[table-format=2.2]@{}l
  S[table-format=2.2]@{}l
  S[table-format=2.2]@{}l
  S[table-format=2.2]@{}l
  S[table-format=2.2]@{}l
@{}}
\toprule
\multirow{2}{*}{Feature}
  & \multicolumn{2}{c}{\pie{0}}
  & \multicolumn{2}{c}{\pie{90}}
  & \multicolumn{2}{c}{\pie{180}}
  & \multicolumn{2}{c}{\pie{270}}
  & \multicolumn{2}{c}{\pie{360}}
\\[-2pt]
  & \multicolumn{2}{c}{\scriptsize None}
  & \multicolumn{2}{c}{\scriptsize Limited}
  & \multicolumn{2}{c}{\scriptsize Partial}
  & \multicolumn{2}{c}{\scriptsize Significant}
  & \multicolumn{2}{c}{\scriptsize Full}
\\
\midrule

% BEGIN inlined from tables/pretty/new-table-summary-P1.tex
Active development         & 45.1 & \% &  0 & \% &  5.88 & \% &  0.00 & \% & 49.02 & \% \\
Commercial support         & 35.29 & \% &  0.00 & \% &  0.00 & \% &  0.00 & \% & 64.71 & \% \\
Live community             & 41.18 & \% &  3.92 & \% & 21.57 & \% &  3.92 & \% & 29.41 & \% \\
Open source                & 39.22 & \% &  0.00 & \% &  3.92 & \% &  0.00 & \% & 56.86 & \% \\
Pricing                    & 55.17 & \% &  0.00 & \% &  1.72 & \% &  0.00 & \% & 31.03 & \% \\
Trendiness                 & 45.1 & \% &  0 & \% & 9.8 & \% &  0 & \% & 45.1 & \% \\
\addlinespace[2pt]
\midrule

Containerization           & 38.00 & \% &  0.00 & \% &  6.00 & \% &  0.00 & \% & 56.00 & \% \\
Work as dedicated instance &  8.00 & \% &  0.00 & \% &  0.00 & \% &  0.00 & \% & 92.00 & \% \\
Work as embedded           & 66.00 & \% &  0.00 & \% &  0.00 & \% &  0.00 & \% & 34.00 & \% \\
Testing in-memory version  & 60.78 & \% &  0.00 & \% &  3.92 & \% &  1.96 & \% & 31.37 & \% \\
Operating on Linux         &  5.88 & \% &  0.00 & \% &  1.96 & \% &  1.96 & \% & 88.24 & \% \\
Operating on Windows       & 28.00 & \% &  0.00 & \% & 12.00 & \% & 12.00 & \% & 48.00 & \% \\
SaaS offering              & 48.78 & \% &  0.00 & \% &  2.44 & \% &  0.00 & \% & 46.34 & \% \\
\addlinespace[2pt]
\midrule

Automatic updates          & 84.00 & \% &  0.00 & \% &  0.00 & \% &  0.00 & \% & 16.00 & \% \\
Client side caching        & 81.63 & \% &  0.00 & \% &  4.08 & \% &  0.00 & \% & 14.29 & \% \\
Data versioning support    & 76.00 & \% &  0.00 & \% &  2.00 & \% &  0.00 & \% & 22.00 & \% \\
Live backups               & 38.00 & \% &  0.00 & \% &  4.00 & \% &  0.00 & \% & 58.00 & \% \\
\addlinespace[2pt]
\midrule

Cluster re-balancing       & 74.00 & \% &  0.00 & \% &  4.00 & \% &  0.00 & \% & 22.00 & \% \\
Data distribution          & 28.00 & \% &  2.00 & \% &  0.00 & \% &  2.00 & \% & 68.00 & \% \\
High-availability          & 28.00 & \% &  0.00 & \% &  6.00 & \% &  0.00 & \% & 66.00 & \% \\
Query distribution         & 25.49 & \% &  1.96 & \% &  7.84 & \% &  1.96 & \% & 60.78 & \% \\
Replication support        & 26.00 & \% &  0.00 & \% &  4.00 & \% &  0.00 & \% & 70.00 & \% \\
\addlinespace[2pt]
\midrule

Data types defined         & 10.00 & \% &  0.00 & \% & 44.00 & \% &  0.00 & \% & 46.00 & \% \\
Logging/Auditing           & 16.00 & \% &  0.00 & \% & 10.00 & \% &  0.00 & \% & 74.00 & \% \\
Object-graph mapper        & 58.00 & \% &  0.00 & \% &  0.00 & \% &  0.00 & \% & 42.00 & \% \\
Reactive programming       & 76.00 & \% &  0.00 & \% &  2.00 & \% &  0.00 & \% & 22.00 & \% \\
Documentation up-to-date   & 13.73 & \% &  1.96 & \% &  7.84 & \% &  5.88 & \% & 70.59 & \%
% END inlined from tables/pretty/new-table-summary-P1.tex
 % body only

\tabularnewline
\bottomrule
\end{tabular}
\end{adjustbox}
\end{table}

%%%%% Summary table P2.

\begin{table}[!t]
\centering
\caption{Summary of feature level of adoption across graph databases and systems (Part II).}
\label{new:table:graph_database_features_summary_P2}

\small
\setlength{\tabcolsep}{4pt}
\renewcommand{\arraystretch}{1.15}

\begin{adjustbox}{max width=\textwidth}
\begin{tabular}{@{}
  >{\raggedright\arraybackslash}p{4.2cm}
  S[table-format=2.2]@{}l
  S[table-format=2.2]@{}l
  S[table-format=2.2]@{}l
  S[table-format=2.2]@{}l
  S[table-format=2.2]@{}l
@{}}
\toprule
\multirow{2}{*}{Feature}
  & \multicolumn{2}{c}{\pie{0}}
  & \multicolumn{2}{c}{\pie{90}}
  & \multicolumn{2}{c}{\pie{180}}
  & \multicolumn{2}{c}{\pie{270}}
  & \multicolumn{2}{c}{\pie{360}}
\\[-2pt]
  & \multicolumn{2}{c}{\scriptsize None}
  & \multicolumn{2}{c}{\scriptsize Limited}
  & \multicolumn{2}{c}{\scriptsize Partial}
  & \multicolumn{2}{c}{\scriptsize Significant}
  & \multicolumn{2}{c}{\scriptsize Full}
\\
\midrule

% BEGIN inlined from tables/pretty/new-table-summary-P2.tex
Binary protocol             & 46.00 & \% & 0.00 & \% &  0.00 & \% & 2.00 & \% & 52.00 & \% \\
CLI                         & 28.00 & \% & 0.00 & \% &  0.00 & \% & 0.00 & \% & 72.00 & \% \\
GUI                         & 38.00 & \% & 0.00 & \% &  2.00 & \% & 0.00 & \% & 60.00 & \% \\
Multi-database              & 64.00 & \% & 0.00 & \% &  0.00 & \% & 0.00 & \% & 36.00 & \% \\
Graph-native data           & 34.00 & \% & 0.00 & \% &  0.00 & \% & 0.00 & \% & 66.00 & \% \\
REST API                    & 34.00 & \% & 0.00 & \% &  2.00 & \% & 0.00 & \% & 64.00 & \% \\
Query language              & 14.00 & \% & 0.00 & \% & 14.00 & \% & 2.00 & \% & 70.00 & \% \\
\addlinespace[2pt]
\midrule

Granular locking             & 72.00 & \% & 0.00 & \% &  2.00 & \% & 0.00 & \% & 26.00 & \% \\
Multiple isolation levels     & 64.00 & \% & 0.00 & \% & 12.00 & \% & 0.00 & \% & 24.00 & \% \\
Read committed transaction    & 48.00 & \% & 2.00 & \% &  6.00 & \% & 0.00 & \% & 44.00 & \% \\
Transaction support           & 12.00 & \% & 0.00 & \% &  4.00 & \% & 0.00 & \% & 84.00 & \% \\
\addlinespace[2pt]
\midrule

Constraints                  & 44.00 & \% & 0.00 & \% & 12.00 & \% & 2.00 & \% & 42.00 & \% \\
Schema support               & 26.00 & \% & 0.00 & \% & 18.00 & \% & 2.00 & \% & 54.00 & \% \\
Secondary indexes            & 26.00 & \% & 0.00 & \% &  0.00 & \% & 0.00 & \% & 74.00 & \% \\
Server side procedures       & 57.14 & \% & 0.00 & \% &  4.08 & \% & 0.00 & \% & 38.78 & \% \\
Triggers                     & 59.18 & \% & 0.00 & \% &  8.16 & \% & 0.00 & \% & 32.65 & \% \\
\addlinespace[2pt]
\midrule

Authentication               & 26.00 & \% & 0.00 & \% &  2.00 & \% & 0.00 & \% & 72.00 & \% \\
Authorization                & 30.00 & \% & 0.00 & \% &  4.00 & \% & 0.00 & \% & 66.00 & \% \\
Data encryption              & 72.00 & \% & 0.00 & \% &  2.00 & \% & 0.00 & \% & 26.00 & \%
% END inlined from tables/pretty/new-table-summary-P2.tex
 % body only

\tabularnewline
\bottomrule
\end{tabular}
\end{adjustbox}
\end{table}

Figure~\ref{fig:triangle_of_dbs_and_features} presents a triangular visualization of database system distribution across the Product-Database-Data space.
Each database is projected onto the three evaluation dimensions (Product, Database, Data).
The left corner corresponds to the Product dimension, the right corner to the Database dimension and the top corner to the Data dimension.
Position shows the relative balance across dimensions; marker size reflects overall feature coverage; shape indicates the supported graph data model; color indicates licensing.
It can be observed that there is a clustering of systems around the center of the triangle. 
This is due to many databases having a relative balance that is similar among the feature categories.
There is a broadly balanced profile across the three dimensions.
Points closer to a triangle vertex represent systems whose feature support is concentrated in that dimension (e.g., stronger product maturity/operational support vs. data-model expressiveness), highlighting specialized design emphases.
Property graph systems exhibit a wider spread in the landscape, indicating higher heterogeneity in the balance between product maturity, operational database features, and data-model/query capabilities. 
In contrast, RDF systems cluster more tightly, suggesting a more uniform capability profile within this family.

\begin{figure}[t]
\centering
\begin{minipage}[c]{0.52\textwidth}
\centering
\begin{tikzpicture}[x=\linewidth,y=\linewidth]%,
  %spy using outlines={circle, magnification=3.0, size=2.4cm, connect spies}]

  % triangle vertices
  \coordinate (P) at (0,0);
  \coordinate (D) at (1,0);
  \coordinate (A) at (0.5,0.8660254);

  % (Optional but recommended) Force bounding box to triangle area only
  % so centering is visually consistent.
  \path[use as bounding box] (-0.02,-0.05) rectangle (1.02,0.90);

  % draw triangle
  \draw[very thick] (P) -- (D) -- (A) -- cycle;

  % corner labels
  \node[below, xshift=16pt]  at (P) {Product};
  \node[below, xshift=-18pt] at (D) {Database};
  \node[above]               at (A) {Data};

  % points
  % BEGIN inlined from figures/dbs_triangle_files/graph_databases_ternary_points.tex
% Auto-generated from ternary CSV. Do not edit by hand.
% Format:
% \ternaryplotpoint{<system_name>}{<x_i>}{<y_i>}{<r_i>}{<graph_model>}{<license>}

\ternaryplotpoint{AgensGraph}{0.420903954802}{0.347388721292}{2.307}{Multiple Data Models}{Mixed}
\ternaryplotpoint{Alibaba Graph Database / TuGraph}{0.529197080292}{0.335031725552}{2.033}{Property Graph Model}{Open source}
\ternaryplotpoint{AllegroGraph}{0.586283185841}{0.302725694243}{2.583}{RDF Data Model}{Proprietary}
\ternaryplotpoint{Altair Graph Lakehouse}{0.532544378698}{0.297215819050}{2.257}{Multiple Data Models}{Proprietary}
\ternaryplotpoint{Amazon Neptune}{0.504807692308}{0.341413861107}{2.487}{Multiple Data Models}{Proprietary}
\ternaryplotpoint{Apache HugeGraph}{0.531250000000}{0.295614440715}{2.487}{Property Graph Model}{Open source}
\ternaryplotpoint{ArangoDB}{0.465909090909}{0.247998183811}{2.552}{Multiple Data Models}{Mixed}
\ternaryplotpoint{ArcadeDB}{0.463133640553}{0.347208341609}{2.536}{Multiple Data Models}{Open source}
\ternaryplotpoint{Azure Cosmos DB}{0.531250000000}{0.295614440715}{2.487}{Alternative Data Models}{Proprietary}
\ternaryplotpoint{BangDB}{0.414364640884}{0.277510902870}{2.332}{Multiple Data Models}{Open source}
\ternaryplotpoint{Blazegraph}{0.506756756757}{0.339388333916}{2.115}{RDF Data Model}{Proprietary}
\ternaryplotpoint{BrightstarDB}{0.500000000000}{0.334863156130}{2.129}{RDF Data Model}{Open source}
\ternaryplotpoint{ByteGraph}{0.688888888889}{0.307920143568}{1.581}{Property Graph Model}{Proprietary}
\ternaryplotpoint{ChronoGraph}{0.433628318584}{0.291229781804}{1.830}{Property Graph Model}{Mixed}
\ternaryplotpoint{Cray Graph Engine}{0.603053435115}{0.383431094805}{1.986}{RDF Data Model}{Proprietary}
\ternaryplotpoint{DataStax Enterprise Graph (DSE)}{0.573964497041}{0.297215819050}{2.257}{Property Graph Model}{Mixed}
\ternaryplotpoint{Dgraph}{0.427860696517}{0.249897877709}{2.449}{Property Graph Model}{Mixed}
\ternaryplotpoint{FaunaDB}{0.533536585366}{0.279874063418}{2.225}{Alternative Data Models}{Open source}
\ternaryplotpoint{Fluree}{0.411764705882}{0.275090422379}{2.264}{RDF Data Model}{Open source}
\ternaryplotpoint{G-Tran}{0.429292929293}{0.183702358379}{1.688}{Property Graph Model}{Open source}
\ternaryplotpoint{Gaffer}{0.613861386139}{0.274384286348}{1.710}{Property Graph Model}{Open source}
\ternaryplotpoint{Galaxybase}{0.511834319527}{0.333086693763}{2.257}{Property Graph Model}{Proprietary}
\ternaryplotpoint{Google Cayley}{0.484472049689}{0.225919670552}{2.205}{Alternative Data Models}{Open source}
\ternaryplotpoint{Graphflow}{0.253521126761}{0.170765572577}{1.258}{Property Graph Model}{Open source}
\ternaryplotpoint{HGraphDB}{0.604895104895}{0.224076503077}{2.079}{Property Graph Model}{Open source}
\ternaryplotpoint{HyperGraphDB}{0.504347826087}{0.286164916033}{1.849}{Alternative Data Models}{Open source}
\ternaryplotpoint{IBM System G}{0.419753086420}{0.342133492853}{1.452}{Multiple Data Models}{Proprietary}
\ternaryplotpoint{JanusGraph}{0.408256880734}{0.293971926055}{2.541}{Property Graph Model}{Open source}
\ternaryplotpoint{KatanaGraph}{0.378571428571}{0.136089706309}{1.232}{Property Graph Model}{Proprietary}
\ternaryplotpoint{Kuzu}{0.391447368421}{0.347549668624}{2.143}{Property Graph Model}{Open source}
\ternaryplotpoint{LiveGraph}{0.296875000000}{0.216506350946}{0.910}{Property Graph Model}{Open source}
\ternaryplotpoint{Memgraph}{0.482926829268}{0.304165019866}{2.471}{Property Graph Model}{Proprietary}
\ternaryplotpoint{MillenniumDB}{0.343750000000}{0.425280332216}{1.821}{Multiple Data Models}{Open source}
\ternaryplotpoint{NebulaGraph}{0.487437185930}{0.287224505778}{2.437}{Property Graph Model}{Open source}
\ternaryplotpoint{Neo4j}{0.451171875000}{0.301079144284}{2.731}{Property Graph Model}{Mixed}
\ternaryplotpoint{ObjectivityDB}{0.475000000000}{0.360843918244}{1.894}{Alternative Data Models}{Proprietary}
\ternaryplotpoint{Ontotext GraphDB}{0.457547169811}{0.269611682310}{2.509}{RDF Data Model}{Proprietary}
\ternaryplotpoint{Oracle Spatial and Graph}{0.561507936508}{0.333351048282}{2.712}{Multiple Data Models}{Proprietary}
\ternaryplotpoint{OrientDB}{0.427906976744}{0.338354111246}{2.525}{Multiple Data Models}{Mixed}
\ternaryplotpoint{PandaDB}{0.554187192118}{0.379686014467}{2.460}{Property Graph Model}{Open source}
\ternaryplotpoint{RedisGraph}{0.409836065574}{0.274477996828}{2.344}{Property Graph Model}{Mixed}
\ternaryplotpoint{SAP Hana Graph}{0.582627118644}{0.326594326003}{2.634}{Property Graph Model}{Proprietary}
\ternaryplotpoint{Sparksee}{0.458715596330}{0.333697862009}{1.792}{Property Graph Model}{Proprietary}
\ternaryplotpoint{Stardog}{0.478947368421}{0.337294104632}{2.386}{Multiple Data Models}{Proprietary}
\ternaryplotpoint{StellarDB}{0.354166666667}{0.288675134595}{1.654}{Property Graph Model}{Proprietary}
\ternaryplotpoint{SurrealDB}{0.456331877729}{0.298759855454}{2.599}{Multiple Data Models}{Mixed}
\ternaryplotpoint{TAO}{0.741666666667}{0.322353900298}{2.326}{Property Graph Model}{Proprietary}
\ternaryplotpoint{TerminusDB}{0.333333333333}{0.200305195433}{2.108}{RDF Data Model}{Open source}
\ternaryplotpoint{TigerGraph}{0.424083769634}{0.299254851569}{2.391}{Property Graph Model}{Proprietary}
\ternaryplotpoint{TypeDB}{0.385572139303}{0.271440798201}{2.449}{Alternative Data Models}{Open source}
\ternaryplotpoint{Ultipa}{0.486559139785}{0.246770679573}{2.362}{Property Graph Model}{Proprietary}
\ternaryplotpoint{Virtuoso}{0.390547263682}{0.340378143776}{2.449}{Multiple Data Models}{Mixed}
\ternaryplotpoint{Weaver}{0.473333333333}{0.150111069989}{1.346}{Property Graph Model}{Open source}
\ternaryplotpoint{ZipG}{0.522727272727}{0.065607985135}{1.096}{Property Graph Model}{Proprietary}
% END inlined from figures/dbs_triangle_files/graph_databases_ternary_points.tex

  % BEGIN inlined from figures/dbs_triangle_files/graph_databases_ternary_centroids_manual.tex
% Auto-generated centroids. Do not edit by hand unless you are fine-tuning label positions.
% Format:
% \ternaryplotcentroidcallout{<label>}{<x>}{<y>}{<modelkey>}{<colorname>}{<label_x>}{<label_y>}

\ternaryplotcentroidcallout{Property Avg.}{0.482242276334}{0.269169333745}{PG}{TernPG}{0.08}{0.58}
\ternaryplotcentroidcallout{RDF Avg.}{0.485534083820}{0.300773654174}{RDF}{TernRDF}{0.66}{0.80}
\ternaryplotcentroidcallout{Multi Avg.}{0.452339069820}{0.328791301400}{Multi}{TernMulti}{0.16}{0.78}
\ternaryplotcentroidcallout{Alternative Avg.}{0.485696433408}{0.286642967861}{Alt}{TernAlt}{1.00}{0.06}
% END inlined from figures/dbs_triangle_files/graph_databases_ternary_centroids_manual.tex

  % BEGIN inlined from figures/dbs_triangle_files/graph_databases_ternary_outlier_labels_manual.tex
% Auto-generated outlier labels. Do not edit by hand.
% Format:
% \ternaryplotlabel{<system_name>}{<x_i>}{<y_i>}{<tikz opts>}

\ternaryplotlabel{Graphflow}{0.323521126761}{0.150765572577}{anchor=east,xshift=-2pt,yshift=-1pt}
\ternaryplotlabel{TAO}{0.6901666666667}{0.352353900298}{anchor=west,xshift=2pt,yshift=1pt}
\ternaryplotlabel{ZipG}{0.522727272727}{0.065607985135}{anchor=west,xshift=2pt,yshift=-1pt}
\ternaryplotlabel{LiveGraph}{0.316875000000}{0.246506350946}{anchor=east,xshift=-2pt,yshift=-1pt}
\ternaryplotlabel{MillenniumDB}{0.343750000000}{0.425280332216}{anchor=east,xshift=-2pt,yshift=1pt}
\ternaryplotlabel{KatanaGraph}{0.528571428571}{0.116089706309}{anchor=east,xshift=-2pt,yshift=-1pt}
\ternaryplotlabel{ByteGraph}{0.628888888889}{0.277920143568}{anchor=west,xshift=2pt,yshift=1pt}
\ternaryplotlabel{TerminusDB}{0.333333333333}{0.200305195433}{anchor=east,xshift=-2pt,yshift=-1pt}
\ternaryplotlabel{StellarDB}{0.354166666667}{0.288675134595}{anchor=east,xshift=-2pt,yshift=1pt}
\ternaryplotlabel{Weaver}{0.623333333333}{0.160111069989}{anchor=east,xshift=-2pt,yshift=-1pt}
% END inlined from figures/dbs_triangle_files/graph_databases_ternary_outlier_labels_manual.tex

  % magnify the central cluster (triangle centroid is roughly (0.50, 0.2887))
  %\spy on (0.50,0.29) in node[draw=black!25, fill=white] at (0.75,0.75);

\end{tikzpicture}
\end{minipage}\hspace{-35pt}
\begin{minipage}[c]{0.50\textwidth}
\centering
\begin{tikzpicture}[x=5.2cm,y=5.2cm] % adjust width scaling here
  % Three horizontal strips at y=0.80, 0.50, 0.20
  \stripaxis{0.80}{Product}{1.0}
  \stripaxis{0.50}{Database}{1.0}
  \stripaxis{0.20}{Data}{1.0}

  % Points (generated by Python; each file places points around the given y level)
  % BEGIN inlined from figures/dbs_triangle_files/graph_databases_strip_product.tex
% Auto-generated strip plot points. Do not edit by hand.
% Format:
% \stripplotpoint{<value>}{<y>}{<graph_model>}{<license>}{<radius>}

\stripplotpoint{0.670}{0.8220}{Multiple Data Models}{Mixed}{1.731}
\stripplotpoint{0.380}{0.8117}{Property Graph Model}{Open source}{1.525}
\stripplotpoint{0.540}{0.7863}{RDF Data Model}{Proprietary}{1.937}
\stripplotpoint{0.500}{0.8102}{Multiple Data Models}{Proprietary}{1.693}
\stripplotpoint{0.620}{0.7984}{Multiple Data Models}{Proprietary}{1.866}
\stripplotpoint{0.620}{0.8012}{Property Graph Model}{Open source}{1.866}
\stripplotpoint{0.860}{0.7863}{Multiple Data Models}{Mixed}{1.914}
\stripplotpoint{0.730}{0.8237}{Multiple Data Models}{Open source}{1.902}
\stripplotpoint{0.620}{0.8263}{Alternative Data Models}{Proprietary}{1.866}
\stripplotpoint{0.770}{0.8283}{Multiple Data Models}{Open source}{1.749}
\stripplotpoint{0.440}{0.8008}{RDF Data Model}{Proprietary}{1.586}
\stripplotpoint{0.460}{0.8188}{RDF Data Model}{Open source}{1.597}
\stripplotpoint{0.120}{0.7708}{Property Graph Model}{Proprietary}{1.185}
\stripplotpoint{0.450}{0.8118}{Property Graph Model}{Mixed}{1.373}
\stripplotpoint{0.230}{0.7774}{RDF Data Model}{Proprietary}{1.490}
\stripplotpoint{0.430}{0.8271}{Property Graph Model}{Mixed}{1.693}
\stripplotpoint{0.860}{0.7841}{Property Graph Model}{Mixed}{1.837}
\stripplotpoint{0.500}{0.7772}{Alternative Data Models}{Open source}{1.668}
\stripplotpoint{0.730}{0.8249}{RDF Data Model}{Open source}{1.698}
\stripplotpoint{0.460}{0.8268}{Property Graph Model}{Open source}{1.266}
\stripplotpoint{0.230}{0.8161}{Property Graph Model}{Open source}{1.282}
\stripplotpoint{0.500}{0.7964}{Property Graph Model}{Proprietary}{1.693}
\stripplotpoint{0.620}{0.8016}{Alternative Data Models}{Open source}{1.654}
\stripplotpoint{0.460}{0.7954}{Property Graph Model}{Open source}{0.944}
\stripplotpoint{0.380}{0.7920}{Property Graph Model}{Open source}{1.559}
\stripplotpoint{0.380}{0.7969}{Alternative Data Models}{Open source}{1.387}
\stripplotpoint{0.310}{0.7850}{Multiple Data Models}{Proprietary}{1.089}
\stripplotpoint{0.920}{0.7906}{Property Graph Model}{Open source}{1.906}
\stripplotpoint{0.380}{0.7719}{Property Graph Model}{Proprietary}{0.924}
\stripplotpoint{0.620}{0.7897}{Property Graph Model}{Open source}{1.607}
\stripplotpoint{0.370}{0.7881}{Property Graph Model}{Open source}{0.683}
\stripplotpoint{0.700}{0.8087}{Property Graph Model}{Proprietary}{1.853}
\stripplotpoint{0.460}{0.7883}{Multiple Data Models}{Open source}{1.366}
\stripplotpoint{0.690}{0.8247}{Property Graph Model}{Open source}{1.828}
\stripplotpoint{0.960}{0.7873}{Property Graph Model}{Mixed}{2.049}
\stripplotpoint{0.380}{0.8200}{Alternative Data Models}{Proprietary}{1.420}
\stripplotpoint{0.820}{0.7746}{RDF Data Model}{Proprietary}{1.882}
\stripplotpoint{0.620}{0.8272}{Multiple Data Models}{Proprietary}{2.034}
\stripplotpoint{0.810}{0.7977}{Multiple Data Models}{Mixed}{1.894}
\stripplotpoint{0.460}{0.8203}{Property Graph Model}{Open source}{1.845}
\stripplotpoint{0.790}{0.8159}{Property Graph Model}{Mixed}{1.758}
\stripplotpoint{0.540}{0.8053}{Property Graph Model}{Proprietary}{1.976}
\stripplotpoint{0.380}{0.7714}{Property Graph Model}{Proprietary}{1.344}
\stripplotpoint{0.620}{0.7730}{Multiple Data Models}{Proprietary}{1.789}
\stripplotpoint{0.460}{0.8001}{Property Graph Model}{Proprietary}{1.240}
\stripplotpoint{0.850}{0.8288}{Multiple Data Models}{Mixed}{1.949}
\stripplotpoint{0.130}{0.8035}{Property Graph Model}{Proprietary}{1.744}
\stripplotpoint{0.810}{0.8160}{RDF Data Model}{Open source}{1.581}
\stripplotpoint{0.770}{0.8000}{Property Graph Model}{Proprietary}{1.794}
\stripplotpoint{0.920}{0.7799}{Alternative Data Models}{Open source}{1.837}
\stripplotpoint{0.690}{0.8057}{Property Graph Model}{Proprietary}{1.772}
\stripplotpoint{0.830}{0.8230}{Multiple Data Models}{Mixed}{1.837}
\stripplotpoint{0.330}{0.8001}{Property Graph Model}{Open source}{1.010}
\stripplotpoint{0.290}{0.8181}{Property Graph Model}{Proprietary}{0.822}
% END inlined from figures/dbs_triangle_files/graph_databases_strip_product.tex

  % BEGIN inlined from figures/dbs_triangle_files/graph_databases_strip_database.tex
% Auto-generated strip plot points. Do not edit by hand.
% Format:
% \stripplotpoint{<value>}{<y>}{<graph_model>}{<license>}{<radius>}

\stripplotpoint{0.390}{0.5077}{Multiple Data Models}{Mixed}{1.731}
\stripplotpoint{0.460}{0.4865}{Property Graph Model}{Open source}{1.525}
\stripplotpoint{0.930}{0.4724}{RDF Data Model}{Proprietary}{1.937}
\stripplotpoint{0.610}{0.5182}{Multiple Data Models}{Proprietary}{1.693}
\stripplotpoint{0.640}{0.4911}{Multiple Data Models}{Proprietary}{1.866}
\stripplotpoint{0.750}{0.4819}{Property Graph Model}{Open source}{1.866}
\stripplotpoint{0.710}{0.4953}{Multiple Data Models}{Mixed}{1.914}
\stripplotpoint{0.570}{0.5003}{Multiple Data Models}{Open source}{1.902}
\stripplotpoint{0.750}{0.5131}{Alternative Data Models}{Proprietary}{1.866}
\stripplotpoint{0.460}{0.4953}{Multiple Data Models}{Open source}{1.749}
\stripplotpoint{0.460}{0.5262}{RDF Data Model}{Proprietary}{1.586}
\stripplotpoint{0.460}{0.4965}{RDF Data Model}{Open source}{1.597}
\stripplotpoint{0.460}{0.5124}{Property Graph Model}{Proprietary}{1.185}
\stripplotpoint{0.300}{0.4797}{Property Graph Model}{Mixed}{1.373}
\stripplotpoint{0.500}{0.4745}{RDF Data Model}{Proprietary}{1.490}
\stripplotpoint{0.680}{0.4825}{Property Graph Model}{Mixed}{1.693}
\stripplotpoint{0.570}{0.5076}{Property Graph Model}{Mixed}{1.837}
\stripplotpoint{0.610}{0.5177}{Alternative Data Models}{Open source}{1.668}
\stripplotpoint{0.430}{0.5156}{RDF Data Model}{Open source}{1.698}
\stripplotpoint{0.320}{0.4848}{Property Graph Model}{Open source}{1.266}
\stripplotpoint{0.460}{0.5225}{Property Graph Model}{Open source}{1.282}
\stripplotpoint{0.540}{0.4738}{Property Graph Model}{Proprietary}{1.693}
\stripplotpoint{0.570}{0.5089}{Alternative Data Models}{Open source}{1.654}
\stripplotpoint{0.110}{0.5247}{Property Graph Model}{Open source}{0.944}
\stripplotpoint{0.680}{0.5238}{Property Graph Model}{Open source}{1.559}
\stripplotpoint{0.390}{0.4809}{Alternative Data Models}{Open source}{1.387}
\stripplotpoint{0.180}{0.4923}{Multiple Data Models}{Proprietary}{1.089}
\stripplotpoint{0.520}{0.4790}{Property Graph Model}{Open source}{1.906}
\stripplotpoint{0.210}{0.5130}{Property Graph Model}{Proprietary}{0.924}
\stripplotpoint{0.290}{0.5291}{Property Graph Model}{Open source}{1.607}
\stripplotpoint{0.110}{0.5007}{Property Graph Model}{Open source}{0.683}
\stripplotpoint{0.630}{0.4750}{Property Graph Model}{Proprietary}{1.853}
\stripplotpoint{0.110}{0.4948}{Multiple Data Models}{Open source}{1.366}
\stripplotpoint{0.640}{0.5292}{Property Graph Model}{Open source}{1.828}
\stripplotpoint{0.710}{0.4757}{Property Graph Model}{Mixed}{2.049}
\stripplotpoint{0.320}{0.4759}{Alternative Data Models}{Proprietary}{1.420}
\stripplotpoint{0.640}{0.5216}{RDF Data Model}{Proprietary}{1.882}
\stripplotpoint{0.930}{0.5019}{Multiple Data Models}{Proprietary}{2.034}
\stripplotpoint{0.500}{0.4773}{Multiple Data Models}{Mixed}{1.894}
\stripplotpoint{0.680}{0.5120}{Property Graph Model}{Open source}{1.845}
\stripplotpoint{0.460}{0.5127}{Property Graph Model}{Mixed}{1.758}
\stripplotpoint{0.930}{0.5113}{Property Graph Model}{Proprietary}{1.976}
\stripplotpoint{0.290}{0.5229}{Property Graph Model}{Proprietary}{1.344}
\stripplotpoint{0.540}{0.5159}{Multiple Data Models}{Proprietary}{1.789}
\stripplotpoint{0.180}{0.5268}{Property Graph Model}{Proprietary}{1.240}
\stripplotpoint{0.650}{0.4712}{Multiple Data Models}{Mixed}{1.949}
\stripplotpoint{1.000}{0.4971}{Property Graph Model}{Proprietary}{1.744}
\stripplotpoint{0.320}{0.5081}{RDF Data Model}{Open source}{1.581}
\stripplotpoint{0.480}{0.4877}{Property Graph Model}{Proprietary}{1.794}
\stripplotpoint{0.460}{0.4918}{Alternative Data Models}{Open source}{1.837}
\stripplotpoint{0.640}{0.5080}{Property Graph Model}{Proprietary}{1.772}
\stripplotpoint{0.390}{0.4757}{Multiple Data Models}{Mixed}{1.837}
\stripplotpoint{0.290}{0.4728}{Property Graph Model}{Open source}{1.010}
\stripplotpoint{0.320}{0.5282}{Property Graph Model}{Proprietary}{0.822}
% END inlined from figures/dbs_triangle_files/graph_databases_strip_database.tex

  % BEGIN inlined from figures/dbs_triangle_files/graph_databases_strip_data.tex
% Auto-generated strip plot points. Do not edit by hand.
% Format:
% \stripplotpoint{<value>}{<y>}{<graph_model>}{<license>}{<radius>}

\stripplotpoint{0.710}{0.1911}{Multiple Data Models}{Mixed}{1.731}
\stripplotpoint{0.530}{0.2213}{Property Graph Model}{Open source}{1.525}
\stripplotpoint{0.790}{0.1802}{RDF Data Model}{Proprietary}{1.937}
\stripplotpoint{0.580}{0.1703}{Multiple Data Models}{Proprietary}{1.693}
\stripplotpoint{0.820}{0.1765}{Multiple Data Models}{Proprietary}{1.866}
\stripplotpoint{0.710}{0.2030}{Property Graph Model}{Open source}{1.866}
\stripplotpoint{0.630}{0.2204}{Multiple Data Models}{Mixed}{1.914}
\stripplotpoint{0.870}{0.2160}{Multiple Data Models}{Open source}{1.902}
\stripplotpoint{0.710}{0.1908}{Alternative Data Models}{Proprietary}{1.866}
\stripplotpoint{0.580}{0.1860}{Multiple Data Models}{Open source}{1.749}
\stripplotpoint{0.580}{0.2211}{RDF Data Model}{Proprietary}{1.586}
\stripplotpoint{0.580}{0.1956}{RDF Data Model}{Open source}{1.597}
\stripplotpoint{0.320}{0.1708}{Property Graph Model}{Proprietary}{1.185}
\stripplotpoint{0.380}{0.1715}{Property Graph Model}{Mixed}{1.373}
\stripplotpoint{0.580}{0.2143}{RDF Data Model}{Proprietary}{1.490}
\stripplotpoint{0.580}{0.2198}{Property Graph Model}{Mixed}{1.693}
\stripplotpoint{0.580}{0.2082}{Property Graph Model}{Mixed}{1.837}
\stripplotpoint{0.530}{0.2290}{Alternative Data Models}{Open source}{1.668}
\stripplotpoint{0.540}{0.1926}{RDF Data Model}{Open source}{1.698}
\stripplotpoint{0.210}{0.2289}{Property Graph Model}{Open source}{1.266}
\stripplotpoint{0.320}{0.2265}{Property Graph Model}{Open source}{1.282}
\stripplotpoint{0.650}{0.1961}{Property Graph Model}{Proprietary}{1.693}
\stripplotpoint{0.420}{0.1749}{Alternative Data Models}{Open source}{1.654}
\stripplotpoint{0.140}{0.2257}{Property Graph Model}{Open source}{0.944}
\stripplotpoint{0.370}{0.1944}{Property Graph Model}{Open source}{1.559}
\stripplotpoint{0.380}{0.2095}{Alternative Data Models}{Open source}{1.387}
\stripplotpoint{0.320}{0.1977}{Multiple Data Models}{Proprietary}{1.089}
\stripplotpoint{0.740}{0.1951}{Property Graph Model}{Open source}{1.906}
\stripplotpoint{0.110}{0.2199}{Property Graph Model}{Proprietary}{0.924}
\stripplotpoint{0.610}{0.1706}{Property Graph Model}{Open source}{1.607}
\stripplotpoint{0.160}{0.2271}{Property Graph Model}{Open source}{0.683}
\stripplotpoint{0.720}{0.1921}{Property Graph Model}{Proprietary}{1.853}
\stripplotpoint{0.550}{0.2116}{Multiple Data Models}{Open source}{1.366}
\stripplotpoint{0.660}{0.1929}{Property Graph Model}{Open source}{1.828}
\stripplotpoint{0.890}{0.2057}{Property Graph Model}{Mixed}{2.049}
\stripplotpoint{0.500}{0.1784}{Alternative Data Models}{Proprietary}{1.420}
\stripplotpoint{0.660}{0.1998}{RDF Data Model}{Proprietary}{1.882}
\stripplotpoint{0.970}{0.1970}{Multiple Data Models}{Proprietary}{2.034}
\stripplotpoint{0.840}{0.2025}{Multiple Data Models}{Mixed}{1.894}
\stripplotpoint{0.890}{0.1811}{Property Graph Model}{Open source}{1.845}
\stripplotpoint{0.580}{0.2047}{Property Graph Model}{Mixed}{1.758}
\stripplotpoint{0.890}{0.1926}{Property Graph Model}{Proprietary}{1.976}
\stripplotpoint{0.420}{0.1800}{Property Graph Model}{Proprietary}{1.344}
\stripplotpoint{0.740}{0.2025}{Multiple Data Models}{Proprietary}{1.789}
\stripplotpoint{0.320}{0.2096}{Property Graph Model}{Proprietary}{1.240}
\stripplotpoint{0.790}{0.2252}{Multiple Data Models}{Mixed}{1.949}
\stripplotpoint{0.670}{0.1876}{Property Graph Model}{Proprietary}{1.744}
\stripplotpoint{0.340}{0.2291}{RDF Data Model}{Open source}{1.581}
\stripplotpoint{0.660}{0.1805}{Property Graph Model}{Proprietary}{1.794}
\stripplotpoint{0.630}{0.1945}{Alternative Data Models}{Open source}{1.837}
\stripplotpoint{0.530}{0.1989}{Property Graph Model}{Proprietary}{1.772}
\stripplotpoint{0.790}{0.2154}{Multiple Data Models}{Mixed}{1.837}
\stripplotpoint{0.130}{0.1995}{Property Graph Model}{Open source}{1.010}
\stripplotpoint{0.050}{0.2221}{Property Graph Model}{Proprietary}{0.822}
% END inlined from figures/dbs_triangle_files/graph_databases_strip_data.tex

\end{tikzpicture}
\end{minipage}

\vspace{2pt}
{\footnotesize%\scriptsize
\begin{tabular}{@{}l@{\quad}l@{\qquad}l@{\quad}l@{}}
  \multicolumn{2}{@{}l@{}}{\textbf{Model (shape)}} &
  \multicolumn{2}{l@{}}{\textbf{License (color)}}\\
  \legIconUpTri & Property Graph Model &
  \textcolor{LicOpen}{\rule{1.6ex}{1.6ex}} & Open\\
  \legIconDownTri & RDF Data Model &
  \textcolor{LicProp}{\rule{1.6ex}{1.6ex}} & Proprietary\\
  \legIconCircle & Multiple Data Models &
  \textcolor{LicMixed}{\rule{1.6ex}{1.6ex}} & Mixed\\
  \legIconSquare & Alternative Data Models & & \\
\end{tabular}
}

\caption{Database system distribution in the Product-Database-Data space.}
\label{fig:triangle_of_dbs_and_features}
\end{figure}

Figure~\ref{fig:xyz_and_pairwise} shows a three-dimensional scatter plot of the database systems across the three dimensions, as well as pairwise 2D projections for each dimension pair.
It can be observed that there is a greater diversity (in terms of graph model focus) of systems that score high on both Data and Product, with the lower-scoring ones being mostly focused on the property graph model (bottom left panel).
The other two dimension combinations show similar patterns, highlighting a wider distribution of property graph focus across the score ranges (both low and high values), with the systems focusing on other models presenting higher scores.

\begin{figure}[t]
\centering
\begin{tikzpicture}[x=1cm,y=1cm]

% ---------- Layout knobs ----------
\def\PW{6.6}     % panel width  (cm units because x=1cm)
\def\PH{4.9}     % panel height
\def\G{0.65}     % gap between panels
\def\showpanelframes{0} % set to 1 to draw panel borders for debugging

% Data file (generated by Python)
\def\XYZFILE{figures/dbs_triangle_files/graph_databases_xyz_points.tex}

% Marker size factor for these panels (use <1 to reduce clutter)
\def\MarkScale{0.85}

% ---------- Helpers ----------
\newcommand{\PanelFrame}{%
  \ifnum\showpanelframes=1
    \draw[black!20] (0,0) rectangle (\PW,\PH);
  \fi
}

% Axes for 2D panels (0..1 range)
\newcommand{\AxesTwoD}[2]{%
  % #1 x-label, #2 y-label
  \draw[black!60, line width=0.5pt] (0,0) -- (\PW,0);
  \draw[black!60, line width=0.5pt] (0,0) -- (0,\PH);

  %%%%% The code below removes the 0 label at the origin for the YY axis.
  % X ticks: keep 0, 0.5, 1
  \foreach \t/\lab in {0/0,0.5/0.5,1/1} {
    \draw[black!60, line width=0.5pt] (\t*\PW,0) -- (\t*\PW,0.12);
    \node[font=\scriptsize, text=black!70, below] at (\t*\PW,0) {\lab};
  }

  % Y ticks: omit the 0 label at origin
  \foreach \t/\lab in {0.5/0.5,1/1} {
    \draw[black!60, line width=0.5pt] (0,\t*\PH) -- (0.12,\t*\PH);
    \node[font=\scriptsize, text=black!70, left] at (0,\t*\PH) {\lab};
  }
  \node[font=\scriptsize\bfseries, text=black!80, below] at (0.5*\PW,-0.35) {#1};
  \node[font=\scriptsize\bfseries, text=black!80, rotate=90] at (-0.55,0.5*\PH) {#2};
}

% ---------- Panel 1: 3D (top-left) ----------
\begin{scope}[shift={(0,\PH+\G)}]
  \PanelFrame
  \clip (-0.5,-0.5) rectangle (\PW,\PH);

  % ---- Scale only the 3D drawing ----
  \begin{scope}[scale=0.91, shift={(0.2,0.1)}] % <-- tune these two numbers
    \begin{scope}[
      x={(5.0cm,0cm)},
      y={(0cm,4.2cm)},
      z={(1.9cm,1.4cm)}
    ]
      \coordinate (O) at (0,0,0);

      \draw[->, black!70, line width=0.6pt] (O) -- (1,0,0) node[font=\scriptsize, anchor=west] {Database};
      \draw[->, black!70, line width=0.6pt] (O) -- (0,1,0) node[font=\scriptsize, anchor=south] {Product};
      \draw[->, black!70, line width=0.6pt] (O) -- (0,0,1) node[font=\scriptsize, anchor=south west] {Data};

      % inside the 3D basis scope, after drawing axes
      \draw[black!10] (0,0,0) -- (1,0,0) -- (1,1,0) -- (0,1,0) -- cycle; % z=0 plane
      \draw[black!10] (0,0,0) -- (1,0,0) -- (1,0,1) -- (0,0,1) -- cycle; % y=0 plane

      % x-axis ticks (Database) - keep 0
      \foreach \t in {0,0.5,1} {
        \draw[black!45, line width=0.4pt] (\t,0,0) -- (\t,0.03,0);
        \node[font=\scriptsize, text=black!70, anchor=north] at (\t,0,0) {\t};
      }
      % y-axis ticks (Product) - omit 0
      \foreach \t in {0.5,1} {
        \draw[black!45, line width=0.4pt] (0,\t,0) -- (0.03,\t,0);
        \node[font=\scriptsize, text=black!70, anchor=east] at (0,\t,0) {\t};
      }
      % z-axis ticks (Data) - omit 0
      \foreach \t in {0.5,1} {
        \draw[black!45, line width=0.4pt] (0,0,\t) -- (0.03,0,\t);
        \node[font=\scriptsize, text=black!70, anchor=west] at (0,0,\t) {\t};
      }

      \begingroup
        \renewcommand{\xyzplotpoint}[7]{%
          \ternarySetColorFromLicense{#7}%
          \ternarySetShapeFromModel{#6}%
          \pgfmathsetmacro{\msnum}{2*(#5)*\MarkScale}%
          \pgfmathsetlengthmacro{\ms}{\msnum pt}%
          \node[\ternaryShapeStyle, minimum size=\ms, inner sep=0pt,
            draw=\ternaryColor!85!black, fill=\ternaryColor!18,
            line width=0.6pt, fill opacity=0.55, draw opacity=0.90
          ] at (#3,#2,#4) {};
          \draw[black!20] (#3,#2,#4) -- (#3,0,#4);
        }%
        \input{\XYZFILE}%
      \endgroup
    \end{scope}
  \end{scope}
  \node[font=\scriptsize\bfseries, text=black!80, anchor=west] at (0.0,\PH-0.2) {3D view (Product/Database/Data)};

\end{scope}

% ---------- Panel 2: Database (x) vs Product (y) (top-right) ----------
\begin{scope}[shift={(\PW+\G,\PH+\G)}]
  \PanelFrame
  \AxesTwoD{Database}{Product}

  \begingroup
  \renewcommand{\xyzplotpoint}[7]{%
    \ternarySetColorFromLicense{#7}%
    \ternarySetShapeFromModel{#6}%
    \pgfmathsetmacro{\xx}{#3*\PW}%
    \pgfmathsetmacro{\yy}{#2*\PH}%
    \pgfmathsetmacro{\msnum}{2*(#5)*\MarkScale}%
    \pgfmathsetlengthmacro{\ms}{\msnum pt}%
    \node[\ternaryShapeStyle, minimum size=\ms, inner sep=0pt,
      draw=\ternaryColor!85!black, fill=\ternaryColor!18,
      line width=0.6pt, fill opacity=0.55, draw opacity=0.90
    ] at (\xx,\yy) {};
  }%
  \input{\XYZFILE}
  \endgroup

  \node[font=\scriptsize\bfseries, text=black!80, anchor=west] at (0.0,\PH-0.2) {DB vs Product};
\end{scope}

% ---------- Panel 3: Data (x) vs Product (y) (bottom-left) ----------
\begin{scope}[shift={(0,0)}]
  \PanelFrame
  \AxesTwoD{Data}{Product}

  \begingroup
  \renewcommand{\xyzplotpoint}[7]{%
    \ternarySetColorFromLicense{#7}%
    \ternarySetShapeFromModel{#6}%
    \pgfmathsetmacro{\xx}{#4*\PW}%
    \pgfmathsetmacro{\yy}{#2*\PH}%
    \pgfmathsetmacro{\msnum}{2*(#5)*\MarkScale}%
    \pgfmathsetlengthmacro{\ms}{\msnum pt}%
    \node[\ternaryShapeStyle, minimum size=\ms, inner sep=0pt,
      draw=\ternaryColor!85!black, fill=\ternaryColor!18,
      line width=0.6pt, fill opacity=0.55, draw opacity=0.90
    ] at (\xx,\yy) {};
  }%
  \input{\XYZFILE}
  \endgroup

  \node[font=\scriptsize\bfseries, text=black!80, anchor=west] at (0.0,\PH-0.2) {Data vs Product};
\end{scope}

% ---------- Panel 4: Data (x) vs Database (y) (bottom-right) ----------
\begin{scope}[shift={(\PW+\G,0)}]
  \PanelFrame
  \AxesTwoD{Data}{Database}

  \begingroup
  \renewcommand{\xyzplotpoint}[7]{%
    \ternarySetColorFromLicense{#7}%
    \ternarySetShapeFromModel{#6}%
    \pgfmathsetmacro{\xx}{#4*\PW}%
    \pgfmathsetmacro{\yy}{#3*\PH}%
    \pgfmathsetmacro{\msnum}{2*(#5)*\MarkScale}%
    \pgfmathsetlengthmacro{\ms}{\msnum pt}%
    \node[\ternaryShapeStyle, minimum size=\ms, inner sep=0pt,
      draw=\ternaryColor!85!black, fill=\ternaryColor!18,
      line width=0.6pt, fill opacity=0.55, draw opacity=0.90
    ] at (\xx,\yy) {};
  }%
  \input{\XYZFILE}
  \endgroup

  \node[font=\scriptsize\bfseries, text=black!80, anchor=west] at (0.0,\PH-0.2) {Data vs DB};
\end{scope}

\end{tikzpicture}

\vspace{2pt}
{\footnotesize%\scriptsize
\begin{tabular}{@{}l@{\quad}l@{\qquad}l@{\quad}l@{}}
  \multicolumn{2}{@{}l@{}}{\textbf{Model (shape)}} &
  \multicolumn{2}{l@{}}{\textbf{License (color)}}\\
  \legIconUpTri & Property Graph Model &
  \textcolor{LicOpen}{\rule{1.6ex}{1.6ex}} & Open\\
  \legIconDownTri & RDF Data Model &
  \textcolor{LicProp}{\rule{1.6ex}{1.6ex}} & Proprietary\\
  \legIconCircle & Multiple Data Models &
  \textcolor{LicMixed}{\rule{1.6ex}{1.6ex}} & Mixed\\
  \legIconSquare & Alternative Data Models & & \\
\end{tabular}
}

\caption{(Top-left) 3D perspective view of Product/Database/Data scores. (Other panels) Pairwise projections. Marker shape encodes model family; color encodes license; marker size encodes overall coverage. The 3D panel provides an overview of the joint distribution; the three 2D projections support accurate pairwise comparison.}
\label{fig:xyz_and_pairwise}
\end{figure}
% END inlined from sections/graph_databases.tex

%%%%%%%%%%%%%%%%%%%%%%%%%%%%%%%%%%%%%%%%%%%%%%%%
%%%%%%%%%%%%%%%%%%%%%%%%%%%%%%%%%%%%%%%%%%%%%%%% RELATED WORK
%%%%%%%%%%%%%%%%%%%%%%%%%%%%%%%%%%%%%%%%%%%%%%%%

% BEGIN inlined from sections/related_work.tex
\section{Related Work}
\label{chapter:publications:survey:sec:related-work}

\normalsize
Graph databases have been analysed and reviewed in the literature before, including specific domains such as health~\citep{graph-db-clinical}, industry~\citep{graph-db-industry}, artificial intelligence~\citep{graph-db-isec}, life sciences~\citep{graph-db-life}, social networks~\citep{graph-db-social}, or software engineering~\citep{graph-db-se}. 
%Although by virtue of not being relational, graph databases are often included in the NoSQL database and Big data domains (on which there is extensive analysis work, e.g.,~\citep{nosql2011,nosql2012,nosql2014,noSQLSurveys2018,nosql2020}), in this survey we limit our analysis to other works exclusively addressing graph databases. 
Graph databases are often included in the NoSQL database and Big data domains, on which there is also  extensive analysis work~\citep{nosql2011,nosql2012,nosql2014,noSQLSurveys2018,nosql2020}. 
%In this survey we limit our analysis to other works exclusively addressing graph databases. 
Specifically, graph databases have been the subject of extensive research, covering various aspects such as system design, user-related features, scalability and transactions, among others. 
This section gives credit to, and addresses, other survey works in such areas.

\paragraph*{Foundational aspects}

Earlier works naturally tend to focus more on theoretical, formal and foundational aspects, with lower emphasis on system, scale and performance issues. 
An example of initial work is found in~\citep{angles2008survey}, laying the groundwork for graph databases. 
It offers a comprehensive overview of graph database modeling, focusing on data structures, query languages and integrity constraints, from foundational aspects to early developments. 
In light of such, it extensively reviews the developments at the time in graph database models, particularly their resurgence due to the need to manage graph-like information.

Another survey~\citep{Kaliyar2015} provides an overview of the different types of graph databases, as well as their typical applications, while assessing and comparing their models based on  a set of varied features, such as API and languages, property graph, hypergraph, query language, graph type.

More recently, the authors of~\citep{angles2008survey} delved in more detail over query languages in~\citep{angles2017foundations,angles2018g}, providing additional insights on  graph pattern matching and navigational expressions, and the study of graph structure, properties and motifs has been addressed in~\citep{survey-subgraphs-networking-21}.

\paragraph*{System design aspects}

Works dedicated to system design aspects tend to address more core aspects, such as architectural and programmatic ones, of each concrete graph database system, e.g., whether it is centralized or distributed, how data is organized, how queries are executed.

A taxonomy of graph database systems is proposed in~\citep{10.1145/3604932}, covering fundamental categories, associated graph models, data organization techniques, and aspects of data distribution and query execution.
In~\citep{Patil2018}, the authors review existing graph data computational techniques and research, offering insights into future research directions in graph database management.

The authors of~\citep{de2014model} propose a model-driven, system-independent methodology for the design of graph databases with emphasis on minimizing the amount of data accesses to answer queries.
Similarly, in~\citep{Erven2019}, the authors propose a diagram for graph databases modeling, demonstrating its effectiveness and compatibility in various case studies. 

\paragraph*{Transactions and Concurrency Models}

While not all graph storage solutions offer transactional semantics, there are several graph databases that do enforce transactional guarantees when manipulating data. This is also often addressed in broader surveys but has also been subject of some specific studies. 

Transaction management in cloud-based graph databases is discussed in \citep{Koloniari2015}, including different levels of transaction support and concurrency control protocols, data distribution issues, and replication protocols.
The work in \citep{conc-mamoth} specifically addresses 
the challenges of efficiently executing large-scale, long-running read-write transactions (termed 'mammoth transactions') in graph databases. These transactions are prevalent in graph workloads but are not adequately supported by many concurrency control protocols and benchmarks. 
%The paper identifies the unique demands of mammoth transactions, and authors suggest leveraging inherent graph properties and specialized protocols to enhance support for these transactions while maintaining strong isolation and high performance.

\paragraph*{Scalability Aspects}

Scalability is one the most important aspects in graph databases and one of the most sought after properties when designing or selecting one, and thoroughly studied. 
Vertical scalability is relevant when processing graphs in machines with large memory and number of cores. 
Horizontal scaling is increasingly relevant both in storage and in processing, because in the Big data era, graph data can grow to huge sizes that surpass the memory and processing power of single, yet powerful, machines. 
Often, specific mechanisms such as partitioning and indexing influence the scalability of graph databases and graph query processing. 

Scalability challenges in graph databases are discussed in \citep{Pokorny2015}, particularly in the context of large graphs and their querying in Big data environments. 
The authors note that when scaling over the network, these challenges are much harder in the case of graphs than in the case of simpler data models.
The authors of~\citep{Dayarathna2012} evaluate the scalability of the \texttt{Sca\-le\-Gra\-ph} library for large-scale graph analysis, providing insights into scalable graph processing.
There is also an in-depth study on the scalability (and performance) of a specific set of graph databases in~\citep{graph-db-isec}, including \dbNeoFRj, \dbJanusGraph, \dbMemgraph, \dbNebulaGraph, and \dbTigerGraph.
Finally, scalability is also globally addressed in ~\citep{besta-thousand-cores} with a comprehensive overview of the proposed \texttt{Gra\-ph} \texttt{Da\-ta\-ba\-se} \texttt{In\-ter\-fa\-ce} (\texttt{GDI}), focusing on how it addresses several key challenges in the development of graph databases such as high performance and scalability, inspired by best practices from the High-Performance Computing (HPC) landscape.

Specifically to partitioning, in~\citep{Pujol2010}, the authors discuss the challenges of scaling online social networks and propose a \texttt{So\-cial} \texttt{Par\-ti\-tion\-ing} \texttt{and} \texttt{Re\-pli\-ca\-tion} middleware (\texttt{SPAR}) to minimize replication overhead and enhance scalability. 
Further, in~\citep{Adoni2019}, the authors review current problems and challenges related to partitioning and processing graph-structured data in distributed systems, focusing on scalability and efficiency.

Indexing has also been surveyed in particular. 
A key graph processing operation is the reachability query, which determines whether a path exists between two vertices in a graph, considering plain and edge-labeled graphs. 
These queries can be resource-intensive, often requiring traversal of extensive graph portions, especially in large graphs. 
To address this and strive for efficient query processing, considerable research has focused on developing indexing techniques, creating what are known as reachability indexes, which have been surveyed in~\citep{graph-db-indexing} and~\citep{scale-graph-index-2}.
    
Finally, the authors of \citep{Roy2018}  present an analysis of various optimization techniques for horizontal scaling in the context of Big data, which is relevant also in the domain of graph databases.

\paragraph*{Other User Relevant Aspects}
 
There are other important aspects when selecting a graph database such as it being open-source, free or commercial, the programming language API, benchmark results, hardware optimizations, etc. 
Thus, we address here survey work dedicated to such analysis.  
 
An initial performance assessment of open-source graph databases is conducted in \citep{McColl2014}, with a study comparing the performance of 12 open-source graph databases using key graph algorithms on networks with up to 256 million edges, emphasizing the need for intuitive interfaces for both experts and novices in graph algorithm implementation. 
Referring back to \citep{graph-db-isec}, it describes the comparative performance evaluation of recent graph database systems such as \dbNeoFRj, \dbJanusGraph, \dbMemgraph, \dbNebulaGraph, and \dbTigerGraph.
The work in~\citep{Vaikuntam2014Evaluation} also evaluates several contemporary graph databases, both commercial and open-source, from a user-based (e.g., procurement, interface, pricing, support and maturity), system-based (e.g., query and storage distribution, indexing and transactional support), and empirical performance-based perspective with benchmarks.
Referring again to the work in \citep{Kaliyar2015}, its study also includes user related features such as freeness, portability, usability.
Furthermore, a  comprehensive evaluation of graph database systems with micro-benchmarks is described in~\citep{lissandrini2018beyond}, with more insights provided than those typically gathered with macro, application-like benchmarks.

Finally, tailoring specific hardware support, such as FPGAs, to non-relational databases (which includes graph databases), and its challenges and design decisions has been addressed in \citep{nosql-fpga}.

\paragraph*{Summary}

While there is extensive previous work on graph databases, some of it addresses very specific features or properties, often only of foundational nature (e.g., query languages, graph models), or just of distributed systems and scalability  aspects (e.g., partitioning, replication, distribution, indexing), only transactional and concurrency issues, or exhaustive performance benchmarks of a very limited set of systems. 
Other issues such as open-source availability, pricing, cloud and containerized offering, community support, schema support, GUI and CLI tools, high availability, automatic updates and live backups, and logging/auditing are seldom analyzed and when so, only for a handful of systems. 

We find that this work has a relevant place in this space by virtue of: i) its variety of features encompassed that cover commercial, architectural, system, programming model, scalability, and user interface topics, ii) the breadth of features and mechanisms addressed in each topic , iii) and the extent and recency of the graph databases that are analysed according to our criteria.
% END inlined from sections/related_work.tex

%%%%%%%%%%%%%%%%%%%%%%%%%%%%%%%%%%%%%%%%%%%%%%%%
%%%%%%%%%%%%%%%%%%%%%%%%%%%%%%%%%%%%%%%%%%%%%%%% CONCLUSION
%%%%%%%%%%%%%%%%%%%%%%%%%%%%%%%%%%%%%%%%%%%%%%%%

% BEGIN inlined from sections/conclusion.tex
\section{Conclusion}
\label{chapter:publications:survey:sec:conclusions}

Top researchers have said the future is big graphs~\citep{10.1145/3434642}, and in recent years we have continued to move towards this vision. 
Graph processing systems and graph databases have been designed, implemented and reinvented across academia and industry. 
As different industries have realized the opportunity of graph-based data manipulation, a sort of arms race was stimulated. 
Existing technology giants incorporated graph-oriented offerings in their cloud services while startups were created, simultaneously catering to business/developer needs while aiming to establish niches within the graph database ecosystem.

Throughout our research and engineering activities along the years, we observed these different initiatives manifesting, with commercial entities such as \dbNeoFRj\ producing educational material to explain graphs and help onboard new clients and developers~\citep{miller2013graph,neo4j_graph_db}.
Other graph databases followed suit, with introductory material being used to ease clients/developers into the concepts and need for graph-focused database solutions.
Parallel to this, the literature saw the addition of surveys on important topics pertaining to graph databases, such as graph database models and data management~\citep{angles2008survey, Kaliyar2015,Patil2018}; the intersection of industry and academia~\citep{graph-db-industry}; applied graph databases~\citep{graph-db-social}; transaction performance and benchmarking~\citep{dominguez2010survey,conc-mamoth,graph-db-bench-2}; feature-based graph database comparison~\citep{Vaikuntam2014Evaluation}; and graph query languages~\citep{Holzschuher:2013:PGQ:2457317.2457351,angles2017foundations,plantikowtowards}, among others.

We analyzed a plethora of graph database systems and surveys~\citep{mhedhbi2021lsqb,lissandrini2022knowledge,10.1145/3604932} with the purpose of highlighting the most relevant intricacies within the ecosystem.
At the same time, we presented an array of features to guide would-be developers and researchers in identifying the specific aspects of their use cases and the technologies available to them. 
Business and user needs call for the development of technology, which then stimulates the research vectors undertaken by researchers worldwide. 
This in turn feeds the entrepreneurial drive towards innovative graph database solutions. 
On an ending note, we believe graph databases will only continue to be more refined as more and more data exists, especially in the context of AI. 
The next ten years will paint an interesting scenario on the tendencies of existing commercial players and features.
% END inlined from sections/conclusion.tex

%%%%%%%%%%%%%%%%%%%%%%%%%%%%%%%%%%%%%%%%%%%%%%%%
%%%%%%%%%%%%%%%%%%%%%%%%%%%%%%%%%%%%%%%%%%%%%%%% DEBUGGING PAGE LAYOUT
%%%%%%%%%%%%%%%%%%%%%%%%%%%%%%%%%%%%%%%%%%%%%%%%

% If we need to print page dimensions and draw a rectangle
% for margin sanity-checking.
% Comment these two lines below if you don't need this debugging info.
%\clearpage
%\input{sections/debug_page_layout.tex}

\appendix%\label{chapter:publications:survey:sec:appendix}
\clearpage
\phantomsection
\addcontentsline{toc}{section}{Appendix}
% BEGIN inlined from tables/new-features-table-annex.tex
% Pretty feature description table (Part I) with light section breaks
\begin{table}[!t]
\centering
\caption{Description of graph database and systems analysis features (Part I).}
\label{new:table:feature_description_P1}

\footnotesize
\setlength{\tabcolsep}{6pt}
\renewcommand{\arraystretch}{1.15}

\begin{tabularx}{\textwidth}{@{}
  >{\raggedright\arraybackslash}p{3.6cm}
  >{\raggedright\arraybackslash}X
@{}}
\toprule
\textbf{Feature} & \textbf{Description} \\
\midrule

Active development &
Score 1 if the GDB has been updated within three months of the reference date ({\FeaturesReferenceDate}).
Score 0.5 if the GDB has been updated within 3--6 months of the reference date ({\FeaturesReferenceDate}).
Otherwise, score 0. \\

Commercial support &
Score 1 if there is a designated team that provides paid support for the GDB.
Otherwise, score 0. \\

Live community &
Score 1 if more than 75\% of the issues addressed by users/customers/developers have been addressed.
Score 0.75 if 50\%--75\% of the issues have been addressed.
Score 0.5 if 25\%--50\% of the issues have been addressed.
Score 0.25 if less than 25\% of the issues have been addressed.
Score 0 if no information is available. \\

Open source &
Score 1 if the source code is freely available with an open-source license (e.g., Apache-2.0, GPLv3, \dots).
Score 0.5 if the source code is available but commercially restrictive.
Otherwise, score 0. \\

Pricing &
Score 1 if at least starting or approximate prices are available on the GDB web page, or the GDB has an open license.
Otherwise, with no price information, score 0. \\

Trendiness &
Score 1 if the overall trend is upward.
Score 0.5 if it is broadly flat but non-zero.
Score 0 if it is downward or flat at zero. \\
% Score 1 if, for July 2022--July 2023, the Google Trends (\url{https://trends.google.com/}) of the GDB have had a constant or growing trend.
% Score 0 if the trend has been downward, or no data are available. \\
\addlinespace[1pt]
\midrule

Containerization &
Score 1 if a container image such as Docker is offered.
Otherwise, score 0. \\

Work as dedicated instance &
Score 1 if the GDB can be run as an individual instance.
Otherwise, if it is available exclusively in embedded mode, score 0. \\

Work as embedded &
Score 1 if it is possible to run the GDB embedded in the application (e.g., in the same JVM).
Otherwise, score 0. \\

Testing in-memory version &
Score 1 if there is an in-memory option of the GDB where data are stored in memory (not on disk) (e.g., for testing purposes).
Otherwise, score 0. \\

Operating on Linux &
Score 1 if Linux is explicitly supported.
Score 0.75 if the GDB code is open source.
Score 0.5 if a Docker image is provided.
Otherwise, score 0. \\

Operating on Windows &
Score 1 if Windows is explicitly supported.
Score 0.75 if the GDB code is open source.
Score 0.5 if a Docker image is provided.
Otherwise, score 0. \\

SaaS offering &
Score 1 if the provider offers its GDB as a service.
Otherwise, score 0. \\
\addlinespace[1pt]
\midrule

Automatic updates &
Score 1 if there are mechanisms to seamlessly update the GDB automatically.
Otherwise, score 0. \\

Client side caching &
Score 1 if the GDB provides a local cache in an application memory on the client side.
Otherwise, score 0. \\

Data versioning support &
Score 1 if the GDB supports versioning of stored data.
Otherwise, score 0. \\

Live backups &
Score 1 if the GDB supports live backups.
Otherwise, score 0. \\
\addlinespace[1pt]
\midrule

Cluster re-balancing &
Score 1 if the GDB is able to rebalance its original data distribution across the cluster efficiently.
Otherwise, score 0. \\

Data distribution &
Score 1 if the GDB enables data sharding (i.e., it can shard the data across several host servers).
Otherwise, score 0. \\

High-availability &
Score 1 if there are mechanisms to achieve high availability (including when the storage layer is graph-unaware, e.g., HBase, but provides HA mechanisms).
Otherwise, score 0. \\

Query distribution &
Score 1 if the GDB enables running a query over distributed data in a distributed manner (assigning the query to the correct partition and/or assembling partial results into a correct result).
Otherwise, score 0. \\

Replication support &
Score 1 if the GDB has replication mechanisms.
Otherwise, score 0. \\
\addlinespace[1pt]
\midrule

Data types defined &
Score 1 if a composite data type is provided (in any form).
Score 0.5 if only basic data types are enabled.
Otherwise, if only one data type is available (e.g., all data as String), score 0. \\

%\bottomrule
\end{tabularx}
\end{table}

% Pretty feature description table (Part II) with light section breaks
\begin{table}[!t]
\centering
\caption{Description of graph database and systems analysis features (Part II).}
\label{new:table:feature_description_P2}

\footnotesize
\setlength{\tabcolsep}{6pt}
\renewcommand{\arraystretch}{1.15}

\begin{tabularx}{\textwidth}{@{}
  >{\raggedright\arraybackslash}p{3.6cm}
  >{\raggedright\arraybackslash}X
@{}}
\toprule
\textbf{Feature} & \textbf{Description} \\
\midrule

Logging/Auditing &
Score 1 if the GDB logs important events.
Otherwise, score 0. \\

Object-graph mapper &
Score 1 if there exists a tool for automatic data conversion between the GDB and an object-oriented programming language.
Otherwise, score 0. \\

Reactive programming &
Score 1 if the GDB supports reactive streams that provide asynchronous processing.
Otherwise, score 0. \\

Documentation up-to-date &
Score 1 if the documentation has been updated for the last version of the GDB.
Score 0.5 if it exists but has not been updated for the last version.
Otherwise, score 0. \\
\addlinespace[1pt]
\midrule

Binary protocol &
Score 1 if the graph database provides an option to communicate via a binary protocol.
Otherwise, score 0. \\

CLI &
Score 1 if the GDB can receive commands via a CLI interface.
Otherwise, score 0. \\

GUI &
Score 1 if the GDB provides a visual interface.
Otherwise, score 0. \\

Multi-database &
Score 1 if the GDB is a multi-model database (i.e., you can also use it as, e.g., a document store, key/value store, or object store).
Otherwise, score 0. \\

Graph-native data &
Score 1 if the GDB is graph-native, i.e., designed to both store and process data as a graph.
Otherwise, if the GDB provides only a graph abstraction over a key-value store, score 0. \\

REST API &
Score 1 if a REST endpoint is offered to interact with the database.
Otherwise, score 0. \\

Query language &
Score 1 if the GDB can be queried with Cypher/Gremlin or any other widely accepted query language for GDBs.
Score 0.5 if a non-standard/proprietary query language is provided.
Otherwise, if no query language is provided, score 0. \\
\addlinespace[1pt]
\midrule

Granular locking &
Score 1 if the GDB allows locking of specific objects (usually nodes).
Otherwise, if no specific-object locking mechanism is implemented, score 0. \\

Multiple isolation levels &
Score 1 if more than one higher-level isolation level is available (i.e., other than ``read uncommitted'').
Score 0.5 if only one higher-level isolation level is available.
Otherwise, if only ``read uncommitted'' is available, score 0. \\

Read committed transaction &
Score 1 if a ``read committed'' transaction is available.
Otherwise, score 0. \\

Transaction support &
Score 1 if the GDB supports transactions.
Otherwise, score 0. \\
\addlinespace[1pt]
\midrule

Constraints &
Score 1 if constraints can be set in the GDB.
Score 0.5 if only uniqueness constraints are available.
Otherwise, score 0. \\

Schema support &
Score 1 if there is a direct tool for defining a schema.
Score 0.5 if only constraints and triggers are available.
Otherwise, score 0. \\

Secondary indexes &
Score 1 if any index beyond the primary index is supported (e.g., single-property, composite, vertex-centric, full-text, etc.).
Otherwise, score 0. \\

Server side procedures &
Score 1 if the GDB enables server-side (stored) procedures.
Otherwise, score 0. \\

Triggers &
Score 1 if the GDB enables server-side events (triggers).
Otherwise, score 0. \\
\addlinespace[1pt]
\midrule

Authentication &
Score 1 if the GDB supports some authentication method.
Otherwise, score 0. \\

Authorization &
Score 1 if the GDB provides any means for authorization (e.g., roles or privileges).
Otherwise, score 0. \\

Data encryption &
Score 1 if the GDB provides a possibility to encrypt stored data.
Otherwise, score 0. \\
\addlinespace[1pt]
%\midrule

\bottomrule
\end{tabularx}
\end{table}
% END inlined from tables/new-features-table-annex.tex

% BEGIN inlined from tables/active_development/2_print_fact_table_latex/2_database_active_dev_fact_table-commands-citations.tex
% Auto-generated by 2_print_fact_table_latex.py

% Required LaTeX packages: booktabs, adjustbox, array

% Optional citation package support: natbib or a compatible \citep fallback

% One row per database; contextual secondary facts are folded into the supporting-fact column when useful.

\begin{table}[!t]
\centering
\caption{Database active-development and lifecycle facts (Part 1 of 5)}
\label{table:annex_fact_1}

\small
\setlength{\tabcolsep}{4pt}
\renewcommand{\arraystretch}{1.12}

\begin{adjustbox}{max width=\textwidth}
\begin{tabular}{@{}>{\raggedright\arraybackslash}p{3.35cm}>{\raggedright\arraybackslash}p{2.25cm}>{\raggedright\arraybackslash}p{1.75cm}>{\raggedright\arraybackslash}p{8.15cm}@{}}
\toprule
Database & Evidence role & Evidence date & Supporting fact \\
\midrule
    \dbAgensGraph\ \citep{agensgraph,agensgraph_github} & Active development & 2025-11-28 & GitHub non-trivial commit 93725359 on branch main: chore: fix warning in preproc build \\
    \dbAlibabaGraphDatabase\ \citep{tugraph_graphdb} & No recent evidence & 2025-03-13 & GitHub release: Version 4.5.2. Newer post-cutoff activity was excluded from snapshot scoring. \\
    \dbAllegroGraph\ \citep{allegrograph} & Active development & 2025-12-17 & The page lists dated release links in an unordered list under 'Archived AllegroGraph Releases', for example 'AllegroGraph 8.4.3 [2025-12-17]'. Also: Documentation: Documentation page for AllegroGraph displaying an update date used as… \\
    \dbAltairGraphLakehouse\ \citep{anzodb} & Active development & 2025 & The page lists version tags together with 'Last pushed' timestamps. Also: Rebranding: This source is used as rebranding evidence, not as active-development evidence. \\
    \dbAmazonNeptune\ \citep{neptune_github_updated,neptune_github,bebee2018amazon} & Active development & 2025-12-18 & This is strong public evidence of ongoing active development and is suitable for refresh-based date extraction relative to a reference date. \\
    \dbHugeGraph\ \citep{apache_hugegraph_github} & Active development & 2025-11-16 & GitHub release: Apache HugeGraph 1.7.0 (Release). Newer post-cutoff activity was excluded from snapshot scoring. \\
    \dbArangoDB\ \citep{arangodb_github,DBLP:conf/data/FernandesB18} & Active development & 2025-12-24 & GitHub non-trivial commit 2c5721f3 on branch devel: Enable vector index tests for Go driver (\#22205). Newer post-cutoff activity was excluded from snapshot scoring. \\
    \dbArcadeDB\ \citep{arcadedb_github} & Active development & 2025-12-09 & GitHub release: 25.11.1. Newer post-cutoff activity was excluded from snapshot scoring. \\
    \dbAzureCosmosDB\ \citep{cosmos_github,paz2018introduction} & No recent evidence & 2023-03-28 & GitHub repository activity timestamp \\
    \dbBangDB\ \citep{bangdb_github} & Active development & 2025-12-06 & GitHub non-trivial commit b9b1630c on branch master: Update install\_bangdb.sh \\
    \dbBlazegraph\ \citep{blazegraph,blazegraph_github} & Discontinuation & -- & Repository is archived. Also: Rebranding: W3 Semantic Web Standards wiki page stating 'Blazegraph (Formerly Bigdata®)' and t… \\
    \dbBrightstarDB\ \citep{brightstardb_github} & No recent evidence & 2023-05-31 & GitHub repository activity timestamp \\
\bottomrule
\end{tabular}
\end{adjustbox}
\end{table}

\begin{table}[!t]
\centering
\caption{Database active-development and lifecycle facts (Part 2 of 5)}
\label{table:annex_fact_2}

\small
\setlength{\tabcolsep}{4pt}
\renewcommand{\arraystretch}{1.12}

\begin{adjustbox}{max width=\textwidth}
\begin{tabular}{@{}>{\raggedright\arraybackslash}p{3.35cm}>{\raggedright\arraybackslash}p{2.25cm}>{\raggedright\arraybackslash}p{1.75cm}>{\raggedright\arraybackslash}p{8.15cm}@{}}
\toprule
Database & Evidence role & Evidence date & Supporting fact \\
\midrule
    \dbByteGraph\ \citep{li2022bytegraph,bytegraph_queries_github} & Active development & 2026 & Useful as indirect active-development evidence for 2026. \\
    \dbChronoGraph\ \citep{haeusler2017chronograph,chronograph_github} & Active development & 2025-11-03 & GitHub non-trivial commit 2f7cbded on branch master: Merge pull request \#3 from Txture/renovate/all-minor-updates. Newer post-cutoff activity was excluded from snap… \\
    \dbCrayGraphEngine\ \citep{rickett2018loading} & Active development & 2021 & Useful as indirect active-development evidence for 2021. \\
    \dbDataStaxEnterprise\ \citep{datastax_github,datastax} & Active development & 2026-01-19 & The page includes release notes for 6.9.18 dated 2026-01-19, providing current public evidence that DSE is still maintained. \\
    \dbDgraph\ \citep{dgraph_graphql,dgraph_github} & Active development & 2025-12-12 & GitHub release: v25.1.0. Newer post-cutoff activity was excluded from snapshot scoring. \\
    \dbFaunaDB\ \citep{faunadb,faunadb_graphdb} & No recent evidence & 2025-05-03 & GitHub repository activity timestamp \\
    \dbFluree\ \citep{fluree,fluree_github} & Active development & 2025-12-17 & GitHub non-trivial commit 9fded201 on branch main: cljfmt fix. Newer post-cutoff activity was excluded from snapshot scoring. \\
    \dbGTran\ \citep{chen2022g,gtran_github} & No recent evidence & 2022-04-06 & GitHub repository activity timestamp \\
    \dbGaffer\ \citep{gaffer_github} & Discontinuation & -- & Repository is archived. \\
    \dbGalaxybase\ \citep{tong2024galaxybase} & Active development & 2023-06-06 & Official Galaxybase release-notes page. \\
    \dbGoogleCayley\ \citep{cayley_github} & Active development & 2025-11-22 & GitHub repository activity timestamp \\
    \dbGraphflow\ \citep{kankanamge2017graphflow,graphflow_github} & Needs review & -- & Github repo check returned HTTP 404. \\
\bottomrule
\end{tabular}
\end{adjustbox}
\end{table}

\begin{table}[!t]
\centering
\caption{Database active-development and lifecycle facts (Part 3 of 5)}
\label{table:annex_fact_3}

\small
\setlength{\tabcolsep}{4pt}
\renewcommand{\arraystretch}{1.12}

\begin{adjustbox}{max width=\textwidth}
\begin{tabular}{@{}>{\raggedright\arraybackslash}p{3.35cm}>{\raggedright\arraybackslash}p{2.25cm}>{\raggedright\arraybackslash}p{1.75cm}>{\raggedright\arraybackslash}p{8.15cm}@{}}
\toprule
Database & Evidence role & Evidence date & Supporting fact \\
\midrule
    \dbHGraphDB\ \citep{hgraphdb_github} & Active development & 2025-12-28 & GitHub non-trivial commit 22085d9f on branch master: Bump org.apache.maven.plugins:maven-release-plugin from 3.1.1 to 3.2.0 (\#452). Newer post-cutoff activity was e… \\
    \dbHyperGraphDB\ \citep{hypergraphdb_not_distributed,iordanov2010hypergraphdb} & Needs review & -- & It lists a latest release labeled 1.3, but no clear recent update date was visible during checking. Manual review is still needed. \\
    \dbIBMSystemG\ \citep{ibm_suite_g} & Discontinuation & 2014 & The URL is no longer reliably accessible; secondary sources list System G with end year 2014 and current survey material treats the product as discontinued. \\
    \dbJanusGraph\ \citep{janus_github,janus,graph-db-isec} & Active development & 2025-11-21 & GitHub repository activity timestamp \\
    \dbKatanaGraph\ \citep{katana_graph_github} & No recent evidence & 2022-08-30 & GitHub release: v0.4.0 \\
    \dbKuzu\ \citep{kuzu_github} & Discontinuation & -- & Repository is archived. \\
    \dbLiveGraph\ \citep{zhu2019livegraph,livegraph_github} & No recent evidence & 2021-04-05 & GitHub repository activity timestamp \\
    \dbMemgraph\ \citep{memgraph_github} & Active development & 2025-12-23 & GitHub release: v3.7.2. Newer post-cutoff activity was excluded from snapshot scoring. \\
    \dbMillenniumDB\ \citep{VrgocRAAABHNRR23,millenniumdb_github} & Active development & 2025-12-16 & GitHub non-trivial commit 87743272 on branch dev: Feat/ssl http websocket (\#101). Newer post-cutoff activity was excluded from snapshot scoring. \\
    \dbNebulaGraph\ \citep{nebulagraph_github,gqlNebula} & Active development & 2025-10-22 & GitHub repository activity timestamp \\
    \dbNeoFRj\ \citep{10.1145/3604932,Holzschuher:2013:PGQ:2457317.2457351,neo4j_graphql_cypher,neo4j_graphql,neo4j_github,neo4j_apoc_lib,DBLP:conf/data/FernandesB18,Pokorny2015,miller2013graph,Webber:2012:PIN:2384716.2384777,graph-db-isec,graph-db-integrity-17,gqlNeo} & Active development & 2025-10-23 & GitHub non-trivial commit 343dcdae on branch 2025.10: CVE-2025-11602 Fix init of self-allocated Bolt BitMask (). Newer post-cutoff activity was excluded from snapsh… \\
    \dbObjectivityDB\ \citep{objectivitydb} & Needs review & -- & This entry is therefore kept as manual-review evidence rather than being labeled active or discontinued. \\
\bottomrule
\end{tabular}
\end{adjustbox}
\end{table}

\begin{table}[!t]
\centering
\caption{Database active-development and lifecycle facts (Part 4 of 5)}
\label{table:annex_fact_4}

\small
\setlength{\tabcolsep}{4pt}
\renewcommand{\arraystretch}{1.12}

\begin{adjustbox}{max width=\textwidth}
\begin{tabular}{@{}>{\raggedright\arraybackslash}p{3.35cm}>{\raggedright\arraybackslash}p{2.25cm}>{\raggedright\arraybackslash}p{1.75cm}>{\raggedright\arraybackslash}p{8.15cm}@{}}
\toprule
Database & Evidence role & Evidence date & Supporting fact \\
\midrule
    \dbOntotextGraphDB\ \citep{ontotext_graphdb} & Active development & -- & The page documents the current 11.3 release series and provides strong public evidence of ongoing active development. \\
    \dbOracleSpatialAndGraph\ \citep{oracledb_graph_features} & Active development & -- & It states that Oracle Graph Server and Client is released four times a year and lists new features for release 26.1. \\
    \dbOrientDB\ \citep{10.1145/3604932,orientdb_graphql_gremlin,DBLP:conf/bncod/0001DLT21,orientdb_github,DBLP:conf/data/FernandesB18,Pokorny2015} & Active development & 2025-12-24 & GitHub release: 3.2.48. Newer post-cutoff activity was excluded from snapshot scoring. \\
    \dbPandaDB\ \citep{zhao2021pandadb,pandadb_github} & No recent evidence & 2022-06-04 & GitHub repository activity timestamp \\
    \dbRedisGraph\ \citep{redis_github,cailliau2019redisgraph} & No recent evidence & 2025-07-21 & GitHub non-trivial commit 5784cb8b on branch master: Update SSPLv1.txt \\
    \dbSAPHanaGraph\ \citep{hwang2018graph,hana_github,rudolf2013graph} & Active development & 2025 & Useful as active-development evidence for graph capabilities delivered as part of SAP HANA Cloud. \\
    \dbSparksee\ \citep{sparksee_former_dex,martinez2011dex} & Active development & 2024-09-04 & This is reasonable active-development evidence, but it is weaker than a dedicated release-notes page. \\
    \dbStardog\ \citep{stardog} & Active development & -- & Official Stardog release-notes page. \\
    \dbStellarDB\ \citep{stellardb} & Active development & 2026-01-21 & Useful as public active-development evidence. \\
    \dbSurrealDB\ \citep{surrealdb_github} & Active development & 2025-12-30 & GitHub non-trivial commit 9ac7269b on branch main: Reduce instrumentation overhead in query record collectors (\#6710). Newer post-cutoff activity was excluded from… \\
    \dbTAO\ \citep{cheng2021ramp,bronson2013tao} & Active development & -- & This Meta Research publication presents transaction support layered on TAO, providing direct evidence that TAO was still being actively evolved. \\
    \dbTerminusDB\ \citep{terminusdb_github} & Active development & 2025-12-16 & GitHub release: TerminusDB Server v12.0.2. Newer post-cutoff activity was excluded from snapshot scoring. \\
\bottomrule
\end{tabular}
\end{adjustbox}
\end{table}

\begin{table}[!t]
\centering
\caption{Database active-development and lifecycle facts (Part 5 of 5)}
\label{table:annex_fact_5}

\small
\setlength{\tabcolsep}{4pt}
\renewcommand{\arraystretch}{1.12}

\begin{adjustbox}{max width=\textwidth}
\begin{tabular}{@{}>{\raggedright\arraybackslash}p{3.35cm}>{\raggedright\arraybackslash}p{2.25cm}>{\raggedright\arraybackslash}p{1.75cm}>{\raggedright\arraybackslash}p{8.15cm}@{}}
\toprule
Database & Evidence role & Evidence date & Supporting fact \\
\midrule
    \dbTigerGraph\ \citep{tigergraph,deutsch2019tigergraph} & Active development & 2026-03-11 & Official TigerGraph Savanna release-notes page. \\
    \dbTypeDB\ \citep{typedb_github} & Active development & 2025-12-13 & GitHub release: TypeDB 3.7.2. Newer post-cutoff activity was excluded from snapshot scoring. \\
    \dbUltipa\ \citep{ultipa_graph} & Active development & 2025-08-11 & Official Ultipa database release-notes page. \\
    \dbVirtuoso\ \citep{10.1145/3604932,virtuoso_github,erling2012virtuoso} & Active development & 2025-10-15 & GitHub release: Virtuoso Open Source Edition v7.2.16.1. Newer post-cutoff activity was excluded from snapshot scoring. \\
    \dbWeaver\ \citep{DBLP:journals/pvldb/DubeyHES16,weaver_github} & No recent evidence & 2016-11-14 & GitHub repository activity timestamp \\
    \dbZipG\ \citep{khandelwal2017zipg,zipg_github} & No recent evidence & 2021-09-01 & GitHub repository activity timestamp \\
\bottomrule
\end{tabular}
\end{adjustbox}
\end{table}
% END inlined from tables/active_development/2_print_fact_table_latex/2_database_active_dev_fact_table-commands-citations.tex

\clearpage

%%%%%%%%%%%%%%%%%%%%%%%%%%%%%%%%%%%%%%%%%%%%%%%%
%%%%%%%%%%%%%%%%%%%%%%%%%%%%%%%%%%%%%%%%%%%%%%%% BIBLIOGRAPHY
%%%%%%%%%%%%%%%%%%%%%%%%%%%%%%%%%%%%%%%%%%%%%%%%
% Helpful sites: 
% How to not mess up your bibliographies with Bibtex
% http://serialmentor.com/blog/2015/10/2/Bibtex
%
% Ignore specific URL fields in biber?
% http://tex.stackexchange.com/questions/133809/ignore-specific-url-fields-in-biber
% 
% How can I use BibTeX to cite a web page?
% http://tex.stackexchange.com/questions/3587/how-can-i-use-bibtex-to-cite-a-web-page

%% Please use bibtex, 

\phantomsection
\addcontentsline{toc}{section}{References}
\bibliography{full_no_howpublished}

\end{document}